\documentclass[aps,pra,twocolumn,superscriptaddress,nofootinbib,floatfix,10pt]{revtex4-2}

\usepackage[utf8]{inputenc}
\usepackage[T1]{fontenc}
\usepackage{amsmath,amssymb,amsthm,bm}
\usepackage{graphicx}
\usepackage{eso-pic}
\usepackage{booktabs}
\usepackage{xcolor}
\usepackage{tikz}
\usetikzlibrary{arrows.meta,positioning,calc,fit,backgrounds}
\usepackage{algorithm}
\usepackage[noend]{algpseudocode}
\usepackage[colorlinks=true,linkcolor=blue!50!black,citecolor=blue!50!black,urlcolor=blue!50!black]{hyperref}

\newtheorem{proposition}{Proposition}

\newcommand{\Can}{\mathrm{Can}}
\newcommand{\cz}{\mathrm{cz}}
\newcommand{\qs}{\texttt{quspect}}

\begin{document}
\AddToShipoutPictureFG*{\AtPageUpperLeft{\put(\LenToUnit{\paperwidth-0.73in-3.3cm},\LenToUnit{-0.543in}){\includegraphics[width=3.3cm]{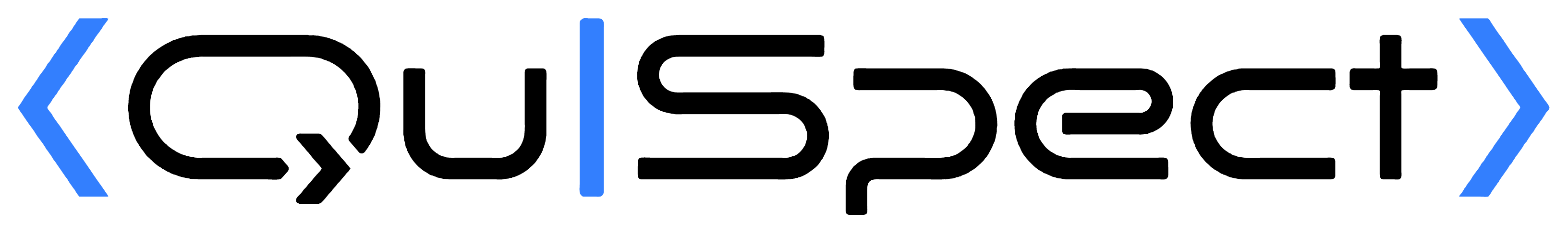}}}}

\title{Circuit compilation in the \qs{} quantum SDK: selecting routings after coupling-constrained resynthesis}

\author{Xusheng Xu}
\affiliation{QuSpect Technology Co., Ltd., Beijing, China}
\author{Runqing Zhang}
\affiliation{QuSpect Technology Co., Ltd., Beijing, China}
\author{Hongtao Xue}
\affiliation{QuSpect Technology Co., Ltd., Beijing, China}
\author{Jize Han}
\affiliation{China Mobile (Suzhou) Software Technology Co., Ltd., Suzhou, China}
\author{Jingwei Wen}
\affiliation{China Mobile (Suzhou) Software Technology Co., Ltd., Suzhou, China}
\author{Ruoyu Zhou}
\affiliation{QuSpect Technology Co., Ltd., Beijing, China}
\author{Lei Jin}
\affiliation{Beijing University of Posts and Telecommunications, Beijing, China}
\author{Xiaoxiao Hu}
\email[Corresponding author: ]{huxiaoxiao@quspect.com}
\affiliation{QuSpect Technology Co., Ltd., Beijing, China}
\author{Gui-Lu Long}
\email[Corresponding author: ]{gllong@tsinghua.edu.cn}
\affiliation{State Key Laboratory of Low-Dimensional Quantum Physics and Department of Physics, Tsinghua University, Beijing 100084, China}
\affiliation{Frontier Science Center for Quantum Information, Beijing 100084, China}
\affiliation{Beijing Academy of Quantum Information Sciences, Beijing 100193, China}

\date{\today}

\begin{abstract}
\qs{} is an open-source quantum computing SDK that provides circuit construction, C++ state-vector simulation with adjoint gradients, support for variational and combinatorial-optimization algorithms, Hamiltonian simulation, control-pulse envelopes and backends for several cloud quantum processors. Here we describe and evaluate its circuit compiler. Routing and synthesis interact: inserting a SWAP can change the circuit structure that later optimization exploits. At its highest optimization level, the compiler evaluates a screened set of routing candidates by their cost after coupling-constrained resynthesis. The pipeline combines parity-state A* search for small CNOT--phase regions, heuristic synthesis for larger regions, and numerical three-qubit instantiation with a per-block $10^{-9}$ check. For a fixed phase polynomial, A* certifies the minimum CX count when it reaches a goal within budget. On an evaluation set of 55 circuits from 41 arXiv papers on five topologies, the compiler uses 0.616 times as many two-qubit gates as Qiskit O3 (95\% circuit-cluster bootstrap interval 0.555--0.679) at 0.592 times its depth. BQSKit level~2 reaches 0.853 and tket 0.930. On a 17-qubit superconducting processor, the compiled circuits had the fewest CZ gates in all six scenarios of the main run. Ablations assess the contributions within the integrated pipeline.
\end{abstract}

\maketitle

\section{Introduction}\label{sec:introduction}

\qs{} is an open-source quantum computing SDK~\cite{quspect2026} (\url{https://github.com/QuSpect/quspect}, Apache-2.0 license). Besides circuit construction, it provides C++ state-vector simulation with adjoint gradients, objectives for variational algorithms, QUBO and Ising tools for combinatorial optimization, Hamiltonian simulation, control-pulse envelopes and backends for several cloud quantum processors. By default, these backends compile every submitted circuit with the SDK's circuit compiler (\texttt{quspect.transpiler}), so the circuits built or generated with the other modules reach a device through this compiler. This paper is about the compiler alone. Compiling a quantum circuit for a superconducting device means choosing an initial placement of logical qubits (layout), inserting SWAP gates so that every two-qubit gate acts on coupled qubits (routing), rewriting the result into the native gate set (synthesis), and ordering it to keep the circuit shallow (scheduling). Layout and routing~\cite{siraichi2018allocation}, the underlying quadratic assignment~\cite{sahni1976pcomplete}, token swapping~\cite{miltzow2016token} and resource-constrained scheduling~\cite{blazewicz1983rcpsp} are NP-hard in general, and exact synthesis searches grow exponentially with the number of qubits~\cite{amy2013mitm,davis2020qsearch}, so production compilers combine heuristics~\cite{li2019sabre,zou2024lightsabre,qiskit2024,sivarajah2020tket,cowtan2019routing}, exact methods for small instances~\cite{wille2019exact,tan2020olsq,lin2023olsq2,molavi2022satmap,zhang2021toqm,shaik2024qsynth}, and numerical synthesis~\cite{davis2020qsearch,smith2023leap,bqskit2021}. On noisy hardware, the two-qubit gate count and circuit depth largely determine output fidelity.

The cost of a routed SWAP depends on its surrounding gates. Next to a two-qubit block, it can merge into that block; next to CNOTs and phase gates, it is another linear map in a CNOT--phase region that can be resynthesized as a whole. The best routing therefore depends on subsequent synthesis. Two measurements illustrate this dependency. On 150 small routing instances, the SWAP-minimal routings found by an exhaustive A* search had higher costs after merging than the routings produced by a previous version of the compiler in 94 cases, by 811 CZ gates in total, although the compiler's routings used 18\% more SWAPs (Sec.~\ref{sec:optimality}). Among cases in which the superoptimizer GUOQ~\cite{xu2025guoq}, applied after Qiskit's routing, still outperformed that version, 65\% of its savings came from rewrites inside CNOT--phase regions of the \emph{routed} circuit, which contain routed SWAPs and Toffolis synthesized for their assigned physical couplings (Sec.~\ref{sec:postroute}). The compiler described here incorporates this dependency: at its highest level, it routes each circuit in several ways, resynthesizes screened routing candidates under the coupling constraints, and selects based on the resulting two-qubit count.

The \qs{} compiler integrates known techniques for each of these stages with a few new ones, and we evaluate it against five other compilers. Its design follows two principles. First, every stage is judged by the cost that remains after the later stages have run. A SWAP that merges into an adjacent two-qubit block is nearly free, so routings are compared by their gate count after merging and resynthesis, not by their SWAP count. The highest optimization level evaluates a screened set of routing candidates by their two-qubit count after coupling-constrained resynthesis. Second, the compiler casts several stages as classical combinatorial problems and searches small pieces using A* with admissible bounds when a budget allows. Short windows of a routing are re-routed under a merge-aware cost. Small CNOT--phase regions of a routed circuit are resynthesized by a shortest-path search over parity states restricted to the coupling graph, which certifies the minimum CX count for the region's fixed phase polynomial when it reaches a goal within its budget. Other formulations are solved heuristically: layered routing on a path becomes a chain of linear ordering problems, solved by local search, and scheduling becomes a resource-constrained project scheduling problem over the commutation structure of the circuit, solved by schedule-generation heuristics.

Our results support a modest reading of where the gains come from. For the previous version of the compiler (commit \texttt{8ec73d8}), a leave-one-out ablation shows that no single component accounts for most of the 17--25\% lead it had over Qiskit O3. A grouped ablation shows why: a reference pipeline with the structure of Qiskit O3 matches Qiskit O3 to within a few percent on three of four suites. The lead comes from the compiler's routing extensions, synthesis passes and candidate search together, with the routing components largely substituting for one another. The final round of development added post-routing resynthesis of CNOT--phase regions and three-qubit blocks; with it, the lead is 23--41\% on the development suites and 38\% on the arXiv evaluation set (previous version: 21\%). In an ablation of the final compiler, removing the final-round components at once increases the two-qubit gate count by 23\% on Standard-42 and 27\% on the arXiv set, more than the sum of their individual effects. These components take over part of the role of the earlier routing and synthesis components. These ablations assess contributions within the integrated pipeline; the marginal effect of the ranking objective on an identical candidate pool was not measured. Where an optimum or a valid bound can be computed, the routings of the previous version were 1.2\% above the same-layout optimum under the merge-aware cost, layered routing on paths reached the optimum of its model in 14 of 15 instances, and the schedules were within 0.3\% of a critical-path bound. The final version ranks routings after post-routing resynthesis, and its selected routings are further from the merge-aware optimum (556 instead of 90 CZ gates above it on the same small instances) while its outputs have fewer gates; this comparison also changes compiler components, so it does not isolate the selection rule. The more specialized techniques give large gains on the circuits whose structure they target (for example, 37\% on QV-100 on a line) and little elsewhere.

The compiler was developed iteratively against development suites, which invites overfitting to the circuits, topologies and opponents used during development. We therefore also evaluate it on 55 circuits extracted from recent arXiv papers, which broaden the source distribution beyond the development suites, against two compilers (Cirq and QMAP) that were not used during development, and in a hardware experiment. The arXiv set is an evaluation set, not an untouched test set: it was evaluated before the final round of development, and its aggregate results informed the final development direction. That round targeted BQSKit and the superoptimizer GUOQ on a separate 157-circuit development set with no circuit identical to one in the arXiv set (Sec.~\ref{sec:devprocess}). The final evaluation therefore includes development-time exposure to aggregate arXiv results. The evaluated version of the compiler (commit \texttt{be4ecda}) was completed on 8 November 2026 and released in \qs{} v0.6.3 on 12 November 2026~\cite{quspect2026}; the release contains later changes, which the reproducibility statement lists.

\paragraph*{Contributions.}
\begin{enumerate}
\item The design of the \qs{} compiler, a complete mapping compiler within a general-purpose SDK (Fig.~\ref{fig:pipeline}, Table~\ref{tab:features}). Numerically computed rewrites are accepted only after the unitary of their emitted gates matches the target up to global phase with a tolerance of $10^{-9}$ per matrix element (for numerically instantiated three-qubit blocks this is a per-block check); symbolic rewrites (phase folding, cancellation) are exact by construction, and large CNOT--phase regions are checked by tolerance-based re-extraction of their affine map and phase polynomial (Sec.~\ref{sec:techniques}).
\item Stage-level techniques, each presented in relation to its closest prior work: a merge-aware window large-neighborhood search (LNS) for routing with admissible A* bounds (Alg.~\ref{alg:lns}); a reduction of layered routing of SU(4) circuits on paths to a chain of linear ordering problems with Kendall-$\tau$ costs, extended to rings (Alg.~\ref{alg:kendall}); trace-monoid depth scheduling with block-level commutation and separate optimization of two-qubit and total depth; and post-routing, coupling-constrained resynthesis of CNOT--phase regions (A* on regions of up to six wires, beam search on larger ones) combined with numerical instantiation of three-qubit blocks under a per-block $10^{-9}$ check, and the selection among screened routing candidates after this resynthesis (Sec.~\ref{sec:postroute}).
\item Gap measurements for these stages against exhaustive A*, OLSQ2, QUEKO and QUBIKOS, exact dynamic programs, integer programs and lower bounds (Sec.~\ref{sec:optimality}), and leave-one-out and grouped ablations of 14 components on four suites and of the ten components added in the final round on five suites (Sec.~\ref{sec:ablation}).
\item An evaluation on 55 circuits from 41 arXiv papers (the arXiv evaluation set), on 42 standard and 20 large circuits and a synthetic suite, under three gate sets and on five topologies plus 127- and 156-qubit devices, against seven configurations of five other compilers (Sec.~\ref{sec:eval}). Gate-level superoptimizers applied after Qiskit's routing, learned transpiler passes and the Benchpress circuits are evaluated in addition. Comparative statements refer to the tested configurations. We include the cases where the compiler loses: on the arXiv set, it loses in 11 of 275 cases against the best tested opponent configuration, mostly to BQSKit (level~2), whose numerical resynthesis of three-qubit blocks to a tolerance $\epsilon$ finds a few more gates; on large devices, the compiler loses to Qiskit on random Clifford circuits and some circuits under the IBM basis.
\item A hardware study on a 17-qubit superconducting chip with two coupling graphs, nine circuit families and three seeds, analyzed with Wilson intervals, a hierarchical bootstrap and a two-sided Benjamini--Hochberg procedure (Sec.~\ref{sec:hw}).
\end{enumerate}
The global beam-search router, which was added to the compiler before the final round and is part of the evaluated version, is described and evaluated in its own section (Sec.~\ref{sec:beam}).

\section{Background and related work}
\label{sec:background}

\paragraph*{Layout and routing.}
Let $G_C=(V_P,E_C)$ be the coupling graph with shortest-path distance $d$, and let a circuit on logical qubits $V_L$ have interaction multigraph $G_I$. A layout is an injection $\pi:V_L\to V_P$. Routing inserts SWAPs, which change $\pi$ over time, until every two-qubit gate $(i,j)$ is executed while $(\pi(i),\pi(j))\in E_C$. SABRE~\cite{li2019sabre} routes greedily. It scores candidate SWAPs on a front layer $F$ and a lookahead (``extended'') set, and it chooses the layout by forward--backward passes. LightSABRE~\cite{zou2024lightsabre} adds a release valve against livelock and many randomized layout and routing trials; it is the default router of Qiskit~\cite{qiskit2024}. tket~\cite{sivarajah2020tket,cowtan2019routing} uses graph placement followed by routing. MQT QMAP~\cite{zulehner2019astar,wille2019exact,wille2023qmap} provides an A*-based heuristic mapper and an exact SAT-based mapper. OLSQ/OLSQ2~\cite{tan2020olsq,lin2023olsq2} and SATMap~\cite{molavi2022satmap} formulate layout synthesis as constraint satisfaction; SATMap also routes slices of a circuit optimally one after another. Token swapping~\cite{miltzow2016token,childs2019circuit} realizes a given permutation with the fewest adjacent swaps. Mapping that exploits gate commutation was studied by Itoko et al.~\cite{itoko2020commutation} and, for two-local Hamiltonian circuits, by 2QAN~\cite{lao20222qan}. Fixed swap networks route all-to-all interaction layers on a line in linear depth~\cite{kivlichan2018linear,ogorman2019swap,maslov2007linear}. Zulehner and Wille exploit SWAP absorption in SU(4) mapping~\cite{zulehner2019su4}; NASSC accounts for subsequent simplifications in its routing costs~\cite{liu2022nassc}, and OLSQ-GA models gate absorption in exact mapping~\cite{tan2021olsqga}. MIRAGE combines mirror gates with decomposition-aware cost estimates to select among routing trials~\cite{mckinney2024mirage}. PAS/PAM couples permutation-aware block synthesis with mapping and post-routing multi-qubit resynthesis~\cite{liu2023pam}. Subgraph monomorphism (VF2/VF2++~\cite{cordella2004vf2,juttner2018vf2pp}) finds layouts that need no SWAP when one exists. QUEKO~\cite{tan2021queko} and QUBIKOS~\cite{ping2025qubikos} are benchmarks with known optimal depth or SWAP count.

\paragraph*{Two-qubit synthesis.}
Every $U\in SU(4)$ has a KAK (Cartan) decomposition~\cite{khaneja2001cartan,kraus2001optimal,zhang2003geometric}
\begin{equation}
\begin{aligned}
U&=(k_1\otimes k_2)\,\Can(a,b,c)\,(k_3\otimes k_4),\\
\Can(a,b,c)&=e^{i(aXX+bYY+cZZ)},
\end{aligned}
\label{eq:kak}
\end{equation}
with $(a,b,c)$ in the Weyl chamber $\pi/4\ge a\ge b\ge |c|$. The number of CZ (equivalently CX) gates needed is 0 at $(0,0,0)$, 1 at $(\pi/4,0,0)$, 2 on the plane $c=0$, and 3 otherwise~\cite{vatan2004optimal,vidal2004universal,shende2004minimal}. Three-qubit and larger blocks can be resynthesized by quantum Shannon decomposition (QSD)~\cite{shende2006qsd}. In CNOT--phase regions, which contain only CX and diagonal gates, the action is $|x\rangle\mapsto e^{if(x)}|Ax\oplus c\rangle$ with a phase polynomial $f$ over affine parities of the inputs and a linear reversible map $A$~\cite{amy2014tpar}; rotations on the same parity can be merged (phase folding~\cite{nam2018automated,amy2014tpar}). Linear reversible circuits can be synthesized by Gaussian elimination, asymptotically optimally by the method of Patel, Markov and Hayes~\cite{patel2008linear}, and under connectivity constraints by Steiner-tree elimination~\cite{kissinger2020steiner,nash2020steiner}. Amy, Azimzadeh and Mosca studied the CNOT complexity of whole CNOT--phase circuits and gave the gray-synth parity-network heuristic~\cite{amy2018graysynth}; architecture-aware variants route the parity network along Steiner trees~\cite{meijer2023phasepoly,vandaele2022phasepoly}. These methods are heuristics for circuits of any size. Exact searches for minimal CNOT or $\{\mathrm{CNOT},T\}$ circuits exist for small instances~\cite{amy2013mitm,meuli2018sat}. HOPPS uses SAT for optimal CNOT count or depth under a coupling graph and resynthesizes blocks from mapped circuits~\cite{li2025hopps}. PhasePoly jointly optimizes phase and output parities using bounded A* and cross-block transformations; its hardware evaluation applies SABRE after logical optimization~\cite{chen2026phasepoly}. The \qs{} compiler uses a budgeted parity-state A*, restricted to CX-plus-phase networks and to the coupling graph, on small regions of already routed candidates (Sec.~\ref{sec:postroute}). BQSKit~\cite{bqskit2021} resynthesizes partitioned blocks numerically (QSearch, LEAP~\cite{davis2020qsearch,smith2023leap}) up to a tolerance~$\epsilon$; QSearch performs an A* search over circuit structures and instantiates the angles of each structure by numerical optimization.

\paragraph*{Superoptimizers.}
Quartz~\cite{xu2022quartz}, QUESO~\cite{xu2023queso} and GUOQ~\cite{xu2025guoq} optimize logical circuits using verified rewrite rules found by search, and GUOQ also uses numerical resynthesis through BQSKit. They do not perform layout or routing, so they complement a mapping compiler rather than replace it.

\paragraph*{Position of this work.}
The \qs{} pipeline selects among screened complete routing candidates after actual coupling-constrained multi-wire resynthesis. The experiments characterize this integration and its components. We did not compare directly with HOPPS, PhasePoly, MIRAGE or PAS/PAM, did not evaluate at equal compute budgets, and did not study the ranking objective on a fixed candidate pool (Sec.~\ref{sec:limitations}).

\paragraph*{Benchmarks.}
Our standard and large suites draw on MQT Bench~\cite{quetschlich2023mqtbench}, QASMBench~\cite{li2023qasmbench} and the Qiskit circuit library~\cite{qiskit2024}, including quantum-volume circuits~\cite{cross2019qv}. The topologies include IBM's heavy-hex lattice~\cite{chamberland2020heavyhex}. Cirq~\cite{cirq} is used as an additional baseline. Table~\ref{tab:features} compares the features of the \qs{} compiler with those of the evaluated compilers and the superoptimizers.

\providecommand{\fyes}{\ensuremath{\bullet}}
\providecommand{\fpart}{\ensuremath{\circ}}
\providecommand{\favail}{\ensuremath{\diamond}}
\providecommand{\fno}{--}
\providecommand{\fn}[1]{\textsuperscript{#1}}
\providecommand{\fhead}[1]{\rotatebox{90}{\parbox{21mm}{\raggedright #1}}}
\providecommand{\ftext}{\raggedright\hspace{0pt}}
\begin{table*}[t]
\caption{Compiler features. \qs{} O3 is level~3 of the \qs{} compiler. \fyes: used in the configuration we evaluate; \fpart: a weaker or partial form is used; \favail: available in the tool but not in the evaluated configuration; \fno: not available; ?: not determined. Superscripts refer to the notes below the table. Configurations: Qiskit \texttt{transpile} at level~3; tket \texttt{FullPeepholeOptimise}, \texttt{DefaultMappingPass}, \texttt{KAKDecomposition}, \texttt{CliffordSimp}, \texttt{SynthesiseTket}; BQSKit \texttt{compile} at level~2; QMAP heuristic mapper (A$^*$, 15-layer lookahead, static initial layout); Cirq \texttt{CZTargetGateset}, \texttt{RouteCQC}, \texttt{CZTargetGateset}. The superoptimizers optimize logical circuits and do not map them.}
\label{tab:features}
\centering
\scriptsize
\setlength{\tabcolsep}{3.2pt}
\begin{tabular}{@{}l c p{21mm} p{24mm} c c c c c c c c c c@{}}
\toprule
 & \fhead{Exact layout (subgraph monomorphism)}
 & \fhead{Layout optimization}
 & \fhead{Router}
 & \fhead{Permutation- / commutation-aware routing}
 & \fhead{Mirror gates, SWAP absorption}
 & \fhead{Exact window re-routing}
 & \fhead{Two-qubit block resynthesis (KAK)}
 & \fhead{Multi-qubit block synthesis}
 & \fhead{Phase folding / phase polynomials}
 & \fhead{Approximation with error budget}
 & \fhead{CX orientation optimization}
 & \fhead{Commutation-aware depth scheduling}
 & \fhead{Noise-aware} \\
\midrule
\qs{} O3
 & \fyes & \ftext SABRE trials + low-dilation QAP (greedy, annealing) & \ftext LightSABRE-style with block nodes; plain SABRE; global beam; layered (Kendall-$\tau$), rollout; candidate search
 & \fyes\fn{a} & \fyes & \fyes\fn{b} & \fyes & \fyes\fn{c} & \fyes\fn{F} & \fyes\fn{d} & \fyes\fn{e} & \fyes & \fno \\
Qiskit 2.6 O3~\cite{qiskit2024}
 & \fyes & \ftext SABRE layout, 20 trials & \ftext LightSABRE, 20 trials
 & \fpart\fn{f} & \fpart\fn{g} & \fno & \fyes & \fpart\fn{h} & \fno\fn{i} & \fpart\fn{j} & \fno\fn{k} & \fno & \favail\fn{l} \\
tket 2.18 O2~\cite{sivarajah2020tket}
 & \fpart\fn{m} & \ftext GraphPlacement (weighted monomorphism) & \ftext LexiRoute (lookahead 10)
 & \fpart\fn{n} & \fpart\fn{g} & \fno & \fyes & \fyes\fn{o} & \favail\fn{p} & \favail\fn{q} & \fno\fn{k} & \fno & \favail\fn{r} \\
BQSKit 1.2 O2~\cite{bqskit2021}
 & \favail\fn{s} & \ftext greedy placement + SABRE layout & \ftext generalized SABRE
 & \favail\fn{t} & \fpart\fn{g} & \fno & \fyes\fn{u} & \fyes\fn{u} & \fno & \fyes\fn{v} & \fno & \fno & \fno \\
QMAP 3.11~\cite{wille2023qmap}
 & \fno & \ftext static / dynamic initial layout & \ftext layer-wise A$^*$, lookahead
 & \fno & \fpart\fn{w} & \favail\fn{x} & \fno & \fno & \fno & \fno & \fno\fn{k} & \fno & \favail\fn{y} \\
Cirq 1.7~\cite{cirq}
 & \fno & \ftext line placement from the device centre & \ftext RouteCQC (greedy, per time step)
 & \fno & \fpart\fn{g} & \fno & \fyes & \favail\fn{z} & \fno & \fno & \fno & \fno & \fno \\
\midrule
Quartz~\cite{xu2022quartz}
 & \multicolumn{6}{c}{no layout or routing}
 & \fno & \fno & \fyes\fn{A} & \fno & \fno & \fno & \fno \\
QUESO~\cite{xu2023queso}
 & \multicolumn{6}{c}{no layout or routing}
 & \fno & \fno & ?\fn{B} & \fno & \fno & \fno & \fno \\
GUOQ~\cite{xu2025guoq}
 & \multicolumn{6}{c}{no layout or routing}
 & \fyes\fn{C} & \fyes\fn{C} & ?\fn{B} & \fyes\fn{D} & \fno & \fno & \fpart\fn{E} \\
\bottomrule
\end{tabular}

\vspace{1mm}
\begin{minipage}{\textwidth}
\tiny\raggedright
\textsuperscript{a}\,$\mathrm{SWAP}\cdot U$ absorbed into the output permutation; a commutation-aware routing family on the CX level.
\textsuperscript{b}\,Window LNS with A$^*$ under the merge-aware SWAP cost.
\textsuperscript{c}\,Analytic QSD of three-qubit blocks; at level~3, numerical instantiation of three-qubit circuits and blocks, checked per block to $10^{-9}$.
\textsuperscript{d}\,Per-block average-infidelity budget and a certified operator-norm sum.
\textsuperscript{e}\,Lagrangian dual decomposition minimizing single-qubit gates.
\textsuperscript{f}\,\texttt{ElidePermutations} removes input SWAPs and permutations before routing; the router is not commutation-aware.
\textsuperscript{g}\,Only after routing: block consolidation and resynthesis merge a SWAP with neighboring gates; the router does not anticipate it.
\textsuperscript{h}\,Quantum Shannon decomposition only for input unitaries on three or more qubits; no three-qubit block collection.
\textsuperscript{i}\,\texttt{CommutativeCancellation} merges rotations across commuting gates, not by parity.
\textsuperscript{j}\,\texttt{approximation\_degree} per gate or block; no global bound is reported.
\textsuperscript{k}\,Gate direction is only fixed for legality on directed couplings.
\textsuperscript{l}\,\texttt{VF2Layout}/\texttt{VF2PostLayout} score layouts by target error rates when the target has them.
\textsuperscript{m}\,\texttt{GraphPlacement} embeds the interaction graph of the first 100 gates (depth 100), so a SWAP-free layout is not guaranteed.
\textsuperscript{n}\,Implicit wire swaps in \texttt{FullPeepholeOptimise} (\texttt{allow\_swaps=True}).
\textsuperscript{o}\,\texttt{ThreeQubitSquash}, before routing only.
\textsuperscript{p}\,\texttt{PauliSimp}, \texttt{GreedyPauliSimp} (Pauli gadgets with phase folding).
\textsuperscript{q}\,\texttt{KAKDecomposition(cx\_fidelity$<$1)}; no global budget.
\textsuperscript{r}\,\texttt{NoiseAwarePlacement}.
\textsuperscript{s}\,\texttt{StaticPlacementPass}, not in the level-2 workflow.
\textsuperscript{t}\,Permutation-aware mapping (PAM)~\cite{liu2023pam}, level~4 only.
\textsuperscript{u}\,Numerical instantiation (QSearch~\cite{davis2020qsearch}, LEAP~\cite{smith2023leap}) and scanning gate removal on blocks of up to three qubits.
\textsuperscript{v}\,\texttt{synthesis\_epsilon} per block; an upper bound on the total error is tracked.
\textsuperscript{w}\,Post-mapping cancellation of adjacent CNOTs.
\textsuperscript{x}\,SAT-based exact mapper for the whole circuit (small circuits only), not windows.
\textsuperscript{y}\,Fidelity-aware heuristic (\texttt{fidelity\_best\_location}) when calibration data are given.
\textsuperscript{z}\,\texttt{three\_qubit\_matrix\_to\_operations} (QSD) exists but is not part of the gateset pipeline.
\textsuperscript{A}\,Rotation merging of Nam et al.~\cite{nam2018automated} as preprocessing.
\textsuperscript{B}\,Rewrite rules include $R_z$ merges; we did not determine whether parity-based merging is applied.
\textsuperscript{C}\,Numerical resynthesis through BQSKit of random subcircuits of at most three qubits.
\textsuperscript{D}\,Global error threshold with additive error tracking.
\textsuperscript{E}\,Fidelity objective from calibration data in the NISQ setting; no placement.
\textsuperscript{F}\,Phase folding before routing; at level~3, coupling-constrained resynthesis of CNOT--phase regions after routing (budgeted parity-state A* on small regions, which certifies the minimum CX count for a fixed phase polynomial when it reaches a goal; beam search on large ones).
\end{minipage}
\end{table*}

\section{The design of the \texttt{quspect} compiler}
\label{sec:techniques}

\begin{figure*}[t]
\centering
\begin{tikzpicture}[
  font=\scriptsize,
  node distance=3.2mm and 3.0mm,
  >={Stealth[length=1.6mm]},
  py/.style={draw=black!55, fill=blue!5, rounded corners=1.2pt, align=center,
             text width=20.6mm, minimum height=11mm, inner sep=1.5pt},
  cc/.style={py, fill=orange!22, draw=orange!70!black, line width=0.7pt},
  wide/.style={text width=29mm},
  lane/.style={draw=black!25, fill=black!2, rounded corners=3pt, inner sep=2.2mm},
  lab/.style={font=\scriptsize\bfseries, anchor=north west, inner sep=1.5pt},
  note/.style={font=\tiny, black!75},
  flow/.style={->, black!70},
  alt/.style={->, black!55, dashed},
]
\node[py] (in) {logical circuit\\{\tiny working basis:}\\{\tiny target 1q + cz, h, rz, x}};
\node[py, right=of in] (hl) {high-level\\cancellation\\{\tiny self-inverse pairs}};
\node[cc, right=of hl] (qsd) {3-qubit blocks\\by QSD\\{\tiny at most 10 or 24}\\{\tiny 2q gates}};
\node[cc, right=of qsd] (pf) {CX level:\\phase folding,\\cancellation};
\node[py, right=of pf] (apx) {approximation\\{\tiny drops vanishing}\\{\tiny interactions only}};
\node[py, right=of apx] (dec) {decompose\\{\tiny keeps ccx}};
\node[cc, right=of dec] (cons) {consolidate\\2q blocks (KAK)};
\foreach \a/\b in {in/hl,hl/qsd,qsd/pf,pf/apx,apx/dec,dec/cons} \draw[flow] (\a) -- (\b);

\node[py, below=15mm of in] (vf2) {exact layout\\{\tiny VF2++ monomorphism,}\\{\tiny $10^5$ states; if found,}\\{\tiny no SWAP}};
\node[cc, right=of vf2] (sab) {SABRE layout\\{\tiny starts: trivial,}\\{\tiny QAP, 6 random;}\\{\tiny 3 fwd/bwd passes}};
\node[cc, right=of sab] (route) {SABRE routing\\{\tiny block nodes, mirror gates,}\\{\tiny SWAP$\cdot U$ absorption,}\\{\tiny Toffoli units}};
\node[cc, right=of route] (layer) {layered routing\\{\tiny Kendall-$\tau$ on path/ring,}\\{\tiny rollout elsewhere}};
\node[cc, right=of layer] (lns) {window LNS + A*\\{\tiny 1--4 SWAPs;}\\{\tiny depth sweep}};
\node[py, right=of lns] (eval) {post-routing\\pipeline on each\\routing};
\node[py, right=of eval] (pick) {select\\{\tiny (2q, 2q depth,}\\{\tiny total gates)}};
\draw[flow] (vf2) -- node[above, note]{none} (sab);
\draw[flow] (sab) -- (route);
\draw[flow] (route) -- node[above, note]{SU(4)} (layer);
\draw[flow] (layer) -- (lns);
\draw[flow] (lns) -- (eval);
\draw[flow] (eval) -- (pick);

\node[py, below=15mm of vf2] (pdec) {decompose,\\consolidate\\{\tiny 3$\to$2 approx.}};
\node[py, right=of pdec] (opt) {peephole opt.,\\decompose,\\consolidate};
\node[py, right=of opt] (tr) {translate to\\target basis,\\{\tiny target-basis peephole}};
\node[py, right=of tr] (swf) {SWAP-face\\templates\\{\tiny 3 variants}};
\node[cc, right=of swf] (sch1) {depth scheduling\\{\tiny trace monoid,}\\{\tiny block level}};
\node[cc, right=of sch1] (gauge) {1q gauge\\{\tiny beam 4; from given and}\\{\tiny from Lagrangian}\\{\tiny CX orientation}};
\node[cc, right=of gauge] (sch2) {reschedule,\\keep shallowest\\$\to$ output};
\foreach \a/\b in {pdec/opt,opt/tr,tr/swf,swf/sch1,sch1/gauge,gauge/sch2} \draw[flow] (\a) -- (\b);

\node[py, text width=34mm, below=15mm of pdec.south west, anchor=north west] (fam) {\textbf{routing families}\\
  {\tiny block + mirror: 20 layouts $\times$ 12 seeds,}\\
  {\tiny keep 2 per layout; block, no mirror;}\\
  {\tiny commutation-aware (CX level); plain}\\
  {\tiny gate-level SABRE; layered / rollout;}\\
  {\tiny global beam search; exact circuit}};
\node[py, right=of fam] (rank) {rank by proxy\\{\tiny estimated 2q count}\\{\tiny after resynthesis}};
\node[py, right=of rank] (top) {evaluate\\{\tiny best + 1 more (budget),}\\{\tiny level-2, layered, beam,}\\{\tiny best comm., 4 plain}};
\node[cc, wide, right=of top] (lns3) {window LNS + A*\\{\tiny 1--4 SWAPs on all evaluated;}\\{\tiny 6, 8, 12 SWAPs on the best;}\\{\tiny depth sweep on the top 2}};
\node[cc, right=of lns3] (full) {post-routing\\pipeline on each\\{\tiny + CNOT--phase}\\{\tiny resynthesis (2 levels)}};
\node[cc, right=of full] (sel3) {select, finish;\\{\tiny 3q instantiation;}\\{\tiny recompile $\le$2 passes}};
\foreach \a/\b in {fam/rank,rank/top,top/lns3,lns3/full,full/sel3} \draw[flow] (\a) -- (\b);

\begin{scope}[on background layer]
  \node[lane, fit={(in)(cons)($(in.north)+(0,4.5mm)$)}] (L1) {};
  \node[lane, fit={(vf2)(pick)($(vf2.north)+(0,4.5mm)$)}] (L2) {};
  \node[lane, fit={(pdec)(sch2)($(pdec.north)+(0,4.5mm)$)}] (L3) {};
  \node[lane, fit={(fam)(sel3)($(fam.north)+(0,4.5mm)$)(sch2.east|-fam.south)},
        fill=green!3, draw=green!40!black, dashed] (L4) {};
\end{scope}
\node[lab] at (L1.north west) {Pre-routing};
\node[lab] at (L2.north west) {Layout and routing (level 2, default)};
\node[lab] at (L3.north west) {Post-routing, then finishing};
\node[lab] at (L4.north west) {Level 3: candidate search replaces the routing lane};

\coordinate (xl) at ($(L1.west)+(-3.0mm,0)$);
\coordinate (xl2) at ($(L1.west)+(-1.6mm,0)$);
\draw[flow] (cons.south) |- ($(L1.south)!0.5!(L2.north)$) -| (xl |- vf2) -- (vf2.west);
\draw[alt] (eval.south) |- ($(L2.south)+(0,-2.6mm)$) -| (xl2 |- pdec) -- (pdec.west);
\node[note, anchor=south east] at ($(eval.south |- L2.south)+(-1mm,-2.6mm)$) {every routing};
\draw[flow] (pick.south) |- ($(L2.south)+(0,-6.2mm)$) -| (swf.north);
\node[note, anchor=south east] at ($(pick.south |- L2.south)+(-1mm,-6.2mm)$) {selected (+ fallbacks)};

\node[py, text width=5mm, minimum height=2.6mm, anchor=north east] (legc) at ($(L4.south east)+(0,-1.5mm)$) {};
\node[note, anchor=east] (legct) at (legc.west) {C++ kernel};
\node[py, fill=blue!5, text width=5mm, minimum height=2.6mm, anchor=east] (legp) at ($(legct.west)+(-4mm,0)$) {};
\node[note, anchor=east] at (legp.west) {Python};
\node[cc, text width=5mm, minimum height=2.6mm] at (legc) {};
\end{tikzpicture}
\caption{Pipeline of the \qs{} compiler (\texttt{quspect.transpiler}) in the evaluated version. Top three lanes: level~2 (the default). Pre-routing runs in a CZ-centered working basis. The pre-routing approximation only removes interactions that vanish within the budget ($10^{-12}$ at the default $\alpha=1$); when it changes the circuit, level~2 also routes the exact circuit and keeps the routing with the smaller estimated overhead. The routing lane yields several routings: SABRE; SABRE with a two-layer lookahead and the layered or rollout routing, both only when every block is a generic SU(4) and kept only if their estimate is lower; and the LNS and depth-sweep refinements, only up to 5000 estimated two-qubit gates. Each routing goes through the post-routing pipeline, and the best by (two-qubit gates, two-qubit depth, total gates) is finished. Finishing tries three SWAP-face variants. It schedules, minimizes single-qubit gates from the given CX orientations and from the Lagrangian ones, reschedules, and keeps the shallowest result. Bottom lane: level~3 replaces the routing lane with a candidate search; it also resynthesizes the CNOT--phase regions of every evaluated candidate (fully for the best three), instantiates three-qubit circuits and blocks numerically, recompiles results with at most 1500 two-qubit gates, and, for small inputs, compiles a second start whose regions were resynthesized before routing (Sec.~\ref{sec:postroute}). Orange boxes are C++ kernels: SABRE and beam routing, the low-dilation (QAP) layout, layered path/ring and rollout routing, LNS/A*, KAK and single-qubit synthesis with block consolidation, the numerical parts of QSD, phase folding, CNOT--phase resynthesis, three-qubit instantiation, gauge, orientation and scheduling. Blue boxes are Python or mixed: VF2++, the high-level cancellation, decomposition, the peephole passes and the orchestration (the post-routing boxes call the C++ consolidation).}
\label{fig:pipeline}
\end{figure*}
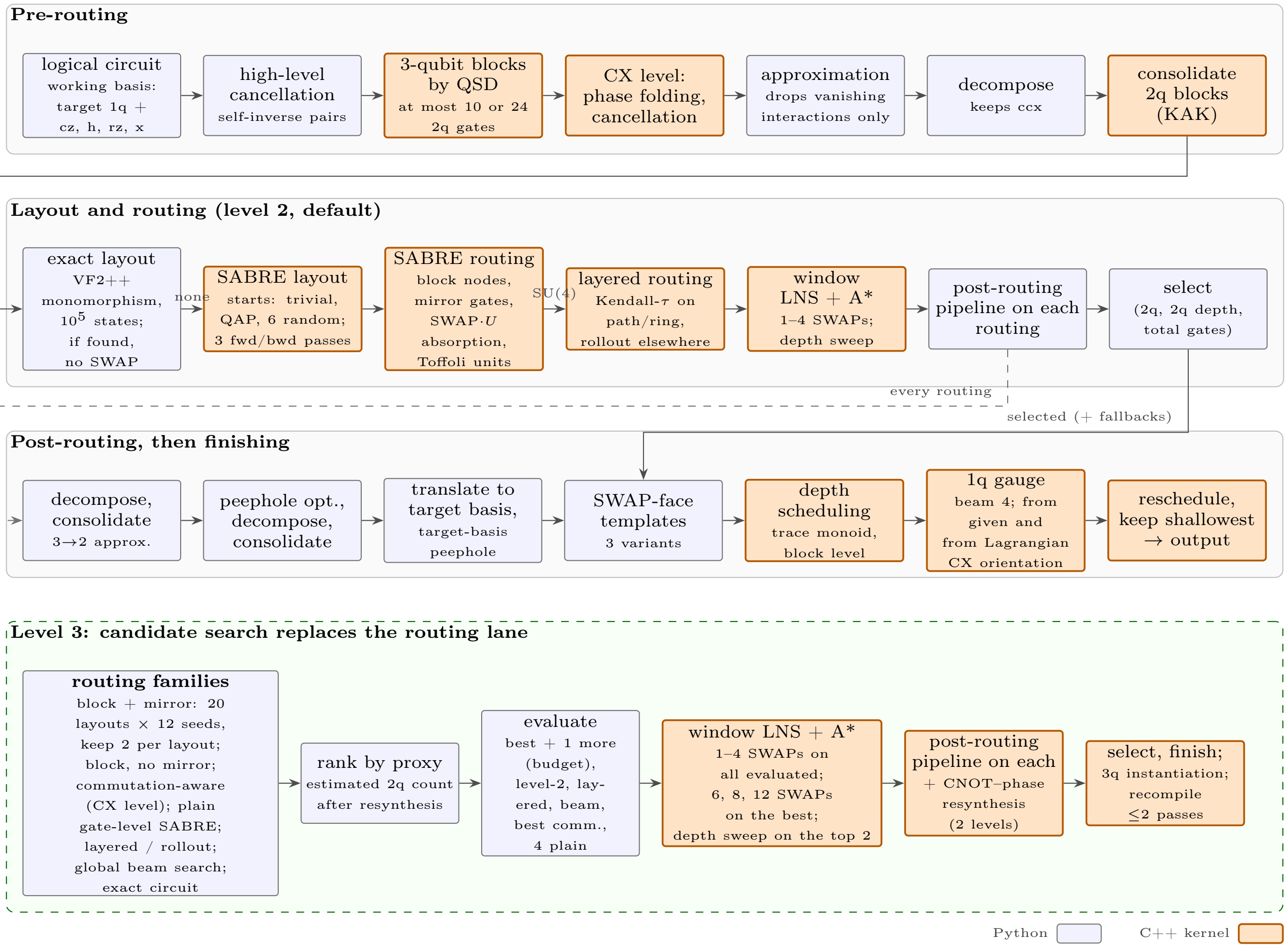

\subsection{Pipeline}
\label{sec:pipeline}
Figure~\ref{fig:pipeline} shows the pipeline of the evaluated version (commit \texttt{be4ecda}). At level~2, the default, the compiler works in a CZ-centered working basis and translates to the target basis at the end. Before routing, it cancels self-inverse pairs, resynthesizes three-qubit blocks by QSD, decomposes to the CX level and applies phase folding (followed by another cancellation if folding has merged any rotation), removes interactions that vanish within the approximation budget, decomposes while keeping Toffolis, and consolidates two-qubit blocks. Layout and routing then produce several routings: an exact or SABRE layout with SABRE routing; for circuits of generic SU(4) blocks, a SABRE routing with a wide lookahead and a layered or rollout routing; and, for circuits with at most 5000 estimated two-qubit gates, refinements by the window LNS and an equal-cost depth sweep. If the pre-routing approximation has changed the circuit, the exact circuit is routed as well and the routing with the smaller estimated overhead is kept (both routings are completed when the estimates are within 10\%). Every routing passes through the post-routing pipeline: decomposition, consolidation (with the optional reduction from 3 to 2 entanglers), peephole optimization, consolidation, translation to the target basis, and a peephole pass in the target basis that is kept only if it does not add two-qubit gates. The best routing by (two-qubit count, two-qubit depth, total gates) is then finished. Finishing tries three variants of the SWAP-face rewrite (Sec.~\ref{sec:synthesis}), schedules the circuit by block-level commutation, minimizes single-qubit gates starting from the given CX orientations and from the Lagrangian ones, reschedules, and keeps the shallowest result. Level~3 replaces the routing lane with a candidate search (Sec.~\ref{sec:candidates}) and adds the post-routing resynthesis of Sec.~\ref{sec:postroute}. The final round of development added only level-3 passes. Between the evaluations of the previous version (commit \texttt{8ec73d8}) and of the final version, the level-2 two-qubit counts changed by at most 0.3\% in geometric mean on every suite and the level-2 depths changed by up to 2.2\% (all lower). Level~0 did not change. These differences come from code merged between the two evaluations, including the beam router and the move of numerical kernels to C++; we did not attribute these differences individually.

The default pipeline drops interactions that vanish to within $10^{-12}$ (Sec.~\ref{sec:approx}) and validates numerical replacements up to global phase. Numerically computed rewrites (KAK resynthesis with stabilizer selection, QSD, SWAP-face templates, three-qubit instantiation and CNOT--phase regions of up to six wires) are accepted only after the unitary of their emitted gate list matches the target to an elementwise, phase-aligned tolerance of $10^{-9}$. Symbolic rewrites (phase folding and the cancellation of self-inverse pairs) are exact by construction. CNOT--phase regions with more than six wires use tolerance-based re-extraction of the affine map and phase polynomial from the emitted gates, supplemented by two-state checks up to 14 wires (Sec.~\ref{sec:postroute}). This checks consistency of the extracted representation rather than independently certifying exact equivalence. For the instantiated circuits, the tolerance applies to each replacement: each replaced block matches its target to $10^{-9}$ per matrix element, and independent checks of the outputs are reported in Sec.~\ref{sec:correctness}. In the evaluated version, the routing, beam routing, low-dilation layout, layered and rollout routing, LNS, gauge, orientation and scheduling kernels, KAK and single-qubit synthesis, block consolidation, the numerical parts of QSD, phase folding, the CNOT--phase resynthesis and the three-qubit instantiation are implemented in C++. The routing kernels and the CNOT--phase resynthesis reproduce their Python reference implementations bit for bit. The synthesis kernels compute on a deterministic numerical layer (fixed-order arithmetic without fused multiply--add, correctly rounded elementary functions, no BLAS or LAPACK); their Python fallbacks, used only when the native module is missing, implement the same algorithms and are numerically equivalent but not bit-identical. VF2++, the high-level cancellation, the peephole passes and the orchestration of the pipeline are implemented in Python.

\subsection{Layout: exact embedding and low-dilation QAP}
\label{sec:layout}
\paragraph*{Exact layout.} If $G_I$ is isomorphic to a subgraph of $G_C$, there is a routing with no SWAP at all. The compiler searches for a subgraph \emph{monomorphism} using a non-recursive VF2++-style search~\cite{cordella2004vf2,juttner2018vf2pp}. Components are ordered by BFS from the vertex of maximum degree, candidates are pruned when a vertex has fewer free neighbors than unmatched neighbors, and candidates are tried in Warnsdorff order. Three fast rejection checks run first: too many vertices or edges, a degree sequence that cannot be dominated, and a non-bipartite $G_I$ on a bipartite $G_C$ such as a grid or heavy-hex. The search stops after a deterministic budget of $10^5$ states, which makes it miss 7 of 248 SWAP-free QUEKO layouts (Sec.~\ref{sec:optimality}).

\paragraph*{Low-dilation layout.} When no embedding exists, SABRE starts from random layouts and refines them locally, which can leave long-range defects. On a 57-qubit heavy-hex device, the longest simple path has 49 vertices, so a 50-qubit chain cannot be embedded, but one SWAP suffices if the chain is placed well. The compiler adds starting layouts that minimize the weighted dilation
\begin{equation}
\min_{\pi}\ \sum_{(i,j)} w_{ij}\,\bigl(d(\pi(i),\pi(j))-1\bigr),
\label{eq:qap}
\end{equation}
where $w_{ij}$ counts routing nodes (blocks) on the pair $(i,j)$. Equation~\eqref{eq:qap} is a Koopmans--Beckmann quadratic assignment problem~\cite{koopmans1957qap}. Its zero-cost instances include Hamiltonian path, so it is NP-hard. For a fixed layout, executing a gate at distance $d$ requires at least $d-1$ SWAPs, which is why dilation measures the routing that a layout forces.

The compiler solves it in two steps.
\begin{enumerate}
\item \emph{Greedy growth.} Logical qubits are placed in maximum-adjacency order, using an approach similar to Prim's algorithm or weighted Cuthill--McKee~\cite{cuthill1969}. Each is placed on the free physical qubit that adds the least dilation. Ties are broken by Warnsdorff's rule~\cite{pohl1967warnsdorff}: prefer the vertex with the fewest free neighbors, so the growth follows the boundary of the free region instead of fragmenting it. A capacity penalty reserves room for branches. Multiple starts are chosen from the pseudo-peripheral BFS endpoints of the Gibbs--Poole--Stockmeyer bandwidth heuristic~\cite{gibbs1976gps}, from low-degree vertices and from centers.
\item \emph{Simulated annealing}~\cite{kirkpatrick1983sa}. Moves are swaps (possibly into an empty site), biased toward the partner's neighborhood, and each move is evaluated incrementally in $O(\deg)$ time. The schedule cools linearly with Metropolis acceptance.
\end{enumerate}
The resulting layouts are used only as \emph{additional} SABRE starting points in every SABRE family, in runs with and without refinement. The routing with the lowest cost is still chosen, so dense or time-varying circuits are not harmed. For GHZ-50 on heavy-hex, the final two-qubit counts are 50 for the \qs{} compiler (the lower bound is 49), 97 for tket and 277 for Qiskit O3. The C++ kernel uses only integer arithmetic and a SplitMix64 generator, so its results are reproducible across platforms.

\subsection{Routing}
\label{sec:routing}

\paragraph*{SABRE v2.}
The router follows LightSABRE~\cite{zou2024lightsabre}. A candidate SWAP $s$ that maps the layout $\pi$ to $\pi_s$ has the score
\begin{equation}
H(s)=\max(\delta_a,\delta_b)\Bigl[\sum_{g\in F} d_{\pi_s}(g)+\frac{W}{|E|}\sum_{g\in E} d_{\pi_s}(g)\Bigr],
\end{equation}
where $F$ is the front layer, $E$ the extended set ($|E|\le 20$, $W=0.5$), and $\delta$ a decay factor applied only to the two swapped qubits (increment $10^{-3}$, reset every 5 SWAPs or whenever a gate becomes executable). Following Qiskit, the front term is not normalized by $|F|$. When every block of the circuit is a generic SU(4), an additional routing uses a wider lookahead: the extended set holds the blocks of the next two layers (at most 100) with $W=1$. A release valve guarantees progress. After $10\,|V_P|$ SWAPs without progress, the router undoes them and routes the nearest front gate along a shortest path. Scores are updated incrementally, touching only gates on the two swapped qubits, and ties are broken by a seeded random generator. Toffolis are routed as three-qubit units. A Toffoli on physical qubits $p_1,p_2,p_3$ is executable when they are connected. Its cost is the size of the smallest tree that spans them minus one, $\min_{m}\sum_{p\ne m}d(m,p)-1$ over the choice of the middle qubit $m\in\{p_1,p_2,p_3\}$. This cost is 1 when the Toffoli is executable. After routing, it is synthesized for the resulting coupling: six CX on a triangle, or an exact eight-CX nearest-neighbor circuit on a path, in which the middle qubit may be either a control or the target.

\paragraph*{Block nodes and mirror gates.}
Consecutive gates on one qubit pair form a single routing node, so a lookahead of 20 nodes covers about three times as many gates. Let $\cz(\cdot)$ denote the minimum CZ count from the Weyl classification. For each maximal block $U$, the router stores the \emph{mirror cost} $\max(0,\cz(U\cdot\mathrm{SWAP})-\min(n_{2q},\cz(U)))$, which is zero for a generic SU(4) block~\cite{mckinney2024mirage,zulehner2019su4}. After the router executes the last node of a block, it may insert a same-pair SWAP after that node if doing so lowers the front-plus-lookahead cost by more than $0.25+0.5\times$(mirror cost). Consolidation later merges $U\cdot\mathrm{SWAP}$ into one block. In the search, a SWAP on a pair whose last operation ended a block is charged its mirror cost instead of 3 CZ, and the fusion bonus scales with the saving. Mirror decisions are greedy and sometimes hurt (on 2D Trotter circuits mapped to a line), so level~3 also keeps a family without mirroring as a fallback.

\paragraph*{Commutation-aware and permutation-aware routing.}
A second candidate family routes the CX-level circuit with \texttt{cx}, \texttt{cp}, \texttt{crz} and \texttt{rzz} kept whole, on a DAG in which consecutive same-axis diagonal gates on a qubit form an unordered group (as in 2QAN~\cite{lao20222qan}). A SWAP that can fuse with the gate just executed on its pair gets a bonus. Before routing, a block $U$ for which $\cz(\mathrm{SWAP}\cdot U)<\cz(U)$ is synthesized as $\mathrm{SWAP}\cdot U$, and the SWAP is absorbed into a relabeling. DCNOT-class blocks at Weyl point $(\pi/4,\pi/4,0)$ drop from 2 CZ to 1. Blocks equal to $\mathrm{SWAP}\cdot(A\otimes B)$ are detected by an SVD rank-1 test of the reshuffled $4\times4$ matrix and elided into the output permutation.

\paragraph*{Window LNS with an admissible A*.}
\phantomsection\label{sec:lns}
Merge-aware SWAP costs are not new: Zulehner and Wille charged zero for a SWAP that follows an SU(4) gate in their A* mapper~\cite{zulehner2019su4}, and NASSC, MIRAGE and OLSQ-GA account for gate absorption in greedy and exact routing~\cite{liu2022nassc,mckinney2024mirage,tan2021olsqga}. Exact routing of whole circuits is out of reach, but exact routing of short windows of an existing routing is not. This is what the compiler adds. We view a routing as an event sequence of node executions and SWAPs. Each physical qubit keeps an ``open'' item, and a SWAP costs
\begin{equation}
c(s)=\min\bigl(c_{\text{merge-before}}(s),\,c_{\text{merge-after}}(s),\,3\bigr),
\end{equation}
where each merge term is the mirror cost of the adjacent block with which the SWAP fuses. Window refinement is a large-neighborhood search~\cite{shaw1998lns} (Alg.~\ref{alg:lns}): cut out $k$ consecutive SWAPs, fix the entry layout and the exit position of every qubit that is used later, and re-route the window exactly. Idle and finished qubits may exit anywhere, since the final permutation absorbs them. A replacement is accepted only if it is \emph{strictly} cheaper than the original window, including merges across the window boundary. Unlike the slicing of SATMap~\cite{molavi2022satmap}, which constructs a routing slice by slice, the LNS improves a complete routing. The search is A*~\cite{hart1968astar} with three admissible components that are combined by max or sum where valid.
\begin{enumerate}
\item \emph{Meet-then-exit.} A pending pair $(a,b)$ must meet on some edge $(m,m')$ and then reach its exits $e_a,e_b$. Its cost is at least $\min_{(m,m')}[d(a,m)+d(m,e_a)+d(b,m')+d(m',e_b)]$ divided by the number of qubits moved per SWAP. The bounds for qubit-disjoint demands add.
\item \emph{Hub-chain DP.} The qubit $h$ with the most pending nodes meets its partners $p_1,p_2,\dots$ in order. Then
$f_i(x)=\min_u[f_{i-1}(u)+d(u,x)]+\min_{y\in N(x)}[d(p_i,y)+d(y,e_{p_i})]$. The first term is a min-plus distance transform, computed by multi-source BFS.
\item \emph{Per-qubit amortization.} Half of the cost of each SWAP is charged to each qubit it moves. In the merge-aware cost model, this term alone roughly halved the number of A* expansions in our measurements.
\end{enumerate}
Lazy evaluation of the expensive bounds, partial-order reduction of commuting SWAPs and a transposition table keep the search small. Under the pure SWAP-count model, the search is exact, and the tests check it against breadth-first search. Level~2 scans windows of 1--4 SWAPs for circuits with at most 5000 estimated two-qubit gates. Level~3 scans windows of 1--4 SWAPs on every fully evaluated candidate, and windows of 6, 8 and 12 SWAPs on the best SABRE candidate after that scan and on any layered or rollout candidate ranked above it. Windows that contain a Toffoli or more than 64 nodes are skipped. The expansion budget is deterministic, scales with the SWAP count and shrinks on devices with more than 27 qubits. A second scan orders solutions lexicographically by (cost, depth), using a per-qubit ``busy-until'' clock and the reverse longest path as a monotone depth proxy, and offers equally cheap but shallower windows as \emph{extra} candidates. Feeding the lexicographic key into the main scan instead changes A*'s budget allocation and was rejected after it increased gate counts.

\begin{figure}[!tbp]
\begin{algorithm}[H]
\footnotesize
\caption{Window large-neighborhood search with A$^*$ under the merge-aware SWAP cost.}
\label{alg:lns}
\begin{algorithmic}[1]
\Require routing $E$ (node executions and SWAPs) from layout $\pi_0$; window sizes $\mathcal K$; total budget $B$; per-window budget $b$
\Ensure routing $E'$ that executes the same nodes, with $c(E')\le c(E)$
\State $c_0\gets c(E)$ \Comment{$c$: each SWAP costs $\min(c_{\text{before}},c_{\text{after}},3)$}
\For{pass $=1,2$}
  \State $\mathit{gain}\gets\textbf{false}$
  \For{$k\in\mathcal K$} \Comment{sizes in increasing order}
    \State $p\gets 0$; $S\gets$ state of $\pi_0$ (layout, open merge items)
    \While{$p<|E|$ \textbf{and} expansions $<B$}
      \State $W\gets E[p..q)$: $k$ SWAPs and the nodes up to the next SWAP
      \If{$W$ has $k$ SWAPs (or is the tail), $\le 64$ nodes, no Toffoli}
        \State $e_t\gets$ position of $t$ after $W$, for every qubit $t$ used after $W$; other qubits exit anywhere
        \State $u\gets$ cost of $W$, including merges with the first operation after $W$
        \State $W'\gets\textsc{WindowAStar}(S,W,e,u,b)$ \Comment{within budget}
        \If{$W'$ found \textbf{and} $c(W')<u$}
          \State replace $W$ by $W'$ in $E$; $\mathit{gain}\gets\textbf{true}$
        \EndIf
      \EndIf
      \State advance $S$ and $p$ past the next SWAP of $E$
    \EndWhile
  \EndFor
  \If{\textbf{not} $\mathit{gain}$} \textbf{break} \EndIf
\EndFor
\If{$c(E)>c_0$} \Return the input routing \Comment{safety check}
\EndIf
\State \Return $E$
\Statex
\Function{WindowAStar}{$S,N,e,u,b$}
  \State root $\gets S$ with all executable nodes of $N$ executed
  \State push root with $f=g+h_{\text{cheap}}$; incumbent cost $\gets u$
  \While{open list nonempty \textbf{and} expansions $<b$}
    \State pop $s$ with least $(f,-g)$
    \If{$s$ is the incumbent goal} \Return its routing \Comment{optimal}
    \EndIf
    \If{$f(s)\ge$ incumbent cost} \textbf{continue}
    \EndIf
    \If{$h_{\text{full}}(s)>h_{\text{cheap}}(s)$} re-push $s$; \textbf{continue} \Comment{lazy A$^*$}
    \EndIf
    \For{edges $(x,y)$ at a qubit with pending nodes or away from its exit}
      \State \textbf{skip} the last SWAP's edge and disjoint edges of smaller index if no node ran since it \Comment{partial-order reduction}
      \State $s'\gets s$ after SWAP $(x,y)$ and draining executable nodes
      \State $g(s')\gets g(s)+$ SWAP cost under the open merge items
      \If{$s'$ not dominated in the transposition table \textbf{and} $g+h_{\text{cheap}}<$ incumbent}
        \State push $s'$; if $s'$ is a goal, it becomes the incumbent
      \EndIf
    \EndFor
  \EndWhile
  \State \Return the incumbent if it is cheaper than $u$, else none
\EndFunction
\end{algorithmic}
\vspace{1pt}
{\raggedright\scriptsize
$h_{\text{cheap}}$: a SWAP-count bound from qubit-disjoint pending demands and exit distances (each SWAP moves two qubits one step), converted to cost by the cheapest remaining merge opportunities, and the per-qubit amortization (half a SWAP per moved qubit). $h_{\text{full}}$ adds the meet-then-exit bound for every pending node and the hub-chain min-plus DP. Level~2: $\mathcal K=(1,2,3,4)$; level~3 additionally $\mathcal K=(6,8,12)$ on the best candidate. $B=\min(10^6,\max(2\cdot10^4,300\,n_{\text{swap}}))$, or $\min(2\cdot10^6,\max(5\cdot10^4,1000\,n_{\text{swap}}))$ for $(6,8,12)$, scaled by $27/|V_P|$ on devices with more than 27 qubits; $b=\max(32,\min(1000,B/(n_{\text{swap}}|\mathcal K|)))$.\par}
\end{algorithm}
\end{figure}

A measurement motivated this design. On small circuits, the level-3 routings of the previous version of the compiler use 18\% more SWAPs than the optimum from the same layout, yet under the merge-aware cost they are optimal in 136 of 158 instances and within 6 CZ in all but two (Sec.~\ref{sec:optimality}; the final compiler, which ranks candidates after post-routing resynthesis, often selects routings that cost more under this measure). This supports judging a router by its post-merge cost, as NASSC argued~\cite{liu2022nassc}, rather than by its SWAP count.

\paragraph*{Kendall-$\tau$ layered routing on paths.}
\phantomsection\label{sec:kendall}
In quantum-volume-like circuits, every two-qubit block is a generic SU(4), so a SWAP on the same pair as an adjacent block is free (its mirror cost is zero). Layer-by-layer routing of such circuits with free SWAPs goes back to Zulehner and Wille~\cite{zulehner2019su4}, who search for each layer's permutation by A*; fixed swap networks~\cite{kivlichan2018linear,ogorman2019swap} realize all-pairs layers without search. Suppose that, in an \emph{aligned schedule}, all blocks of layer $\ell$ are executed simultaneously on adjacent pairs of a path. The arrangement of layer $\ell$ is then determined by the \emph{order} $\sigma_\ell$ of its blocks along the path, because the order inside a block is free. The following is classical; inversion counts are the standard SWAP cost of reordering on a line~\cite{wille2014reorder,hirata2011lnn}.

\begin{proposition}
On a path with an aligned schedule, the minimum number of paid SWAPs between layers $\ell-1$ and $\ell$ equals the Kendall-$\tau$ distance (inversion number)~\cite{kendall1938tau} between the qubit sequences induced by $\sigma_{\ell-1}$ and $\sigma_\ell$, minimized over the free in-block orientations. It is realized by odd--even transposition sort~\cite{habermann1972oddeven,knuth1998taocp3}, which swaps only inverted adjacent pairs and therefore uses exactly that many SWAPs, in at most $n$ parallel rounds.
\end{proposition}
\noindent\emph{Proof sketch.} An adjacent transposition changes the inversion number by exactly one, which gives the lower bound. Odd--even transposition sort is a sorting network that exchanges only inverted pairs, so it attains the bound. \qed

Wille et al.~\cite{wille2014reorder} reorder the lines of a gate-level circuit exactly to reduce its nearest-neighbor SWAPs, and Zulehner and Wille~\cite{zulehner2019su4} choose each layer's permutation separately. What we add is a joint optimization of the orders of all layers. Routing reduces to $\min_{\sigma_1,\dots,\sigma_L}\sum_\ell d_\tau(\sigma_{\ell-1},\sigma_\ell)$, a chain of median-ordering problems in the spirit of Kemeny~\cite{kemeny1959}. With both neighbors fixed, the cost of placing block $A$ before block $B$ in layer $\ell$ depends only on the pair $(A,B)$, so the subproblem is a linear ordering problem (LOP)~\cite{marti2011lop}. The compiler solves it by insertion local search, uses block-coordinate descent along the chain, and adds iterated local search~\cite{lourenco2003ils} that perturbs runs of consecutive layers and keeps a perturbation only if it improves the total (Alg.~\ref{alg:kendall}). For all 35 six-qubit instances that we solved exhaustively, whose layers are perfect matchings, the asynchronous optimum equals the aligned one; we have not proved that this holds in general. When merged blocks leave slack in the layers, an asynchronous routing can be one SWAP cheaper, and the window LNS exploits this (Sec.~\ref{sec:optimality}). The method is enabled automatically when the coupling graph is a path and all blocks are generic SU(4). On QV-100 on a 100-qubit line, it gives 249{,}641 two-qubit gates, against 349{,}725 for Qiskit O3 and 443{,}601 for tket O2 (Large-20 data).

\begin{figure}[!tbp]
\begin{algorithm}[H]
\footnotesize
\caption{Layered routing of generic SU(4) blocks on a path: block-coordinate descent and iterated local search over the chain of block orders.}
\label{alg:kendall}
\begin{algorithmic}[1]
\Require dependency layers $L_1,\dots,L_m$ (disjoint qubit pairs; idle qubits are singleton blocks); seed
\Ensure initial arrangement and SWAPs between consecutive layers
\State $d(\ell)\gets$ inversions between the qubit sequences of $\sigma_{\ell-1}$ and $\sigma_\ell$, pairs in one block excluded \Comment{Fenwick tree}
\State $D\gets\sum_{\ell=2}^m d(\ell)$
\Statex \textit{Greedy start}
\State $\sigma_1\gets$ any order of the blocks of $L_1$
\For{$\ell=2,\dots,m$}
  \State sort the blocks of $L_\ell$ by mean position in $\sigma_{\ell-1}$
  \State $\sigma_\ell\gets\textsc{Insertion}(\sigma_\ell,\,w$ from $\sigma_{\ell-1}$ only$)$
\EndFor
\Statex \textit{Block-coordinate descent}
\For{$r=0,\dots,63$}
  \For{$\ell$ in increasing order if $r$ is even, else decreasing}
    \State $w(A,B)\gets\#\{(u,v)\in A\times B:\pi_{\ell-1}(u)>\pi_{\ell-1}(v)\}+\#\{(u,v)\in A\times B:\pi_{\ell+1}(u)>\pi_{\ell+1}(v)\}$
    \State $\sigma_\ell\gets\textsc{Insertion}(\sigma_\ell,w)$ \Comment{LOP with both neighbors fixed}
  \EndFor
  \If{$D$ did not decrease} \textbf{break} \EndIf
\EndFor
\Statex \textit{Iterated local search}
\For{$P=\min(2\cdot10^4,\,1000\,m)$ kicks}
  \State draw $lo$; $hi\gets lo+6$; $[a,b)\gets[lo-1,hi+1)\cap[1,m]$
  \State save $\sigma_a,\dots,\sigma_{b-1}$; $D_{\text{loc}}\gets\sum_{\ell=a}^{b} d(\ell)$
  \State in each layer of $[lo,hi)$ swap 1--4 random adjacent blocks
  \State up to 4 alternating sweeps of block-coordinate descent on $[a,b)$, while $\sum_{\ell=a}^{b} d(\ell)$ decreases
  \If{$\sum_{\ell=a}^{b} d(\ell)\ge D_{\text{loc}}$} restore the saved orders \EndIf
\EndFor
\Statex \textit{Emit SWAPs}
\State arrangement $\gets$ qubits in the block order $\sigma_1$
\For{$\ell=1,\dots,m-1$}
  \State flip each block of $L_\ell$ whose two qubits are inverted relative to $\sigma_{\ell+1}$ \Comment{free: merges into the block}
  \State odd--even transposition sort to the block order $\sigma_{\ell+1}$, swapping only inverted neighbors \Comment{$d(\ell+1)$ paid SWAPs}
\EndFor
\Statex
\Function{Insertion}{$\sigma,w$}
  \Repeat \ (at most 64 rounds)
    \For{each block $x$ of $\sigma$}
      \State evaluate all insertion points of $x$ in $\sigma\setminus x$ incrementally ($+w(y,x)-w(x,y)$ per step)
      \State move $x$ to the cheapest point if strictly cheaper
    \EndFor
  \Until{no move}
  \State \Return $\sigma$
\EndFunction
\end{algorithmic}
\vspace{1pt}
{\raggedright\scriptsize
Used when every two-qubit block is a generic SU(4) (mirror cost 0) and the circuit has no Toffoli. On a ring of at most 24 qubits the same chain is optimized with the cyclic crossing cost (block order and rotation per layer, $\min(500,25m)$ kicks). On every graph other than a path, rings included, a truncated rollout of SABRE is also tried. These routings are extra candidates, kept only when their estimated two-qubit count is lower.\par}
\end{algorithm}
\end{figure}

\paragraph*{Rings, grids and heavy-hex: cyclic chains and truncated rollout.}
On a ring of at most 24 qubits, we lift positions to the universal cover. For a given lift, the minimum number of adjacent transpositions equals the number of crossings between the lifted copies~\cite{jerrum1985}. With fixed in-block orientations, an optimal lift follows a threshold rule: the $K=\sum_i d_i/n$ qubits with the largest forward displacement travel backward. We checked this rule exhaustively on small instances. A periodic odd--even sort realizes the lift with exactly that many swaps. Block-coordinate descent with iterated local search then optimizes the cyclic block order and rotation of each layer.

On grids, we first tried a direct generalization (Hungarian assignment of per-layer positions followed by token swapping), and it failed. The chain proxy (half the total displacement) became small, but realizing the transitions by token swapping~\cite{miltzow2016token,childs2019circuit}, including a three-phase row--column--row sort~\cite{annexstein1990}, required 1.7--1.9 times as many SWAPs and performed worse than SABRE. Fixing the meeting positions in advance is the bottleneck. On every graph other than a path, including rings, where the rollout competes with the cyclic chain, the compiler therefore also uses a \emph{truncated rollout}~\cite{bertsekas1997rollout,bertsekas2020rollout} of SABRE's greedy policy. A candidate SWAP is scored by $H(s)+\min_{s'}H(s\circ s')$, and only the 16 best first-level candidates are expanded. Free mirror swaps are taken whenever they bring both qubits closer to their next partners, and ties go to the SWAP whose qubits are idle earliest, which reduces depth. Random restarts and forward/backward layout iterations are ranked by paid SWAPs $+\,0.5\times$ the depth proxy. On QV-100 the final counts are 60{,}364 two-qubit gates on a $10\times10$ grid and 89{,}787 on heavy-hex, against 69{,}165 and 98{,}910 for Qiskit O3.

\paragraph*{Candidate selection.}
\phantomsection\label{sec:candidates}
Level~3 routes each circuit using several families: block nodes with mirror gates (20 layout trials, 12 routing seeds per promising layout, keeping the 2 with the fewest SWAPs, and the wide lookahead on SU(4) circuits); block nodes without mirroring; the commutation-aware family, run when the routing DAG has at most 500 nodes or commutation reduces its depth below 0.95 of the original, and the main family needs SWAPs; the layered and rollout routings; when the pre-routing approximation changed the circuit, a level-2 routing of the exact circuit; the global beam-search router (Sec.~\ref{sec:beam}); and, added in the final round, a plain gate-level SABRE family on the CX-level circuit without block nodes, mirror gates, commutation or fusion bonuses, which supplies routings of the shape that Qiskit-style routers produce. Every SABRE family includes the low-dilation layouts among its starts. All candidates are ranked by a cheap proxy that estimates the two-qubit count after block resynthesis (charging $\min(3,\cdot)$ per consolidated pair block), then by SWAP count and depth. The full post-routing pipeline runs on the top-ranked candidate, on at most one more candidate in rank order if its estimated two-qubit count is at most 2500 gates, on the level-2 routing, on the layered candidate, the beam candidate and the best commutation-aware candidate, on the four best plain-SABRE candidates, and, if all candidates chosen by rank start from an embedding, on up to two non-embedding candidates within the same 2500-gate budget. In the final version, the post-routing pipeline for every evaluated candidate includes the CNOT--phase resynthesis of Sec.~\ref{sec:postroute}, so level~3 selects among screened complete routing candidates after actual coupling-constrained multi-wire resynthesis; candidates that the proxy screens out are not resynthesized. The evaluated candidates are compared by (two-qubit count, two-qubit depth, total gates); at the end, the selected candidate and its fallbacks are compared by (two-qubit count, depth, two-qubit depth, total gates). The work budget is deterministic.
\emph{Design invariant.} For a fixed seed, level~3 is meant never to return more two-qubit gates than level~2. Level~3 reproduces level~2's routing bit-identically as one of its candidates (its first layout trials coincide with level~2's and use the same seeds), always evaluates that candidate fully, and minimizes two-qubit count among the evaluated candidates. Because later stages are budgeted and not monotone, the invariant holds empirically rather than as a theorem. In the final data, quspect O3 never uses more two-qubit gates than quspect O1 (levels~3 and~2 of the compiler, named as explained in Sec.~\ref{sec:setup}; Table~\ref{tab:wtl}); in the previous version, there was one such case (2D Heisenberg-64 on heavy-hex, 2325 against 2289).

\subsection{Synthesis}
\label{sec:synthesis}

\paragraph*{KAK resynthesis.}
Maximal two-qubit blocks, including adjacent single-qubit gates and routed SWAPs, are consolidated into $4\times4$ unitaries. Each is decomposed as in Eq.~\eqref{eq:kak} by simultaneously diagonalizing the real and imaginary parts of $M=U_B^{T}U_B$ in the magic basis. The native kernel does this using Jacobi rotations; it needs no random numbers and is stable on degenerate spectra. The Python fallback diagonalizes a random real combination with \texttt{eigh}, verifies the result and retries on failure. The Weyl coordinates are then canonicalized with an explicit local correction for every symmetry step, and the block is resynthesized with the minimum number of entanglers. A replacement is accepted if it has fewer two-qubit gates, with total gate count as the tie-breaker, so the two-qubit count never increases. Requiring the total count not to grow either would have cost 2.5--7.2\% in two-qubit gates. Consolidation runs before routing, after routing and after optimization. The entangler is abstracted: each CZ is rewritten as $(A_0\otimes A_1)\,E\,(B_0\otimes B_1)$ for any CZ-class gate $E$ (CX in both orientations, CY, ECR), so the pipeline is basis-agnostic.

\paragraph*{Stabilizer selection, including mirrored degenerate points.}
If $S$ commutes with $\Can(a,b,c)$, then $U=(k_{12}S)\,\Can\,(S^\dagger k_{34})$ is another valid factorization. At degenerate points ($a=b=c$, $a=b$, $b=c$, $b=c=0$) the stabilizer is continuous, and an eigendecomposition returns an essentially arbitrary element. The outer single-qubit gates then become generic rotations; for instance, a Clifford-dressed Heisenberg block needed about 25 single-qubit gates in the IBM basis. The compiler scores candidate stabilizers by the Euler length of the outer factors. The candidates are the identity, Pauli pairs $P\otimes P$, and members of the continuous family that turn one outer factor into the identity, multiplied by the discrete part. The best few are emitted and verified to $10^{-9}$, so the result is never longer than the default choice. Because the Weyl chamber keeps the sign of $c$, the point $b=-c$ is a degenerate point as well. Using $(I\otimes Z)\,XX\,(I\otimes Z)=-XX$ and $(I\otimes Z)\,YY\,(I\otimes Z)=-YY$, we get
\begin{equation}
\Can(a,b,-b)=(I\otimes Z)\,\Can(a,b,b)^{\dagger}\,(I\otimes Z),
\end{equation}
and $\Can^\dagger$ has the same stabilizer as $\Can$. The stabilizer at the mirrored point is therefore the $Z$-conjugate of the stabilizer at $(a,b,|c|)$. Blocks on the SWAP face $a=b=\pi/4$, which typically come from SWAPs merged with interactions, admit a template $\mathrm{CX}\cdot\mathrm{CX}'\cdot R_z(\pi/2-2|c|)\cdot\mathrm{CX}$ with a single inner rotation. Using this template everywhere interfered with later CX cancellation, so the final stage instead completes the selected routing in three ways (as is, with a direct SWAP-face rewrite, and with a re-run with templates enabled) and keeps the lexicographically best result. The two-qubit count therefore cannot regress.

\paragraph*{Three-qubit QSD and Toffolis.}
Maximal three-qubit blocks are resynthesized by QSD~\cite{shende2006qsd}. If one qubit acts as a multiplexer once the outer single-qubit gates on it are stripped, the block is demultiplexed using an eigendecomposition of $AB^\dagger$ at a cost of at most 10 two-qubit gates. Otherwise a cosine--sine decomposition costs at most 24. A block is replaced only if the result is strictly cheaper and the emitted gates pass verification. This targets phase estimation, where the repeated controlled-$U$ gates collapse into one multiplexed block. Before decomposition, a high-level pass cancels self-inverse pairs (including Toffolis with permuted controls), even across gates that commute on the shared qubits.

\paragraph*{Phase folding.}
On CX-level circuits, the compiler tracks for each wire an affine parity $x_{i_1}\oplus\dots\oplus x_{i_k}\oplus c$ of path variables~\cite{amy2014tpar,nam2018automated}. A diagonal rotation ($r_z$, $p$, $t$, $s$, $z$) adds its angle to the term corresponding to the current parity of its wire. Gates such as $h$, $r_y$ and $u_3$ introduce fresh variables, as do the targets of Toffolis; measurements, barriers and symbolic parameters are boundaries. Rotations on the same parity are merged into one, multiples of $2\pi$ are deleted, and, if any rotation was merged, a follow-up cancellation pass removes the CX pairs that the merging exposes. The analysis is linear in the gate count times the parity width, and the rewrite is exact.

\paragraph*{Bounded approximate synthesis.}
\phantomsection\label{sec:approx}
By default Qiskit deletes any gate whose average gate infidelity is below $10^{-12}$ and drops merged rotations with $|\theta|<10^{-5}$. In comparisons with an exact-only compiler, these defaults account for large differences on QFT-like circuits. The compiler adopts an explicit budget instead. Each replacement may have an average gate infidelity of at most $\max(1-\alpha,10^{-12})$, where $\alpha$ is the approximation degree (default $\alpha=1$). The replacement projects the Weyl point onto the nearest point reachable with $k$ entanglers, namely $(0,0,0)$, $(\pi/4,0,0)$ or the plane $c=0$, while keeping the local factors. For $\mathrm{CP}(\theta)$ only the $ZZ$ part is dropped and the local rotations are kept, which incurs infidelity $\theta^2/20$ instead of $0.15\,\theta^2$ for deleting the whole gate. The phase-optimal operator-norm distances are summed into a certified upper bound on the distance of the whole circuit (up to global phase). Approximations that remove an interaction completely run before routing (with the default budget, only interactions that vanish to $10^{-12}$); reductions from 3 to 2 entanglers run only after routing, because applying them earlier perturbs routing and made QV-100 and Draper adders worse.

\subsection{Post-routing resynthesis, recompilation and terminal permutations}
\label{sec:postroute}
The final round of development (Sec.~\ref{sec:devprocess}) targeted the small circuits on which GUOQ and BQSKit still used fewer two-qubit gates than the \qs{} compiler. On its development set, 65\% of GUOQ's savings came from rewrites inside CNOT--phase regions of the \emph{routed} circuit: a routed SWAP is a linear map, a Toffoli synthesized for the coupling on which it landed is a CX--$T$ network, and both often lie in one region with neighboring CX gates and phases. Before that round, the compiler folded phases and resynthesized three-qubit blocks (QSD) only before routing. After routing, it resynthesized only two-qubit blocks and SWAP-face blocks, so no pass worked across blocks. None of the losses to GUOQ that we diagnosed required numerical resynthesis. All passes in this section run only at level~3.

\paragraph*{Region resynthesis under coupling constraints.} A region $C$ of CX and diagonal gates acts as $C|x\rangle=e^{if(x)}|Ax\oplus c\rangle$ with $f(x)=\sum_{y}\theta_y\,(y\cdot x)$~\cite{amy2014tpar,amy2018graysynth}. After the affine offset is handled as described below, a CX network with interleaved diagonal gates implements $C$ if the network implements $A$ and every parity $y$ with $\theta_y\not\equiv0 \pmod{2\pi}$ appears on some wire at some point, where $R_z(\theta_y)$ is placed. The compiler searches only over such networks for the phase polynomial computed from the region. Within this model, the minimum CX count is a shortest-path cost: states are (current parity rows of the wires, set of required parities already produced), a step is one CX allowed by the coupling graph ($\text{row}_t\mathrel{\oplus}=\text{row}_c$), the start is the identity, and the goal is $A$ with all required parities produced. The compiler searches these states using A*~\cite{hart1968astar} with the heuristic $h=$ (missing parities) $+$ (rows that differ from their target and whose target is not a missing parity). This heuristic is admissible because a CX writes one row and creates one new row value, and the CX gates that create missing parities and the last writes of different rows are distinct. The original region's two-qubit count minus one bounds the emitted CX count. A* returns a solution only when the goal is removed from its priority queue, certifying the minimum CX count for this fixed phase polynomial and coupling graph. If the budget is exhausted, no A* solution is returned; four-wire partitions use 200{,}000 expansions per search, and five-wire and larger partitions use 20{,}000. Affine offsets are absorbed into phase signs during extraction and restored by final $X$ gates, which do not affect the CX objective. The result is not optimal in a wider sense: a phase polynomial modulo $2\pi$ is not unique, so other parity sets can implement the same region, and in this model a CZ read as a diagonal gate costs two CX gates. For example, CCZ (seven parities) needs 6 CX on a triangle and 8 on a three-qubit path. The CNOT complexity of CNOT--phase circuits was studied by Amy, Azimzadeh and Mosca~\cite{amy2018graysynth}, and exact searches for small circuits were conducted by Amy et al.\ and Meuli et al.~\cite{amy2013mitm,meuli2018sat}. We use a direct search restricted to small regions and to the coupling graph, whereas gray-synth, Steiner--Gauss and their architecture-aware variants~\cite{kissinger2020steiner,meijer2023phasepoly,vandaele2022phasepoly} are heuristics for circuits of any size. HOPPS~\cite{li2025hopps} and PhasePoly~\cite{chen2026phasepoly} are the closest methods; we did not compare with them directly. Each extracted region is optimized jointly, allowing routed SWAPs and coupling-specific Toffoli decompositions to merge with surrounding logic.

\paragraph*{Finding regions in the routed circuit.} The working basis is CZ-centered, and each CZ can be read as CX in either direction or as a diagonal gate. With the readings fixed, the single-qubit segment between two two-qubit gates on a wire becomes $H^{t'}RH^{t}$, and a region can pass through it only if this segment is identified as a monomial (diagonal or antidiagonal) matrix using a $10^{-10}$ zero-entry tolerance. A local search over the readings minimizes the number of non-monomial segments, and convex regions are then grown greedily in circuit order up to a wire limit of 4, 5 or 16; each limit gives one partition, and the partition with the fewest two-qubit gates is kept. Regions of at most six wires are solved by A*. Regions of five or more wires (up to 16) are also solved by a layer-wise beam search of width 64 over the same states. This search ranks children by a weighted count of missing parities, their Hamming distance to the nearest row, and the distance of the rows to the target map, and is pruned by the admissible bound; on 5--6 wires both searches run and the shorter solution is kept. Beam-only results are heuristic. A region may use the coupling edges of the device and the qubit pairs already used, so the output is always legal. Every replacement is checked on its emitted gates: regions of at most six wires by their full unitary ($10^{-9}$, phase-aligned), larger regions by tolerance-based re-extraction of the affine map and phase polynomial from the emitted gates (mod $2\pi$, to $10^{-9}$), supplemented up to 14 wires by checks on two pseudo-random states. For the larger regions, this checks consistency of the extracted representation rather than independently certifying exact equivalence. The pass is accepted only if it reduces the two-qubit count after the following consolidation, and it runs only on circuits with at most 3000 two-qubit gates; 44 of the 80 Large-20 outputs are above this limit.

\paragraph*{Two-level scoring and a second start.} Every fully evaluated routing candidate (Sec.~\ref{sec:candidates}) undergoes the 4- and 5-wire resynthesis, which ranks the candidates. The three best then undergo the full resynthesis, with 16-wire partitions grown from both ends of the circuit and a second pass on the routed circuit before two-qubit block consolidation, because KAK resynthesis of adjacent blocks breaks structures such as $R_x$ on CX targets in Trotter chains. For inputs with at most 300 two-qubit gates and no gates on three or more qubits, level~3 also resynthesizes the regions of the \emph{logical} circuit before routing, using only qubit pairs that already interact, compiles both versions and keeps the better one. Using only the rewritten circuit made 168 development cases worse, because its structure can be less suited to block resynthesis and routing. The largest gains come from all-to-all entangling layers: each layer of 66 CX gates on 12 qubits becomes a ladder of 11.

\paragraph*{Numerical instantiation of three-qubit unitaries.} The template has a $U_3$ on each wire followed by CZ gates on coupling edges, each followed by $R_zR_y$ on both of its qubits, so $k$ CZ gates give $9+4k$ angles. For a fixed structure, the angles and a global phase are fitted using Levenberg--Marquardt~\cite{marquardt1963lm} with the elementwise residual and an analytic Jacobian, and a fit succeeds if $\max|V(\theta)-e^{i\varphi}U|\le10^{-12}$. Structures are searched using A* in the style of QSearch~\cite{davis2020qsearch} with $f=$ (CZ count) $+\,20\sqrt{1-|\mathrm{tr}(U^\dagger V)|^2/64}$. The budget counts optimizer iterations, not wall-clock time, and the kernel uses only basic IEEE operations, correctly rounded trigonometric functions and an integer pseudo-random generator, so the result is identical across platforms and thread counts. The search is applied in two places: to whole three-qubit circuits, with the layout and the output permutation free (one A* over the 18 or 6 targets $P_{\mathrm{end}}UP_{\mathrm{start}}^{T}$, the output permutation being absorbed into the final layout); and to maximal three-qubit blocks with at least four two-qubit gates in routed circuits with at most 400 two-qubit gates (at most 10 blocks per compilation, fixed layout, the block's own coupling edges, and a free output permutation for blocks at the end of the circuit).

\paragraph*{Checking of instantiated blocks.} The angles come from a numerical fit, and the replacement passes the same check as every other numerical rewrite: the $8\times8$ unitary is recomputed from the emitted gates and, after phase alignment, compared elementwise with the target to within $10^{-9}$ (the fit itself reaches $10^{-12}$, typically $10^{-15}$--$10^{-13}$). A block is replaced only if the result has strictly fewer two-qubit gates. Each replaced block is therefore within $10^{-9}$ of its target per matrix element. This bound is per block, not per circuit: one compilation may replace several blocks, and recompilation instantiates again on its own output, so the deviations add up. In an independent check of the outputs (Sec.~\ref{sec:correctness}), the whole-circuit elementwise deviation had a median of $1.6\times10^{-14}$ and a maximum of $1.02\times10^{-9}$. Instantiation does not use the approximation budget of Sec.~\ref{sec:approx}. Blocks on four or more qubits are not instantiated.

\paragraph*{Recompilation and terminal permutations.} Recompiling the routed output of Qiskit O3 with the \qs{} compiler gave fewer gates than compiling the original circuit in some routing-dominated cases (QAOA-10 on a ring: 84 instead of 96 two-qubit gates), and recompiling the compiler's own output helped as well. Level~3 therefore recompiles any result with at most 1500 two-qubit gates, taking its physical circuit as input with the identity layout and composing the layouts. There are at most two more passes, and the best result is kept. Together with the plain-SABRE family, this supplies the routing shapes that the other routers produce without calling them. Output permutations of whole three-qubit circuits and of terminal three-qubit blocks are absorbed into the final layout, which has the effect of Qiskit's elision of final permutations in these cases. On the 157-circuit development set of the final round, applying region resynthesis, then the larger regions with candidate ranking and the second start, and then instantiation with the plain-SABRE family and recompilation reduced the two-qubit ratio to Qiskit O3 from 0.783 to 0.721, 0.666 and 0.621; these increments are stacked, and Sec.~\ref{sec:ablation} reports an ablation.

\subsection{Single-qubit gauge and CX orientation}
\label{sec:gauge}
After translation to the target basis, the compiler minimizes single-qubit gates without touching the two-qubit structure. A rotation may be pushed through a two-qubit gate on a wire where the two commute ($r_z$ through a CX control or either side of a CZ, $r_x$ through a CX target). With that freedom, the problem decomposes into one chain per wire, and a beam search of width 4 decides which angle to carry into the next segment. The score is 2 per gate plus 1 per non-empty segment; counting gates alone made IBM-basis segment counts 4--8\% worse. The minimization also reverses individual CX gates using a per-wire local search (two sweeps).

\paragraph*{CX orientation.}
Because $\mathrm{CX}(a,b)=(H\otimes H)\,\mathrm{CX}(b,a)\,(H\otimes H)$, each CX carries a binary orientation $x_g$. Existing compilers fix orientation only to ensure legality on directed couplings~\cite{qiskitgatedirection} or charge reversals as a mapping cost~\cite{wille2019exact,zulehner2019astar}. On symmetric hardware, the compiler also treats the orientations as one global problem. The single-qubit cost of wire $w$ depends only on the orientations of the CX gates on $w$, so the total is $\sum_w f_w(x|_w)$, and every CX appears in exactly two terms. We duplicate each variable into $x_g^{a},x_g^{b}$ and relax the coupling constraint~\cite{fisher1981lagrangian}:
\begin{equation}
L(\lambda)=\sum_w \min_{x^{w}} \Bigl[f_w(x^{w})+\sum_{g\ni w}\pm\lambda_g x^{w}_g\Bigr].
\end{equation}
Each wire's subproblem is solved by dynamic programming along the chain, with state (orientation, beam of four carried rotations), and the multipliers are updated using a Polyak subgradient step~\cite{polyak1969minimization} (Alg.~\ref{alg:orientation}). In every iteration, each of the two copies is taken as a full orientation and evaluated, and the best orientation seen is kept; the unflipped and the all-flipped orientations are evaluated before the first iteration. Because of the beam, $L(\lambda)$ is not a lower bound, and the method is a heuristic. We regard it as an implementation detail: when evaluated against exact integer-programming solutions, it is about as good as greedy hill climbing and fails on Clifford-like circuits (Sec.~\ref{sec:optimality}). The global flip of QAOA-SK-10 on an all-to-all device (379 $\to$ 214 single-qubit gates) comes from the explicit all-flipped candidate, while the dual iterations contribute after routing (Sec.~\ref{sec:ablation}). The cost is about one second per 4000 CX gates in C++. An oriented variant is kept only if, after rescheduling, it is best in (depth, two-qubit depth, gate count).

\begin{figure}[!tbp]
\begin{algorithm}[H]
\footnotesize
\caption{CX orientation by Lagrangian dual decomposition over the wires.}
\label{alg:orientation}
\begin{algorithmic}[1]
\Require CX gates $g=1,\dots,G$ with control wire $a_g$ and target wire $b_g$; on each wire $w$ the single-qubit runs $R_{w,0},\dots,R_{w,n_w}$ between its CX gates
\Ensure orientation $x\in\{0,1\}^G$ ($x_g=1$: write $\mathrm{CX}(a_g,b_g)$ as $(H\!\otimes\!H)\,\mathrm{CX}(b_g,a_g)\,(H\!\otimes\!H)$)
\State $F(x)\gets\sum_w \textsc{Chain}_w(x|_w,0)$ \Comment{primal cost: $2\cdot$gates $+\,1\cdot$nonempty runs}
\State $x^\ast\gets\arg\min\{F(\mathbf 0),F(\mathbf 1)\}$ \Comment{no flip, global flip}
\State $\lambda\gets 0$; $\theta\gets 1$; $L^\ast\gets-\infty$
\For{$t=1,\dots,40$}
  \For{each wire $w$ whose multipliers changed}
    \State $(L_w,x^{w})\gets\textsc{Chain}_w(\text{free},\,\ell)$ with $\ell_g=+\lambda_g$ on the control wire, $-\lambda_g$ on the target wire
  \EndFor
  \State $L\gets\sum_w L_w$ \Comment{dual value}
  \State $x^{c}_g\gets x^{a_g}_g$; $x^{t}_g\gets x^{b_g}_g$ \Comment{control and target copies}
  \For{$y\in\{x^c,x^t\}$} \Comment{repair: each copy is a full orientation}
    \If{$F(y)<F(x^\ast)$} $x^\ast\gets y$ \EndIf
  \EndFor
  \If{$x^c=x^t$} \textbf{break} \Comment{copies agree}
  \EndIf
  \If{$L>L^\ast$} $L^\ast\gets L$ \textbf{else if} no dual gain for 3 iterations \textbf{then} $\theta\gets\theta/2$
  \EndIf
  \State $\lambda\gets\lambda+\theta\,\dfrac{\max(F(x^\ast)-L,\,1)}{\|x^c-x^t\|^2}\,(x^c-x^t)$ \Comment{Polyak step}
\EndFor
\State \Return $x^\ast$
\Statex
\Function{Chain$_w$}{orientations (fixed or free), linear costs $\ell$}
  \State states $\gets\{(\text{cost }0,\ \text{no carried rotation})\}$
  \For{$j=0,\dots,n_w$}
    \For{each state and each orientation $o$ of the $j$-th CX on $w$ (one if fixed; none after the last)}
      \State $M\gets$ run $R_{w,j}$ with the Hadamards of both neighboring orientations and the carried rotation
      \State options: emit $M$; or split off a leading rotation about the axis that commutes with the next CX ($Z$ on its control, $X$ on its target) and carry it forward
      \State cost $+{=}\ 2\cdot\#\text{gates}(M')+[\,M'\neq I\,]+\ell_j\,[o=1]$
    \EndFor
    \State keep the 4 cheapest states per orientation (beam), dropping equal carried angles
  \EndFor
  \State \Return the cheapest state and its orientations
\EndFunction
\end{algorithmic}
\vspace{1pt}
{\raggedright\scriptsize
$\#\text{gates}$ is the length of the run in the target's single-qubit basis. With an exact chain DP, $L$ would bound the optimum from below; with the beam the method is a heuristic. The result seeds the single-qubit gauge minimization, and the oriented variant is kept only if, after rescheduling, it is best in (depth, two-qubit depth, gate count) (Fig.~\ref{fig:pipeline}).\par}
\end{algorithm}
\end{figure}

\subsection{Scheduling on the commutation trace monoid}
\label{sec:sched}
Commutation-aware scheduling with qubits as resources was formulated by Itoko and Imamichi as job-shop scheduling and solved by constraint programming~\cite{itoko2020scheduling}; commutation DAGs~\cite{iten2022pattern} and gate reordering for QAOA~\cite{venturelli2018temporal,alam2020qaoa,lao20222qan} are related, and list scheduling was used by Guerreschi and Park~\cite{guerreschi2018twostep}. We use the equivalent trace-monoid view and standard heuristics, and add two refinements. Two gates \emph{strictly commute} if, on every qubit they share, both lie in the algebra generated by $\{I,P\}$ for the same Pauli $P$. Grouping maximal same-axis runs per qubit gives a dependency order, and every linear extension of that order is an equivalent circuit; equivalently, the circuit is an element of a trace monoid~\cite{cartier1969,diekert1995traces}. Minimizing depth over all linear extensions is a resource-constrained project scheduling problem (RCPSP) with unit durations, in which qubits are unary resources~\cite{blazewicz1983rcpsp}. The compiler uses the serial schedule-generation scheme with insertion~\cite{kolisch1996serial}, Hu levels~\cite{hu1961parallel} combined with per-qubit resource bounds as priorities, and double justification~\cite{valls2005justification}.

The two refinements are the following.
\begin{enumerate}
\item \emph{Decoupled objectives.} The two-qubit depth depends only on the relative order of the two-qubit gates on each qubit. The compiler builds three schedules (total depth only; two-qubit order preserved; two-qubit depth first, then total depth with that order fixed) and keeps the shallowest one that does not increase two-qubit depth.
\item \emph{Block-level commutation.} For each maximal two-qubit block, the compiler uses the $4\times4$ unitary to compute the Pauli axes with which the block commutes on each qubit, so blocks are scheduled like gates. A ZZ chain $\mathrm{cx}\cdot r_z\cdot\mathrm{cx}$ appears serial when viewed gate by gate. At the block level, it is an edge-coloring problem and needs at most $\Delta+1$ layers (Vizing~\cite{vizing1964}; see also~\cite{majumdar2021qaoa}). For example, a 5-layer ZZ chain fits in 2 layers.
\end{enumerate}
Reordering blocks before gauge minimization preserves their commutation, since the gauge never changes the order of two-qubit gates. A scheduling pass never changes any gate count (recompilation, which starts from the scheduled circuit, can do so; see Sec.~\ref{sec:ablation}). For 350 cases compiled by the previous version of the compiler, the resulting depth is within 0.3\% of a critical-path lower bound of the commutation DAG (Sec.~\ref{sec:optimality}), so the remaining depth gaps come from the gates themselves rather than from their order.

\subsection{Correctness and complexity}
\label{sec:correctness}
Every numerically computed rewrite is verified on its \emph{emitted} gate list, not on intermediate factors, with a phase-aligned elementwise tolerance $\max|A-e^{i\varphi}B|<10^{-9}$; the symbolic rewrites are discussed in Sec.~\ref{sec:pipeline}. Both choices matter: a trace-overlap test is only quadratically sensitive to a rotation error $\theta$, because $1-|\mathrm{tr}(A^\dagger B)|/2\approx\theta^2/8$, whereas the elementwise test is linear in it. With $\sim10^4$ rewrites per circuit, the linear-scale tolerance keeps the accumulated infidelity well below $10^{-6}$. For exact rewrites, the actual deviation is at the rounding level; the numerically instantiated three-qubit blocks use more of the tolerance. We therefore independently re-checked all 420 outputs that use instantiation in the arXiv, synthetic and \texttt{guoq\_dev} runs of the final compiler. QCEC~\cite{burgholzer2021qcec} proved 357 of them equivalent (90 only at a relaxed threshold of $10^{-10}$), one 33-qubit output was inconclusive, and the other 62 were checked densely. QCEC reported 30 of these 62 as non-equivalent even at $10^{-10}$ (26 by its simulation checker), but their full unitaries (16 outputs, elementwise deviation at most $1.0\times10^{-9}$) or dense state vectors with random product inputs ($1-F\le1.2\times10^{-12}$) agree with the ideal circuit, so we attribute these reports to numerical error in the decision-diagram checks on the generic angles produced by instantiation (as for \texttt{qram\_20} in limitation (ii)). Dense state vectors with random product inputs gave $1-F\le1.2\times10^{-12}$ on all 417 outputs with at most 27 qubits. For the 324 outputs small enough for a full unitary (at most 12 qubits after compaction), the elementwise deviation from the ideal circuit after phase alignment has a median of $1.6\times10^{-14}$; it exceeds $10^{-10}$ in 7 outputs and $10^{-9}$ in one, \texttt{small\_vqe\_uccsd\_n4\_4} on a $2\times2$ grid ($1.02\times10^{-9}$, with $1-|\mathrm{tr}|/d=1.3\times10^{-14}$), where deviations from several instantiated blocks and recompilation accumulate. These error measurements apply to the outputs checked. They are far below the $10^{-6}$ infidelity threshold of the benchmark's check, but the outputs are not exact to rounding: each replaced block is verified to $10^{-9}$ per element on its emitted gates, and the whole circuit is not bounded by that tolerance. The costs of the main components are as follows. SABRE is linear in the number of routing steps times the size of the front and lookahead sets. VF2 stops at its state budget. One annealing move of the QAP costs $O(\deg)$. LNS is exponential only in the window length ($k\le12$) and runs under an expansion budget. A layered-routing sweep costs $O(m^2)$ per layer with $m$ blocks. The trace-monoid scheduler is near-linear in practice.

\section{Evaluation methodology}
\label{sec:method}

\subsection{Development process}
\label{sec:devprocess}
The \qs{} compiler was developed iteratively using the development suites of Sec.~\ref{sec:suites}. The final round of development targeted BQSKit and GUOQ on a separate development set of 157 circuits with 3--16 qubits (\texttt{guoq\_dev}, Sec.~\ref{sec:suites}), which contains no circuit identical to an arXiv circuit: we compared normalized gate sequences, unitaries up to global phase (up to 10 qubits) and unitaries up to qubit permutation and global phase (up to 9 qubits), and found no match. On this set, region resynthesis reduced the two-qubit ratio to Qiskit O3 from 0.783 to 0.721. The larger regions with candidate ranking and the second start then reduced it to 0.666, and instantiation with the plain-SABRE family and recompilation then reduced it to 0.621. These increments are stacked, and Sec.~\ref{sec:ablation} reports an ablation. The arXiv evaluation set (Sec.~\ref{sec:evalprotocol}) was evaluated before the final round, and its aggregate results informed the final development direction. The final evaluation therefore includes development-time exposure to aggregate arXiv results.

The evaluated version is commit \texttt{be4ecda}, completed on 8 November 2026; it was released in \qs{} v0.6.3 on 12 November 2026~\cite{quspect2026}. All \qs{} results in Sec.~\ref{sec:eval} were computed with this commit on 8 November 2026. Measurements that refer to the previous version of the compiler (commit \texttt{8ec73d8}, preserved as tag \texttt{paper-r20}) or to the earlier release v0.6.0 are marked as such.

\subsection{Benchmark design and metrics}
\label{sec:suites}
\textbf{Inputs.} Every circuit is represented as a single OpenQASM~2 text that all compilers read. Final measurements and barriers are removed.
\textbf{Targets.} Each target is a gate set plus a coupling graph. The gate sets are \emph{CZ} $=\{\mathrm{cz},h,r_z,s,s^\dagger,t,t^\dagger,x,y,z\}$, \emph{CX} (the same with cx), and \emph{IBM} $=\{\mathrm{cx},r_z,\sqrt{x},x\}$. The topologies are \emph{line}, \emph{ring}, \emph{grid}, \emph{heavy-hex} and \emph{full} (all-to-all), each the smallest instance that holds the circuit.
\textbf{Validity.} An output is \texttt{invalid} if it contains a gate outside the target set, contains a two-qubit gate on an uncoupled pair, or fails the state-vector check. The check prepares random product states, aligns the initial and final layouts, and requires fidelity $\ge 1-10^{-6}$. It is run on outputs with up to 20 physical qubits; larger outputs are counted as successful but unverified. Outputs of \qs{} O3 with more than 20 physical qubits were checked separately with MQT QCEC~\cite{burgholzer2021qcec} and stabilizer simulation (limitation (ii)).
\textbf{Metrics.} All outputs are converted to one circuit representation and counted by the same code: two-qubit gate count (primary), depth, two-qubit depth and single-qubit gate count.
\textbf{Aggregation.} A \emph{case} is a (circuit, topology instance) pair. For each case and compiler, we take the median over seeds of successful runs. We then report the geometric mean over cases of the ratio to Qiskit O3, so lower is better. A win, tie or loss is determined by comparing per-case medians; equality is a tie. Pairwise ratios use the cases where both compilers succeeded.

\textbf{Suites.}
\emph{Synthetic}: seven parameterized families (GHZ, Bernstein--Vazirani~\cite{bernstein1997}, QFT, QAOA on random 3-regular graphs with $p=2$, Heisenberg-chain Trotter steps, random circuits, and the Cuccaro adder~\cite{cuccaro2004adder}). Circuits with widths 6, 10, 16 and 20 run on all five topologies with three seeds; those with widths 32 and 50 run on four topologies with one seed and a 900~s limit, for 196 cases in total.
\emph{Standard-42}: 42 circuits comprising circuits from MQT Bench, QASMBench and the Qiskit library as well as two custom circuits, with 6--25 qubits, run on five topologies with three seeds (210 cases) under each of the three gate sets.
\emph{Large-20}: 20 circuits with 40--100 qubits and up to 15{,}000 two-qubit gates, run on four topologies with one seed and a 900~s limit (80 cases).
All three are development suites. The final round of development (Sec.~\ref{sec:devprocess}) also used \texttt{guoq\_dev}, 157 circuits with 3--16 qubits on the five topologies under CZ (785 cases): 37 from Standard-42, 33 from the QASMBench circuits of Benchpress, 21 instances of the synthetic families (6, 10 and 16 qubits), 14 Toffoli benchmarks of Nam et al.~\cite{nam2018automated}, 28 RevLib circuits from GUOQ's benchmark collection, and 24 small MQT Bench circuits. It overlaps with Standard-42 and the synthetic suite and shares no circuit with the arXiv set.

\subsection{Evaluation protocol}
\label{sec:evalprotocol}
Development may have overfitted to the suites above, to the opponents used during development (Qiskit and tket, and in the final round GUOQ and BQSKit), or to the benchmark itself. The final evaluation therefore adds three things to the development suites. None of them is an untouched test of the final version; the exposure is stated below and in Sec.~\ref{sec:beam-caveats}.

\begin{figure}[t]
\centering
\includegraphics[width=\linewidth]{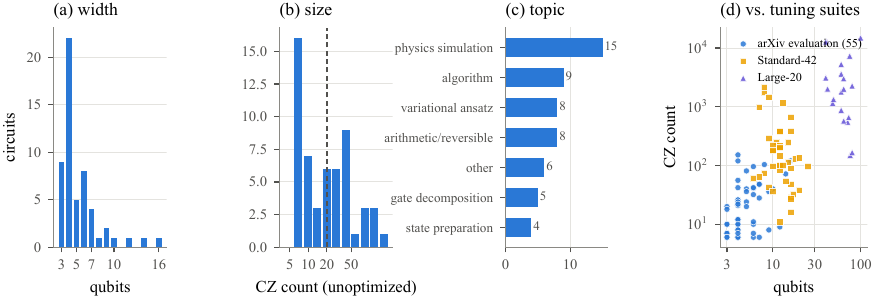}
\caption{The arXiv evaluation set (55 circuits, 41 papers). (a) Width. (b) CZ count after unoptimized translation (dashed line: median, 20). (c) Topic. (d) Width and size compared with the development suites.}
\label{fig:dataset}
\end{figure}

\emph{arXiv evaluation set.}
A pipeline collected key circuits from the \LaTeX{} sources of quant-ph papers submitted between 2026-06-28 and 2026-10-28. It started from 8158 papers, of which 4983 passed a keyword screen. Sources were downloaded for 3411 papers, and 448 papers with 1514 \texttt{quantikz}/\texttt{qcircuit}/OpenQASM figures were found. Of those figures, 441 contained no abstract boxes and could be converted exactly, and 141 met the size floor of at least 3 qubits and at least 6 CX gates. After review, 55 circuits from 41 papers were kept. The acceptance rules were strict. A circuit had to be transcribed gate by gate from the source; circuits with abstract boxes, ellipses or bundled wires were rejected, as were textbook gadgets. Each paper could contribute at most three structurally distinct circuits. Every circuit was checked for unitary equivalence with the parsed figure and functionally checked against the paper's claims. For the earlier two months (2026-06-28 to 2026-08-27) we downloaded only the 688 keyword-screened papers with ``circuit'' in the title or abstract, because downloads were slow and the yield of the other papers was very low; 24 of the 25 accepted papers from that period are among them. Every decision was logged against a written review brief. Where a paper gives no concrete angles, the angles are fixed pseudo-random values derived from the symbol names (in 21 circuits), which does not change the circuit structure. Measurements were removed; in 8 circuits, these included mid-circuit measurements, and circuits whose meaning depends on mid-circuit measurement, reset or classical feedback were rejected. The circuits were submitted between 2026-07-01 and 2026-10-27. They have 3--16 qubits (median 4) and 6--152 CZ gates after unoptimized translation (median 20). Figure~\ref{fig:dataset} and Appendix~\ref{app:arxiv} give the statistics and the full list; the circuits, the manifest, the review log and the extraction tools are public at \url{https://github.com/QuSpect/arxiv-circuits}. These circuits broaden the source distribution beyond the development suites. The set was evaluated before the final round of development (with the previous version of the compiler), the per-case results of that run were inspected, and its aggregate results informed the final development direction. The final round targeted BQSKit and GUOQ on the separate 157-circuit development set, which contains no circuit identical to an arXiv circuit (Sec.~\ref{sec:devprocess}). The final evaluation therefore includes development-time exposure to aggregate arXiv results. The final version was evaluated on the set once, on 2026-11-08, on all five topologies with three seeds; later runs on it (ablations and bit-identity checks, Secs.~\ref{sec:ablation} and~\ref{sec:correctness}) reproduced that evaluation and were not used for development. All arXiv numbers in this paper are from that run unless marked as referring to the previous version.

\emph{Compilers not used during development.}
Cirq~1.7.0~\cite{cirq} and MQT QMAP~3.11~\cite{wille2023qmap} (heuristic A* mapper~\cite{zulehner2019astar}, plus the exact SAT mapper~\cite{wille2019exact} with a 120~s limit and seed~0 only) were not used during development. BQSKit~1.2.1~\cite{bqskit2021} (optimization level~2, $\epsilon=10^{-10}$, mapping enabled; its PAS/PAM workflow~\cite{liu2023pam} is not part of level~2 and was not tested) was one of the two targets of the final round, together with GUOQ (Sec.~\ref{sec:competitors}). The outputs of these compilers are rebased onto the target gate set by an exact, gate-by-gate translation that performs no optimization across two-qubit gates. QMAP does not resynthesize two-qubit blocks, so its numbers measure a mapper, not a full optimizing compiler.

\emph{Hardware.} The hardware experiment (Sec.~\ref{sec:hw}) uses a chip and coupling graphs that were never used during development.

The ablation and gap experiments on the previous version, the first scaling experiment and the beam-router experiments were run before the final round, after the arXiv set had first been evaluated. The beam router and the methods of the final round were therefore designed with knowledge of the previous version's results on the arXiv set (Sec.~\ref{sec:beam-caveats}).

\subsection{Setup}
\label{sec:setup}
All results in Sec.~\ref{sec:eval} use the final version of the \qs{} compiler (commit \texttt{be4ecda}), run once on 2026-11-08. Only the \qs{} compiler was run on that day, and the results of the other compilers, computed earlier with the same circuits, seeds, limits and checks, are reused per case. The reference opponents used during development are Qiskit~2.6.1 (\texttt{transpile}, levels 1 and 3) and pytket~2.18.5 (O2: FullPeepholeOptimise, default mapping, KAK, CliffordSimp and SynthesiseTket). The other opponents are listed in Secs.~\ref{sec:evalprotocol} and~\ref{sec:competitors}; ``BQSKit O2'' in tables and figures denotes BQSKit at level~2. Comparative statements in this paper refer to these tested configurations. The \qs{} rows (``quspect O0'', ``quspect O1'' and ``quspect O3'' in all tables and figures) represent optimization levels of the \qs{} compiler, named after the \texttt{transpile} level they use, with one exception kept for continuity with the development records: ``quspect O1'' denotes level~2, the default pipeline (not level~1), ``quspect O3'' denotes level~3, and ``quspect O0'' denotes level~0 (decomposition and SABRE routing without optimization). Most runs were executed on one Apple M3 Ultra workstation with 24 parallel jobs. Some ran on separate Linux nodes: Cirq and QMAP on Standard-42 (IBM), part of the BQSKit runs on Standard-42, BQSKit on the large devices (Sec.~\ref{sec:scale}), QUESO (Sec.~\ref{sec:competitors}) and the BQSKit baselines of \texttt{guoq\_dev}.

\section{Results}
\label{sec:eval}

\begin{table*}[t]
\caption{Final comparison: geometric-mean ratio to Qiskit O3 (two-qubit count / depth; lower is better). Each cell uses the cases where that compiler and Qiskit O3 both succeeded. QMAP (exact) ran only on the arXiv set, with seed 0 and a 120~s limit, and solved 231 of 275 cases; BQSKit finished 36 of 80 large cases within 900~s (failures in Table~\ref{tab:fail}). \qs{} rows: final version (commit \texttt{be4ecda}); other compilers: the runs made for the evaluation of the previous version, reused per case. ``BQSKit O2'' is BQSKit at level~2. ``--'' means not run in that evaluation. BQSKit O2 was run under the CX and IBM bases afterwards (0.968 / 2.330 and 1.054 / 2.219); these results and those for further opponents are reported in Sec.~\ref{sec:competitors} and Table~\ref{tab:competitors}.}
\label{tab:main}
\footnotesize
\begin{tabular}{lcccccc}
\toprule
Compiler & arXiv & Std CZ & Std CX & Std IBM & Synth. & Large \\
\midrule
quspect O3 & \textbf{0.616} / \textbf{0.592} & \textbf{0.607} / \textbf{0.496} & \textbf{0.593} / \textbf{0.518} & \textbf{0.648} / \textbf{0.504} & \textbf{0.770} / \textbf{0.678} & \textbf{0.758} / \textbf{0.600} \\
quspect O1 & 0.805 / 0.677 & 0.806 / 0.605 & 0.808 / 0.617 & 0.879 / 0.622 & 0.893 / 0.716 & 0.848 / 0.636 \\
quspect O0 & -- & 1.147 / 1.199 & 1.151 / 1.150 & 1.251 / 1.357 & 1.132 / 1.172 & 1.172 / 1.231 \\
Qiskit O3 & 1.000 / 1.000 & 1.000 / 1.000 & 1.000 / 1.000 & 1.000 / 1.000 & 1.000 / 1.000 & 1.000 / 1.000 \\
Qiskit O1 & 1.036 / 1.061 & 1.128 / 1.159 & 1.132 / 1.171 & 1.229 / 1.043 & 1.171 / 1.020 & 1.159 / 1.219 \\
tket O2 & 0.930 / 1.003 & 0.946 / 0.980 & 0.949 / 1.087 & 1.029 / 0.936 & 1.029 / 1.062 & 1.179 / 1.340 \\
Cirq & 1.119 / 2.018 & 1.157 / 2.043 & -- & 1.260 / 2.339 & 1.239 / 2.312 & 1.441 / 2.817 \\
QMAP (heuristic) & 1.284 / 1.798 & 1.403 / 1.983 & -- & 1.532 / 1.208 & 1.671 / 2.646 & 1.461 / 2.675 \\
QMAP (exact) & 1.013 / 1.447 & -- & -- & -- & -- & -- \\
BQSKit O2 & 0.853 / 1.552 & 0.955 / 1.671 & -- & -- & 1.066 / 2.038 & 1.146 / 1.928 \\
\bottomrule
\end{tabular}

\end{table*}

\subsection{Main results}

\begin{figure*}[t]
\centering
\includegraphics[width=\textwidth]{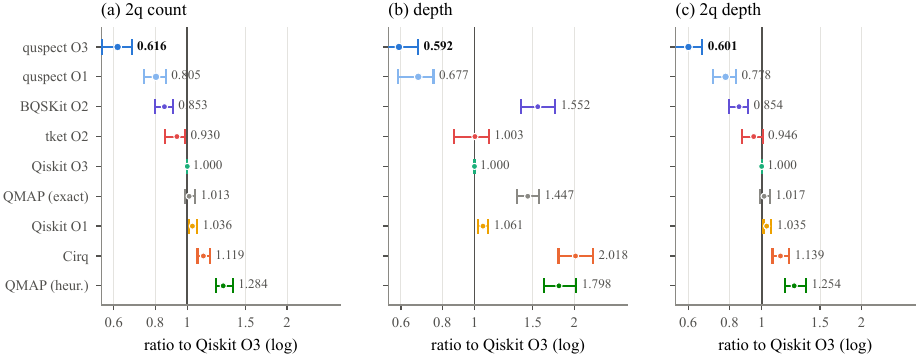}
\caption{arXiv evaluation set, 275 cases (55 circuits $\times$ 5 topologies, CZ basis, median of 3 seeds). Dots are geometric-mean ratios to Qiskit O3; bars are 95\% bootstrap intervals that resample the 55 circuits together with all their topologies (circuit-cluster bootstrap, 2000 resamples). \qs{}: final version, evaluated once on this set. Cirq and both QMAP modes were not used during development; BQSKit was a target of the final round on a separate development set. The evaluation includes prior exposure to aggregate results on this set (Sec.~\ref{sec:evalprotocol}).}
\label{fig:forest}
\end{figure*}

\begin{figure*}[t]
\centering
\includegraphics[width=0.8\textwidth]{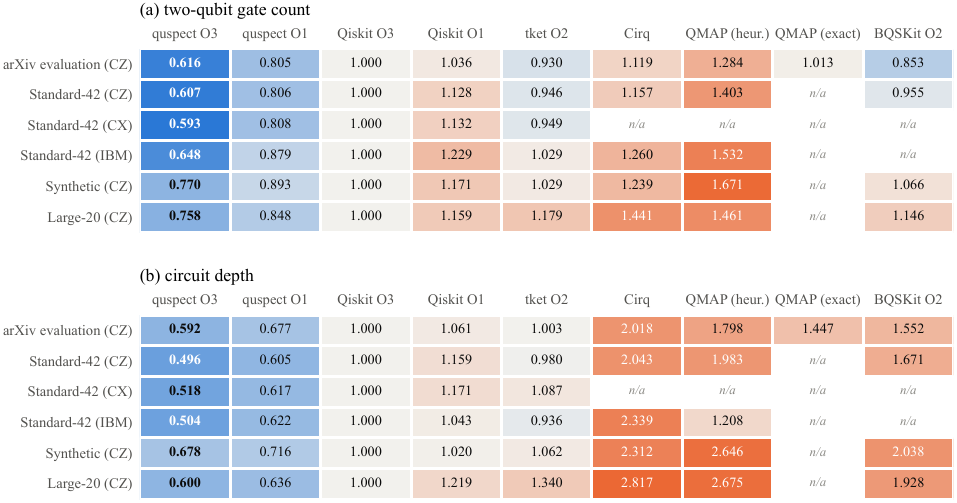}
\caption{Geometric-mean ratio to Qiskit O3 for every suite and compiler: (a) two-qubit count, (b) depth. Blue is better than Qiskit O3 and orange worse. \qs{}: final version. ``n/a'' marks combinations for which the other compilers were not run (for BQSKit under CX and IBM, see Table~\ref{tab:competitors}).}
\label{fig:heatmap}
\end{figure*}

Table~\ref{tab:main} and Fig.~\ref{fig:heatmap} summarize all suites, and Fig.~\ref{fig:forest} shows the arXiv evaluation set. On the arXiv set, quspect O3 reaches 0.616 [0.555, 0.679] in two-qubit count, 0.592 [0.507, 0.676] in depth and 0.601 [0.541, 0.662] in two-qubit depth (95\% intervals from a bootstrap over the 55 circuits; resampling the 41 papers instead gives [0.548, 0.687] for the two-qubit count, and resampling the 275 cases independently would give the narrower [0.587, 0.645]). The previous version reached 0.787, 0.669 and 0.768. The final version uses fewer two-qubit gates than the previous one in 187 of the 275 cases and more in one, by one gate. BQSKit at level~2 has the next-lowest two-qubit ratio (0.853 [0.800, 0.904]), with 1.552 times the depth and 2.271 times the single-qubit count of Qiskit O3. tket O2 follows at 0.930, and the exact QMAP mapper is at 1.013 on the cases it solved. Restricted to the 274 cases on which all compilers except QMAP (exact) succeeded, the ratios are 0.615 for the \qs{} compiler and 0.853 for BQSKit. Across topologies (computed from the per-case data), the advantage of the \qs{} compiler over BQSKit is largest on grid and heavy-hex (0.603 and 0.600 against 0.879 and 0.876) and smallest on the all-to-all target (0.686 against 0.891). The two-qubit ratio on the arXiv set (0.616) lies between those on Standard-42 (0.607) and the synthetic suite (0.770) and is close to the 0.621 reached on the development set of the final round. Because the evaluation includes prior exposure to aggregate arXiv results (Sec.~\ref{sec:evalprotocol}), we do not interpret this agreement as evidence about overfitting.

On the development suites the advantage holds across gate sets. The two-qubit ratios are 0.607, 0.593 and 0.648 under the CZ, CX and IBM bases (0.747, 0.749 and 0.815 for the previous version), and the depth ratios are 0.496, 0.518 and 0.504. The ratios are 0.770/0.678 on the synthetic suite and 0.758/0.600 on Large-20; the final-round components help less on these suites, whose circuits are larger, because recompilation, three-qubit instantiation and CNOT--phase resynthesis are limited to circuits with at most 1500, 400 and 3000 two-qubit gates. Compared with the previous version, quspect O3 uses more two-qubit gates in two cases in the development suites, both on Large-20: the HRS multiplier-41 on heavy-hex (16{,}902 against 16{,}869) and inexact QPE-60 on the grid (3779 against 3777). Among the other compilers in Table~\ref{tab:main}, BQSKit (0.955) and tket O2 (0.946) on Standard-42 (CZ) and tket O2 under CX (0.949) beat Qiskit O3 on these suites in the geometric mean; of the later opponents, BQSKit under CX (0.968) and the AI transpiler passes under CZ (0.953) do so narrowly (Sec.~\ref{sec:competitors}). On Standard-42 (CZ) the two-qubit ratios of Cirq, QMAP and BQSKit are 1.157, 1.403 and 0.955; BQSKit's depth ratio there is 1.671. Under the IBM basis Cirq and QMAP reach 1.260 and 1.532. On the synthetic suite Cirq, QMAP and BQSKit reach 1.239, 1.671 and 1.066. On Large-20 tket, Cirq, QMAP and BQSKit reach 1.179, 1.441, 1.461 and 1.146. The last ratio is for BQSKit's 36 completed cases (\qs{}: 0.788 on the same cases).

\begin{table*}[t]
\caption{Two-qubit-count win/tie/loss record of quspect O3 against each compiler, per case. The italic row compares quspect O3 with the \emph{per-case best tested opponent configuration}, i.e.\ the minimum over all non-\qs{} compilers in Table~\ref{tab:main}, including QMAP (exact) where it ran. The record including the opponents added in Sec.~\ref{sec:competitors} is given in Table~\ref{tab:competitors}.}
\label{tab:wtl}
\footnotesize
\begin{tabular}{lcccccc}
\toprule
vs. & arXiv & Std CZ & Std CX & Std IBM & Synth. & Large \\
\midrule
quspect O3 & \emph{195/71/9} & \emph{169/36/5} & \emph{175/35/0} & \emph{175/35/0} & \emph{121/71/4} & \emph{66/11/3} \\
quspect O1 & 194/81/0 & 168/42/0 & 175/35/0 & 175/35/0 & 114/82/0 & 61/19/0 \\
Qiskit O3 & 228/46/1 & 176/34/0 & 180/30/0 & 180/30/0 & 131/65/0 & 70/10/0 \\
Qiskit O1 & 234/41/0 & 184/26/0 & 189/21/0 & 189/21/0 & 158/38/0 & 73/7/0 \\
tket O2 & 216/59/0 & 171/39/0 & 176/34/0 & 176/34/0 & 133/59/4 & 71/7/1 \\
Cirq & 240/35/0 & 187/23/0 & -- & 184/19/0 & 147/49/0 & 70/7/1 \\
QMAP (heuristic) & 250/25/0 & 194/16/0 & -- & 195/13/0 & 172/23/0 & 73/5/0 \\
QMAP (exact) & 200/31/0 & -- & -- & -- & -- & -- \\
BQSKit O2 & 213/53/8 & 184/21/5 & -- & -- & 137/48/0 & 27/8/1 \\
\bottomrule
\end{tabular}

\end{table*}

\begin{figure*}[t]
\centering
\includegraphics[width=0.78\textwidth]{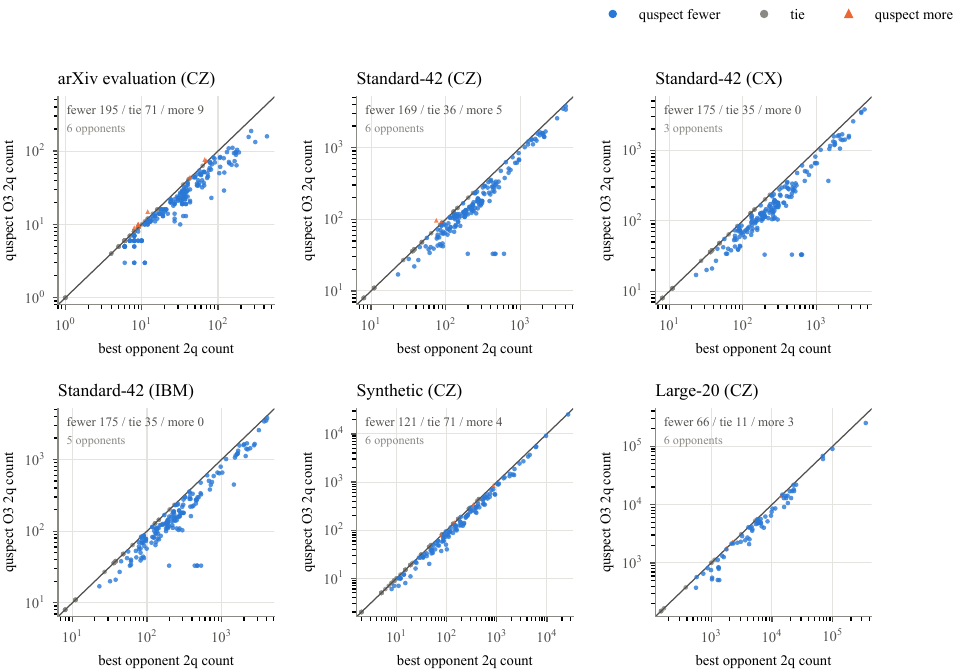}
\caption{Two-qubit count of quspect O3 compared with that of the best tested opponent configuration for each case (minimum over all external compilers, excluding the seed-0-only QMAP (exact)). Points below the diagonal are cases in which the \qs{} compiler uses fewer gates.}
\label{fig:scatter}
\end{figure*}

\begin{figure}[t]
\centering
\includegraphics[width=\linewidth]{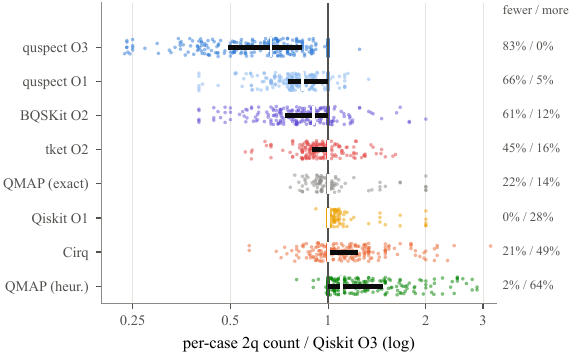}
\caption{arXiv evaluation set: per-case two-qubit ratio to Qiskit O3. Thick bars are interquartile ranges and white ticks medians. The right column gives the share of cases with fewer or more gates than Qiskit O3.}
\label{fig:dist}
\end{figure}

\paragraph*{Per-case view.}
Table~\ref{tab:wtl} and Fig.~\ref{fig:scatter} compare cases one at a time. Against the best tested opponent configuration per case among the compilers in Table~\ref{tab:main}, quspect O3 wins or ties in 266 of 275 cases on the arXiv set (195/71/9; previous version: 98/110/67), 205 of 210 on Standard-42 (CZ; 169/36/5), all 210 under CX (175/35/0; only Qiskit and tket ran there) and under IBM (175/35/0), 192 of 196 on the synthetic suite (121/71/4) and 77 of 80 on Large-20 (66/11/3). With the opponents added in Sec.~\ref{sec:competitors} (BQSKit under CX and IBM, the AI transpiler passes and, on the arXiv set, the superoptimizers), the records become 175/89/11 on the arXiv set and 169/36/5, 174/35/1 and 174/35/1 on Standard-42 under CZ, CX and IBM. Against Qiskit O3, quspect O3 loses one case on the arXiv set (a 6-qubit circuit on a line, 15 against 12) and none on the other suites, under any gate set. The remaining losses on the synthetic suite and Large-20 are close: on random circuits on the all-to-all target the \qs{} compiler uses one more gate than tket (four cases), and on Clifford+T-60 (all-to-all) it uses 1776 two-qubit gates against 1773 for tket. Figure~\ref{fig:dist} shows the distribution on the arXiv set. quspect O3 uses fewer two-qubit gates than Qiskit O3 in 83\% of the cases and more in 0.4\%; for BQSKit the shares are 61\% and 12\%, and for tket 45\% and 16\%.

\subsection{Failure analysis}
\label{sec:failure}
On the arXiv set the per-case record against the best tested opponent configuration among the compilers of Table~\ref{tab:main} is 195 wins, 71 ties and 9 losses (175/89/11 when the superoptimizers and AI passes of Sec.~\ref{sec:competitors} are included). For the previous version it was 98/110/67, with BQSKit O2 the best tested configuration in 66 of the 67 losses; the record against BQSKit alone has moved from 129/79/66 to 213/53/8. Eight of the nine losses are to BQSKit, on four circuits: a 4-qubit boson-sampling circuit on all five topologies (10 against 9 CZ, all-to-all 9 against 8), a 5-qubit boson-sampling circuit on the ring (74 against 70), a 4-qubit three-photon beam-splitter circuit on the all-to-all target (77 against 67) and a 9-qubit multi-controlled gate with a dirty ancilla on heavy-hex (44 against 43). BQSKit saves a median of 1 CZ there (maximum 10); its circuits are on average 1.87 times as deep as those of the \qs{} compiler in these cases. BQSKit's level~2 resynthesizes partitioned three-qubit blocks numerically to $\epsilon=10^{-10}$ before and after its SABRE mapping and deletes gates within $\epsilon$, whereas the \qs{} compiler instantiates only selected three-qubit blocks after routing (Sec.~\ref{sec:postroute}); we did not attribute the remaining savings to a particular step. The ninth loss is to Qiskit O3, on a 6-qubit circuit on a line (15 against 12 CZ). Three BQSKit runs failed because its local runtime crashed and are counted as errors.

On the development suites, too, the losses that remain are few and small. On Standard-42 (CZ), all 5 losses against the best tested opponent configuration are to BQSKit: the BMW quark-cardinality circuit on four topologies (90 against 87 CZ) and QPE-8 with random $U$ on the all-to-all target (97 against 75). BQSKit's circuits are 1.25--1.69 times as deep as those of the \qs{} compiler. The previous version had 23 losses on 9 circuits, 19 of them to BQSKit with a median saving of 28 CZ. On the synthetic suite the 4 losses are on random circuits on the all-to-all target, each by one gate to tket, and the \qs{} compiler no longer loses a case to BQSKit there (137/48/0, against 118/54/13 for the previous version, for which all Cuccaro adders and QFT-32/50 were losses). On Large-20 the 3 losses occur for Clifford+T-60 on the all-to-all target (1776 against 1773 for tket) and on the grid (14{,}247 against 14{,}107 for Cirq) and for the Draper adder-64 on the all-to-all target (2289 against 2246 for BQSKit). Under the CX and IBM bases the \qs{} compiler loses no case to the compilers of Table~\ref{tab:main}; with BQSKit added it loses one, QPE-8 on the all-to-all target (Sec.~\ref{sec:competitors}). On average BQSKit uses more two-qubit gates than the \qs{} compiler on both Standard-42 and the synthetic suite (0.955 and 1.066 against 0.607 and 0.770, relative to Qiskit O3), and against BQSKit alone the records are 184/21/5 and 137/48/0. Table~\ref{tab:fail} lists every non-success; it includes 21 Cirq and 8 QMAP time-outs on Standard-42 (IBM), and on the synthetic suite 14 BQSKit runtime crashes (all BV on heavy-hex) and 5 BQSKit time-outs (QFT-50, random-50). The \qs{} compiler had no failures.

\begin{table}[t]
\caption{Non-successful compilations (runs count individual seeds; cases ok counts cases with at least one successful seed). QMAP (exact) errors are 120~s time-outs of the SAT subprocess. The \qs{} compiler had no failures in any suite in this table, in either the previous or the final version. All other compiler/suite pairs succeeded everywhere. Failures of the opponents added later are given in Sec.~\ref{sec:competitors}.}
\label{tab:fail}
\scriptsize
\setlength{\tabcolsep}{1.8pt}
\begin{tabular}{llrrrrr}
\toprule
Suite & Compiler & runs & error & timeout & invalid & cases ok \\
\midrule
arXiv & QMAP (exact) & 275 & 44 & 0 & 0 & 231/275 \\
arXiv & BQSKit O2 & 825 & 3 & 0 & 0 & 274/275 \\
Std IBM & Cirq & 630 & 0 & 21 & 0 & 203/210 \\
Std IBM & QMAP (heuristic) & 630 & 0 & 8 & 0 & 208/210 \\
Synth. & tket O2 & 476 & 0 & 0 & 2 & 196/196 \\
Synth. & QMAP (heuristic) & 476 & 0 & 1 & 0 & 195/196 \\
Synth. & BQSKit O2 & 476 & 14 & 5 & 0 & 185/196 \\
Large & tket O2 & 80 & 0 & 1 & 0 & 79/80 \\
Large & Cirq & 80 & 0 & 2 & 0 & 78/80 \\
Large & QMAP (heuristic) & 80 & 0 & 2 & 0 & 78/80 \\
Large & BQSKit O2 & 80 & 0 & 44 & 0 & 36/80 \\
\bottomrule
\end{tabular}

\end{table}

Some depth weaknesses remain, because level~3 selects by two-qubit count first. In the previous version the candidate with the fewest gates could be much deeper than Qiskit's on large grids (VQE-two-local-60 on a $7\times9$ grid: depth 13{,}492 against 4932, because that candidate chained its SWAPs; AQFT-80 on a grid: 5044 against 3921 for tket). In the final version a different candidate wins in both cases, with depths of 3421 and 2194 and fewer two-qubit gates (12{,}887 and 4010). The largest remaining depth deficits occur for QFT-50 on the all-to-all target (980 against 632 for Qiskit O3), the hidden-linear-function-60 circuit on the grid (2122 against 1431) and Draper adder-64 and inexact QPE-60 on the all-to-all target (about 1.4 times). The output depth of the \qs{} compiler exceeds that of Qiskit O3 in 7 of the 80 Large-20 cases, 7 of the 196 synthetic cases and 1 of the 210 Standard-42 cases, and that of both Qiskit O3 and tket O2 in 2, 4 and 0 of them. An optional depth weight trades two-qubit count for depth in such cases but is off by default, in keeping with the metric order.

\subsection{Superoptimizers and further compilers}
\label{sec:competitors}
The comparison above leaves out the superoptimizers (Sec.~\ref{sec:background}), learned routing and synthesis passes, the Benchpress circuits and BQSKit under the CX and IBM bases. We ran them after the evaluation of the previous version, following the protocol of Sec.~\ref{sec:suites}: cases are (circuit, topology) pairs, the median over successful seeds is taken, ratios are geometric means relative to Qiskit O3 over the cases where both succeeded, and every output passes the gate-set, coupling and (up to 20 physical qubits) state-vector checks. The opponents were not re-run for the final version; the \qs{} entries are the final results of Table~\ref{tab:main}, matched per case. Table~\ref{tab:competitors} summarizes the new opponents. GUOQ and BQSKit were also the targets of the final round, on its development set, which contains no arXiv circuits but includes 37 circuits from Standard-42.

\begin{table*}[t]
\caption{Further opponents: geometric-mean ratio to Qiskit O3 (two-qubit count / depth) and the two-qubit-count win/tie/loss of quspect O3 (final version) against each, per case. GUOQ, QUESO and Quartz run on the routed output of Qiskit O3 (seed~0, 300~s per case; Quartz 120~s), only on the arXiv set and with additional edge restrictions for GUOQ and QUESO (see text). Non-equivalent outputs are excluded, so Quartz covers 257 of 275 cases and the AI transpiler passes cover 274, 198 and 199 cases; the \qs{} ratios on the same cases differ by at most 0.010. The superoptimizers ran with seed~0 only, whereas the \qs{} entries are medians over three seeds; with seed~0 only, quspect O3 reaches 0.618 on the arXiv set. The BQSKit entries under arXiv and Std CZ repeat Table~\ref{tab:main}. The last row is the record against the per-case best tested opponent configuration over \emph{all} external compilers, those of Table~\ref{tab:main} included; the record over the compilers of Table~\ref{tab:main} alone (Table~\ref{tab:wtl}) is given in parentheses. ``--'': not run.}
\label{tab:competitors}
\footnotesize
\setlength{\tabcolsep}{2.2pt}
\begin{tabular}{lcccccccc}
\toprule
 & \multicolumn{2}{c}{arXiv (CZ)} & \multicolumn{2}{c}{Std CZ} & \multicolumn{2}{c}{Std CX} & \multicolumn{2}{c}{Std IBM} \\
\cmidrule(lr){2-3}\cmidrule(lr){4-5}\cmidrule(lr){6-7}\cmidrule(lr){8-9}
Compiler & ratio & W/T/L & ratio & W/T/L & ratio & W/T/L & ratio & W/T/L \\
\midrule
quspect O3 & \textbf{0.616} / \textbf{0.592} & -- & \textbf{0.607} / \textbf{0.496} & -- & \textbf{0.593} / \textbf{0.518} & -- & \textbf{0.648} / \textbf{0.504} & -- \\
BQSKit O2 & 0.853 / 1.552 & 213/53/8 & 0.955 / 1.671 & 184/21/5 & 0.968 / 2.330 & 190/19/1 & 1.054 / 2.219 & 190/19/1 \\
Qiskit O3 $\to$ GUOQ & 0.881 / 1.423 & 208/65/2 & -- & -- & -- & -- & -- & -- \\
Qiskit O3 $\to$ QUESO & 0.932 / 1.477 & 218/57/0 & -- & -- & -- & -- & -- & -- \\
Qiskit O3 $\to$ Quartz & 0.922 / 1.515 & 192/65/0 & -- & -- & -- & -- & -- & -- \\
Qiskit AI passes O3 & 0.964 / 1.769 & 218/56/0 & 0.953 / 1.850 & 165/33/0 & -- & -- & 1.035 / 0.956 & 171/28/0 \\
\midrule
\emph{best tested} & & \emph{175/89/11} & & \emph{169/36/5} & & \emph{174/35/1} & & \emph{174/35/1} \\
 & & (195/71/9) & & (169/36/5) & & (175/35/0) & & (175/35/0) \\
\bottomrule
\end{tabular}
\end{table*}

\paragraph*{BQSKit under the CX and IBM bases.} With the configuration used under CZ (BQSKit 1.2.1, \texttt{compile} at level~2 with mapping, $\epsilon=10^{-10}$, three seeds, 1200~s limit), BQSKit O2 completed all 630 runs on Standard-42 under each basis. It reaches 0.968 under CX and 1.054 under IBM, with depth ratios of 2.33 and 2.22. The \qs{} compiler wins, ties and loses 190/19/1 against it under both bases (184/21/5 under CZ; previous version: 164/27/19 under CX and IBM and 162/29/19 under CZ), and the one loss is on QPE-8 with random $U$ on the all-to-all target (98 and 86 against 75), where BQSKit's numerical resynthesis of the multi-qubit controlled blocks saves gates. Under the IBM basis BQSKit's ratio exceeds 1 mainly because Qiskit O3 itself routes better under $\{\mathrm{cx},r_z,\sqrt{x},x\}$ than under CZ. BQSKit~1.2.1 is not reproducible under a fixed seed: among 216 runs that were repeated during the final round with the same version, adapter, circuit, topology and seed, 43 gave a different two-qubit count, with differences in both directions (for example, QFT-16 on a ring with seed~0: 598 against 542). Its numbers are therefore samples from a randomized compiler, and results from different runs or machines (some of its Standard-42 runs used a second node) are comparable only statistically.

\paragraph*{Superoptimizers.} GUOQ~\cite{xu2025guoq}, QUESO~\cite{xu2023queso} and Quartz~\cite{xu2022quartz} optimize at the gate level and do not lay out or route. We therefore apply them to the routed output of Qiskit O3 (same seed, identical to the Qiskit O3 seed-0 results), in the Nam gate set $\{\mathrm{cx},h,r_z,x\}$ that all three support, with connectivity preserved. GUOQ and QUESO use their \texttt{-{}-preserve-mapping} option, under which rewrites use only qubit pairs already present in the matched pattern. GUOQ's numerical resynthesis is restricted to each input block's existing edges, and Quartz, which uses the device couplings, rejects or penalizes rewrites that create uncoupled CX gates. The restrictions on GUOQ and QUESO also apply on the all-to-all target and are stricter than the device couplings available to the \qs{} compiler. The result is translated exactly to the CZ basis, which leaves the two-qubit count and two-qubit depth unchanged. GUOQ and QUESO minimize the two-qubit count first (\texttt{-opt TWO\_Q}); QUESO uses the beam search recommended by its authors without resynthesis, GUOQ uses its default search with BQSKit resynthesis, and Quartz uses its default cost with the pre-generated Nam ECC set of size (5,3). On the arXiv set with seed~0 and a per-case limit of 300~s (120~s for Quartz), GUOQ reduces Qiskit O3's two-qubit count by 12\% (ratio 0.881), QUESO by 7\% (0.932) and Quartz by 8\% (0.922), against 0.616 for the \qs{} compiler (previous version: 0.787). GUOQ's search converges within about 60~s on these circuits (ratio to its input 0.883 at 60~s and 0.882 at 300~s), whereas QUESO, which ran on a slower x86 node, was still improving slowly (0.954 at 60~s, 0.933 at 300~s). The \qs{} compiler uses fewer two-qubit gates than GUOQ in 208 cases, ties in 65 and uses more in 2 (previous version: 150/87/38, with a loss margin of 117 CZ in total). Both losses are on one 4-qubit angle-encoding circuit with an all-pairs entangler, on the all-to-all target (22 against 18) and on the ring (26 against 24). It uses fewer than QUESO and Quartz in 218 and 192 cases and never more. The depth ratios of the three tools are 1.42--1.52, since none of them targets depth. Quartz returned non-equivalent circuits in 18 of 275 cases (fidelity 0.02--0.82); the same inputs are optimized correctly when its rotation-merging preprocessing is disabled, so we attribute the errors to that step. These cases are excluded. None of the outputs violated the coupling graph. Because the superoptimizers start from Qiskit's layout and routing, these numbers measure the pipeline ``Qiskit O3 followed by superoptimization'', not the superoptimizers alone; applying them to the output of the \qs{} compiler or before routing was not evaluated.

\paragraph*{Learned transpiler passes.} The AI passes of qiskit-ibm-transpiler~0.18.0~\cite{qiskitibmtranspiler,kremer2024rl} run locally without an IBM account; their reinforcement-learning models are public. We used the stages of the package's level-3 AI pass manager (Qiskit level-3 initialization, AI layout and routing, AI linear-function synthesis, then Qiskit level-3 translation and optimization) in a separate environment with Qiskit~2.4. We called the local routing function directly and recorded its layout ourselves because the package's wrapper stores the inverse of the initial layout that the router returns, causing even Qiskit's own equivalence check to reject the wrapper's output. The passes reach 0.964 on the arXiv set and 0.953 and 1.035 on Standard-42 under the CZ and IBM bases, with depth ratios of 1.77, 1.85 and 0.96. In 8 of 825, 55 of 630 and 54 of 630 runs, the output is not equivalent to the input under the layout that the tool reports. These runs are concentrated on Grover-7, quantum walk-8 and the RGQFT multiplier-12 and occur only on topologies that need routing; they are excluded. The AI routing does not follow the transpiler seed, so its runs are not reproducible.

\paragraph*{Benchpress.} Benchpress~\cite{nation2025benchpress} drives each SDK through its own test harness. Instead of writing a \qs{} harness, we exported the circuits from its abstract-topology transpilation test on QASMBench~\cite{li2023qasmbench} (123 files) and compiled them with our own driver. We kept the 84 circuits with at most 100 qubits and at most 15{,}000 two-qubit gates (counting a Toffoli as six); 9 circuits with mid-circuit measurement, 2 with classically controlled gates, 1 without multi-qubit gates, 21 with more than 100 qubits and 6 above the gate limit were dropped. They were compiled to the line, grid, heavy-hex and all-to-all targets that correspond to Benchpress's abstract topologies (sized to the circuit), in the CZ basis with seed~0 and a 900~s limit. We excluded the four cases of one circuit that contains a reset (rejected by all three compilers), one 8-qubit UCCSD case on heavy-hex on which Qiskit O3 and tket timed out, and the cases of the 100-qubit, 15{,}000-gate circuit on which tket timed out (four topologies). The previous version of the \qs{} compiler had also timed out on three of the latter, whereas the final version compiled every case within the limit. The 33 QASMBench circuits in \texttt{guoq\_dev} (Sec.~\ref{sec:suites}) were taken from this set, so part of this set was development data for the final round. On the 327 cases where all three compilers succeeded, the \qs{} compiler reaches 0.714 in two-qubit count and 0.605 in depth relative to Qiskit O3, with 218 wins, 109 ties and no loss (previous version: 0.821 and 0.678 with 188/139/1). The corresponding values for tket O2 are 0.993 and 1.021. The \qs{} compiler wins, ties and loses 191/134/2 against tket, losing on a 5-qubit error-correction circuit on the grid and on the line (17 against 13 and 24 against 20). The 141 outputs with more than 20 physical qubits were checked for gate set and coupling only. The device-specific tests of Benchpress (133-qubit Heron) are not included; Sec.~\ref{sec:scale} covers devices of that size.

\paragraph*{Per-case record with all tested opponents.} Over all tested external configurations, the \qs{} compiler wins, ties and loses 175/89/11 on the arXiv set (195/71/9 without the added opponents), and 169/36/5, 174/35/1 and 174/35/1 on Standard-42 under the CZ, CX and IBM bases (Table~\ref{tab:competitors}). On the arXiv set, 9 of the 11 losses are those in Table~\ref{tab:wtl}, and the other 2 are the two GUOQ cases above; together, these losses amount to 29 CZ, with a median of 1 and a maximum of 10. For the previous version, the record was 76/114/85, with a median margin of 3 CZ (maximum 34). Under CX and IBM, the single loss is to BQSKit. In terms of the geometric mean, the \qs{} compiler has the lowest two-qubit ratio among the tested configurations on every suite.

\subsection{Scaling to 127- and 156-qubit devices}
\label{sec:scale}

The suites above use the smallest topology that accommodates each circuit. To test current device sizes, we compiled the 20 Large-20 circuits and 28 new circuits of 120, 140 and 156 qubits (``XL-wide'': GHZ, BV, random Clifford and Clifford+T layers on random perfect matchings, approximate QFT, QFT-120, QAOA on random 3-regular graphs, 1D Heisenberg and 2D transverse-field Ising Trotter steps, and depth-5 quantum volume) on the coupling graphs of IBM Eagle (127 qubits, 144 edges), IBM Heron (156 qubits, 176 edges) and a $12\times12$ grid, with seed~0 and an 1800~s limit. The devices are used as undirected graphs with the benchmark's CZ and IBM gate sets. The \qs{} results are from the final compiler (\texttt{be4ecda}); Qiskit, tket and BQSKit were run once for the same experiment with the previous version, and their results are reused per case.

\begin{table}[t]
\caption{Large devices: geometric-mean ratio to Qiskit O3 (two-qubit count / depth) on the cases where all three compilers succeeded, and two-qubit W/T/L of quspect O3 against Qiskit O3. \qs{}: final version. Data: benchmark commits \texttt{0620898} and \texttt{d350f87} (\texttt{paper/scale/}).}
\label{tab:scale}
\scriptsize
\setlength{\tabcolsep}{2.2pt}
\begin{tabular}{llrccc}
\toprule
Basis & Circuits & cases & quspect O3 & tket O2 & W/T/L \\
\midrule
CZ  & Large-20 & 59 & 0.755 / 0.681 & 1.186 / 1.458 & 55/4/0 \\
CZ  & XL-wide  & 57 & 0.787 / 0.657 & 1.336 / 1.347 & 39/6/12 \\
CZ  & all      & 116 & 0.771 / 0.669 & 1.258 / 1.402 & 94/10/12 \\
IBM & all, heavy-hex & 77 & 0.844 / 0.736 & 1.425 / 1.499 & 55/4/18 \\
\bottomrule
\end{tabular}
\end{table}

\paragraph*{Results.} All 117 CZ-basis and 78 IBM-basis compilations by quspect O3 finished within the limit, the largest output being QV-100 on Heron with about 91{,}400 two-qubit gates. Under the CZ basis, quspect O3 uses 0.771 times as many two-qubit gates as Qiskit O3 at 0.669 times the depth (previous version: 0.794 and 0.697; 0.770 on Eagle, 0.769 on Heron and 0.773 on the grid), and it loses no Large-20 case against Qiskit O3 (Table~\ref{tab:scale}); tket O2 is at 1.258. The final round gains only about 3\% here, compared with about 20\% on small circuits, because its passes are capped by circuit size: recompilation, three-qubit instantiation and CNOT--phase resynthesis are limited to circuits with at most 1500, 400 and 3000 two-qubit gates. For the previous version, moving the Large-20 circuits from the smallest fitting topology to the large devices increased the compiler's two-qubit count by 0.5--2.4\%, while Qiskit was nearly unchanged and tket gained 1--3\%. Our guess, not tested by an ablation, is that the layout search budget of the compiler scales with the circuit rather than with the device and covers a smaller fraction of the candidate embeddings on a larger device. Under the IBM basis the lead narrows to 0.844 (55/4/18 against Qiskit O3).

\paragraph*{New weaknesses.} The lead disappears on the random Clifford and Clifford+T circuits, whose entangling layers act on random qubit pairings. On these circuits, routing dominates and there is little structure to exploit: the \qs{} compiler has a geometric-mean two-qubit count 3--5\% higher than that of Qiskit O3 and loses 19 of these 20 cases (0.3--7\%). On Eagle, GHZ-120 needs 179 two-qubit gates against 151 for Qiskit O3 (previous version: 184). Under the IBM basis, VQE-two-local-60 on Heron needs 15{,}639 against 10{,}983 (1.42$\times$); QAOA-SK-40 and QPE-42 lose by 1--13\%, and inexact QPE-60 on Eagle loses by 8.2\% (4508 against 4165). On these circuits Qiskit routes much better under the IBM basis than under CZ (VQE: 17{,}839 $\to$ 10{,}983), whereas the \qs{} compiler produces nearly the same routing under both.

\paragraph*{Other compilers and verification.} tket O2 produced an invalid output for hidden-linear-function-60 on Eagle under both bases (two two-qubit gates on uncoupled qubits); the case is excluded from the comparison. BQSKit O2, tested on 11 smaller circuits on the three devices, completed 12 of 33 runs within 900~s on a shared node; on the completed runs it used 2.39 times as many two-qubit gates as Qiskit O3. For the 11 Clifford inputs (118 compilations by the three compilers under both bases), we verified the outputs exactly by propagating the stabilizers of the ideal state through the compiled circuit; all were equivalent, and so were the 40 \qs{} outputs of the final version for these inputs. The outputs of the other large circuits are not verified by the benchmark (Sec.~\ref{sec:limitations}).

\section{Where the advantage comes from}
\label{sec:ablation}

\paragraph*{Scope.} Ablations assess the contributions within the integrated pipeline. The first ablations in this section were run on the previous version of the compiler (commit \texttt{8ec73d8}, tag \texttt{paper-r20}), so they measure the contributions of the components of Secs.~\ref{sec:layout}--\ref{sec:sched} within that pipeline, whose lead over Qiskit O3 was 17--25\% on the development suites. The final round added the components of Sec.~\ref{sec:postroute}, and the beam router was added before that round; the contributions of these components in the final compiler (\texttt{be4ecda}) are given at the end of the section. Because the final-round components act after routing, on the routed candidates, the contributions of the earlier components may differ in the final compiler; we also re-measure them there at the end of the section.

\paragraph*{Protocol.} We added an ablation switch for each component to the previous version. With all switches off, the output is bit-identical to that version on every case in the four CZ suites (SHA-256 of the output circuit and of the initial and final layouts; Standard-42 at levels 1--3, the synthetic suite, the arXiv set and Large-20), and the full unit-test suite passes. Each switch removes one component and leaves the rest of the pipeline unchanged; ``select'' replaces the level-3 candidate search with the level-2 routing, so it measures level~3 against level~2. For every suite, the baseline (full quspect O3) and all variants ran in the same job on the same machine. Every variant passed the state-vector check on every case with at most 20 qubits. Table~\ref{tab:ablation} reports the geometric mean over cases of the ratio variant/baseline, so a value above one means that the component helps.

\begin{table*}[t]
\caption{Leave-one-out ablation of quspect O3 under the CZ gate set: geometric-mean ratio of the variant without the component to the full compiler (above 1: the component helps). In parentheses: cases in which removing the component gives fewer / equal / more two-qubit gates. The four suites have 210, 196, 80 and 275 cases (the arXiv two-qubit ratio is over 274 cases with a nonzero baseline). Pipeline of the previous version. Data: quspect \texttt{8ec73d8} with ablation switches (\texttt{46b640a}; tag \texttt{paper-r20-ablation}), benchmark commit \texttt{f64c9ae}.}
\label{tab:ablation}
\scriptsize
\setlength{\tabcolsep}{2.6pt}
\begin{tabular}{llccccrrrr}
\toprule
 & & \multicolumn{4}{c}{two-qubit count} & \multicolumn{4}{c}{depth} \\
\cmidrule(lr){3-6}\cmidrule(lr){7-10}
Removed component & Sec. & Std-42 & Synthetic & Large-20 & arXiv & Std-42 & Synth. & Large & arXiv \\
\midrule
VF2 exact layout & \ref{sec:layout} & 1.000 (0/210/0) & 1.000 (0/196/0) & 1.000 (0/80/0) & 1.000 (0/274/1) & 1.000 & 1.000 & 1.000 & 1.000 \\
low-dilation (QAP) starts & \ref{sec:layout} & 1.003 (1/191/18) & 1.010 (2/183/11) & 1.014 (3/63/14) & 1.000 (1/270/4) & 1.002 & 1.010 & 1.001 & 1.001 \\
mirror gates & \ref{sec:routing} & 1.003 (9/173/28) & 1.001 (22/161/13) & 1.017 (4/53/23) & 0.999 (12/248/15) & 1.008 & 1.003 & 1.052 & 1.000 \\
commutation-aware family & \ref{sec:routing} & 1.044 (4/122/84) & 1.028 (0/155/41) & 1.020 (1/58/21) & 1.012 (2/234/39) & 1.032 & 1.017 & 0.961 & 0.998 \\
window LNS (and depth sweep) & \ref{sec:lns} & 1.014 (0/126/84) & 1.004 (0/152/44) & 1.005 (0/36/44) & 1.007 (0/250/25) & 1.014 & 1.009 & 1.012 & 1.006 \\
Kendall-$\tau$ layered routing & \ref{sec:kendall} & 1.000 (0/209/1) & 1.000 (0/196/0) & 1.006 (0/78/2) & 1.000 (0/275/0) & 1.001 & 1.000 & 1.004 & 1.000 \\
truncated rollout & \ref{sec:kendall} & 1.000 (0/209/1) & 1.000 (0/196/0) & 1.002 (0/76/4) & 1.000 (0/275/0) & 1.000 & 1.000 & 1.002 & 1.000 \\
candidate search (level 3 $\to$ 2) & \ref{sec:candidates} & 1.081 (0/83/127) & 1.074 (0/98/98) & 1.063 (1/31/48) & 1.025 (0/207/68) & 1.058 & 1.065 & 1.027 & 1.006 \\
stabilizer selection & \ref{sec:synthesis} & 1.001 (5/187/18) & 1.000 (2/185/9) & 1.000 (2/61/17) & 1.006 (0/251/24) & 1.040 & 1.033 & 1.030 & 1.040 \\
three-qubit QSD & \ref{sec:synthesis} & 1.070 (6/183/21) & 1.000 (0/196/0) & 1.000 (6/66/8) & 1.027 (1/250/24) & 1.070 & 1.000 & 0.997 & 1.023 \\
phase folding & \ref{sec:synthesis} & 1.007 (35/143/32) & 1.001 (5/177/14) & 1.052 (5/49/26) & 1.010 (0/265/10) & 1.044 & 1.032 & 1.160 & 1.013 \\
bounded approximation & \ref{sec:approx} & 1.000 (0/210/0) & 1.005 (0/189/7) & 1.019 (0/61/19) & 1.000 (0/275/0) & 1.000 & 1.002 & 1.010 & 1.000 \\
CX orientation & \ref{sec:gauge} & 1.000 (0/210/0) & 1.000 (0/196/0) & 1.000 (0/80/0) & 1.000 (0/275/0) & 1.000 & 1.000 & 1.000 & 1.000 \\
trace-monoid scheduling & \ref{sec:sched} & 1.000 (0/210/0) & 1.000 (0/196/0) & 1.000 (0/80/0) & 1.000 (0/275/0) & 1.142 & 1.101 & 1.129 & 1.092 \\
\bottomrule
\end{tabular}
\end{table*}

\paragraph*{Single-component effects are small.} The largest single contribution is the level~3 candidate search: without it, two-qubit counts rise by 2.5--8.1\% and almost no case improves. It is followed by the commutation-aware routing family (1.2--4.4\%). Every other component changes the geometric mean of the two-qubit count by at most a few percent on any suite, and several change it by less than 1\%. Compared with the 17--25\% lead of the previous version over Qiskit O3, these single-component effects are small. In the grouped ablation below, the components overlap.

\begin{figure*}[t]
\centering
\includegraphics[width=\textwidth]{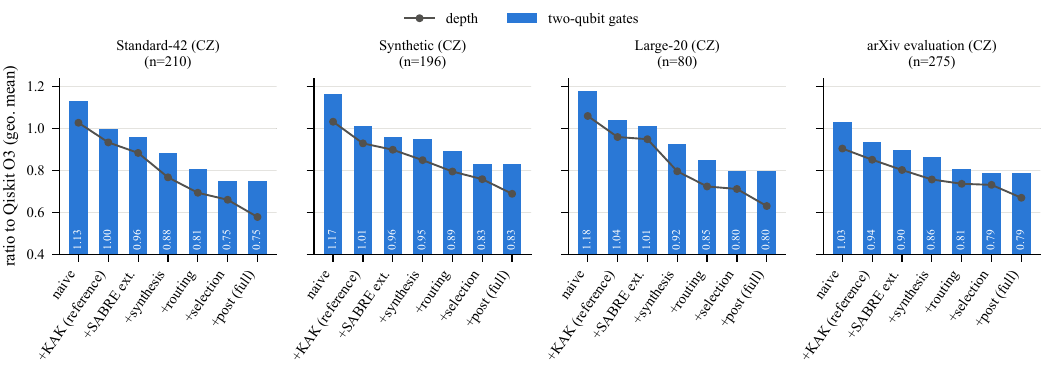}
\caption{Contribution ladder of the previous version (\texttt{8ec73d8}): two-qubit count (bars) and depth (line) relative to Qiskit O3 (dashed line at 1) as parts of the \qs{} compiler are enabled one at a time, from the naive pipeline (without KAK resynthesis and without the \qs{} extensions of the SABRE core) through the reference pipeline (+KAK) and +SABRE ext.\ to the full level-3 compiler of that version.}
\label{fig:ladder}
\end{figure*}

\paragraph*{Grouped ablation and contribution ladder.} We grouped the components into synthesis (QSD, phase folding, approximation, stabilizer selection), routing (VF2, low-dilation layout, mirror gates, the commutation-aware family, LNS, layered and rollout routing), candidate selection and post-processing (scheduling, orientation). We also added switches that remove KAK block resynthesis and the \qs{} extensions of the SABRE core (block nodes, Toffoli units, SWAP absorption). With all switches off, the outputs are bit-identical to the previous version on all four suites. Figure~\ref{fig:ladder} shows the parts being enabled one at a time. We call the pipeline without KAK resynthesis and without the \qs{} extensions of the SABRE core \emph{naive}, the naive pipeline with KAK resynthesis the \emph{reference} pipeline (the structure of Qiskit O3), and the reference pipeline with the SABRE extensions \emph{+SABRE ext.} The reference pipeline has two-qubit ratios of 0.997, 1.009 and 1.037 on Standard-42, the synthetic suite and Large-20, and 0.936 on arXiv. On the three development suites, nearly the whole lead therefore comes from the components added beyond this reference pipeline; on arXiv the reference pipeline already accounts for 28\% of the log-gap from Qiskit O3 to the full level~3 of the previous version. On a log scale, 18\% of the gap to Qiskit O3 on arXiv comes from the SABRE extensions, 17\% from synthesis, 27\% from routing and 10\% from candidate selection (the remaining 28\% is already in the reference pipeline); on Standard-42 the shares are 13\%, 28\%, 31\% and 27\%. Removing all seven routing components at once costs 4.9--13.8\% more two-qubit gates, about twice the product of their individual effects, because they substitute for one another; the synthesis components are additive. Candidate selection and routing are complementary: the level-3 routing families run only inside the candidate search, so on Standard-42 the routing group gains 13\% with selection and 9\% without it. Scheduling accounts for 22--26\% of the depth advantage and none of the gate-count advantage. With all four groups removed but KAK resynthesis and the SABRE extensions kept (+SABRE ext.), the \qs{} compiler is still 4--10\% ahead of Qiskit O3 on three suites and 1.3\% behind it on Large-20. The ``SABRE extensions'' switch keeps the compiler's own LightSABRE kernel rather than Qiskit's, so the reference pipeline still differs from Qiskit O3 in implementation details. For groups that substitute for one another, a ladder share is an attribution in one order, not an independent contribution. Data: benchmark commit \texttt{f3bbd3b}; switches \texttt{16fbb94} (tag \texttt{paper-r20-ablation}).

\paragraph*{Structure-specific components: small on average, large on specific circuits.} Several components act only on circuits with a particular structure, and their suite averages understate their effects. Without the low-dilation layout, GHZ-50 on heavy-hex needs 148 instead of 50 two-qubit gates, the 50-qubit adder 612 instead of 499, and the 73-qubit VBE adder 53\% more. Without Kendall-$\tau$ layered routing, QV-100 on a line needs 37\% more and QV-48 15\% more; without the rollout, QV-100 needs 6.6\% more on a grid and 4.7\% more on heavy-hex. Three-qubit QSD acts on phase estimation and multi-controlled gates (7.0\% on Standard-42, 2.7\% on the arXiv set, nothing on the synthetic suite). Phase folding saves 5.2\% of the two-qubit gates and 16\% of the depth on Large-20, mostly on QFT-based multipliers (1.8--2.3$\times$ without it); on Standard-42, 35 cases are better without it, so a folded circuit is not always easier to route. Bounded approximation matters only for circuits with small-angle controlled phases (QPE and Draper adders, 18--36\% on Large-20).

\paragraph*{Depth and single-qubit components.} The trace-monoid scheduler does not change any gate count. Removing it increases the depth by 9--14\%, with an increase in almost every case. Stabilizer selection changes the two-qubit count by at most 1\% but the depth by 3--4\% and the single-qubit count by 5--14\% under CZ. The CX orientation has no effect under the CZ basis, which contains no CX. Table~\ref{tab:ablation-ibm} gives the single-qubit components under the IBM and CX bases. Removing the orientation increases the single-qubit count by 1.5--2.7\% and the depth by 0.8--4.2\%. Removing only the dual iterations, but keeping the two initial candidates (no flip and all flipped), loses nearly as much of the benefit, so on these suites the dual iterations, not the global-flip candidate, provide most of the effect. The example often quoted for global flips, QAOA-SK-10 on an all-to-all device, is the exception: the reduction in its number of single-qubit gates (379 $\to$ 214) comes entirely from the all-flipped candidate. After routing on a line, ring, heavy-hex or grid, the all-flipped candidate no longer helps and the dual iterations give 399 $\to$ 334, 401 $\to$ 329, 406 $\to$ 386 and 320 $\to$ 306.

\begin{table}[t]
\caption{Single-qubit components under the IBM and CX bases (ratios as in Table~\ref{tab:ablation}; two-qubit ratios are 1.000--1.014). ``dual off'' keeps the two initial orientations (none and all flipped) and skips the dual iterations. Data: benchmark commit \texttt{f64c9ae}.}
\label{tab:ablation-ibm}
\footnotesize
\setlength{\tabcolsep}{3pt}
\begin{tabular}{llcc}
\toprule
Removed & Suite & depth (better/equal/worse) & 1q count \\
\midrule
orientation & Std-42 IBM & 1.042 (0/128/82) & 1.027 \\
dual off & Std-42 IBM & 1.039 (5/130/75) & 1.024 \\
stabilizer & Std-42 IBM & 1.056 (33/42/135) & 1.127 \\
orientation & arXiv IBM & 1.014 (0/215/60) & 1.019 \\
dual off & arXiv IBM & 1.014 (0/215/60) & 1.019 \\
stabilizer & arXiv IBM & 1.038 (28/133/114) & 1.065 \\
orientation & Std-42 CX & 1.008 (0/164/46) & 1.015 \\
dual off & Std-42 CX & 1.008 (0/164/46) & 1.015 \\
stabilizer & Std-42 CX & 1.064 (20/40/150) & 1.196 \\
\bottomrule
\end{tabular}
\end{table}

\paragraph*{Components without an independent effect.} Removing VF2 changes the two-qubit count in one of 761 cases. When a SWAP-free embedding exists, the SABRE trials with the low-dilation starts find it as well, so on these suites VF2 saves time and provides a guarantee rather than reducing the gate count. The window LNS never makes a case worse when present (zero cases improve without it on all four suites), but its average effect is 0.4--1.4\%. Mirror gates help on average only on Large-20, and on every suite some cases are better without them, which is why level~3 keeps a family without mirroring. Components can substitute for one another (VF2 and the low-dilation starts, mirror gates and the fallback family, LNS and the candidate search), so removing one at a time can understate the effect of removing a group.

\paragraph*{Final-round components.} We repeated the ablation on the final compiler (\texttt{be4ecda}), with all earlier switches ported and one switch for each of the ten components added in or just before the final round (Table~\ref{tab:ablation-final}). With all switches off, the outputs are bit-identical to \texttt{be4ecda} on every case in Standard-42 (levels 1--3), the synthetic suite, the arXiv set, Large-20 and \texttt{guoq\_dev}, and every variant passed the state-vector check on every case with at most 20 qubits. The baseline and variants for each suite again ran in the same job on one machine. The coverage is smaller than for the previous version. Only seed~0 was run (previous version: median over three seeds). On Large-20, only the switches for the final-round components and the key group configurations were run. The \texttt{guoq\_dev} run was stopped at the deadline after 13{,}884 of 21{,}195 compilations; its column uses the 513 cases for which all configurations finished. This set excludes all 272 cases with 3--5 qubits, so whole-circuit instantiation has no effect there. The stabilizer-selection switch has no effect in the final compiler, because KAK synthesis moved to C++ after the previous version and the switch acts only on the Python fallback; for this component, the numbers in Table~\ref{tab:ablation} still apply.

\begin{table*}[t]
\caption{Ablation of the final compiler, quspect O3 under CZ, seed~0. Top: leave-one-out ablation of the final-round components, two-qubit ratio without/with the component (above 1: the component helps) and cases with fewer / equal / more two-qubit gates without it. Middle: groups removed at once, two-qubit / depth ratio on the cases where all group configurations and Qiskit O3 succeeded. Bottom: earlier components in the final compiler, two-qubit ratio, with the value for the previous version (Table~\ref{tab:ablation} and Sec.~\ref{sec:ablation}) in parentheses. --: not run or not applicable. \texttt{guoq\_dev}: 513 cases with 6--16 qubits. Data: quspect \texttt{be4ecda} with ablation switches (\texttt{c82472d}), benchmark commit \texttt{76a5256}.}
\label{tab:ablation-final}
\scriptsize
\setlength{\tabcolsep}{3pt}
\begin{tabular}{llccccc}
\toprule
Removed & Sec. & Std-42 (210) & Synthetic (196) & Large-20 (80) & arXiv (275) & \texttt{guoq\_dev} (513) \\
\midrule
CNOT--phase regions, $\le5$ wires (A*) & \ref{sec:postroute} & 1.019 (20/102/88) & 1.016 (8/124/64) & 1.002 (4/62/14) & 1.007 (19/218/38) & 1.018 (58/275/180) \\
CNOT--phase regions, 16 wires (beam) & \ref{sec:postroute} & 1.013 (9/116/85) & 1.008 (9/144/43) & 1.011 (0/58/22) & 1.013 (11/229/35) & 1.012 (40/309/164) \\
two-level scoring & \ref{sec:postroute} & 1.002 (13/180/17) & 1.002 (7/174/15) & 1.000 (1/77/2) & 0.999 (5/263/7) & 1.004 (26/429/58) \\
pass before block consolidation & \ref{sec:postroute} & 1.008 (15/137/58) & 1.005 (11/145/40) & 1.001 (1/72/7) & 1.007 (7/248/20) & 1.008 (40/348/125) \\
pre-routing second start & \ref{sec:postroute} & 1.062 (3/181/26) & 1.000 (1/193/2) & -- & 1.019 (0/263/12) & 1.024 (15/463/35) \\
instantiation of 3-qubit circuits & \ref{sec:postroute} & 1.000 (0/210/0) & 1.000 (0/196/0) & -- & 1.034 (0/244/31) & 1.000 (0/513/0) \\
instantiation of routed 3-qubit blocks & \ref{sec:postroute} & 1.013 (2/143/65) & 1.016 (1/154/41) & 1.000 (0/80/0) & 1.034 (4/189/82) & 1.023 (8/324/181) \\
plain-SABRE family & \ref{sec:postroute} & 1.010 (8/165/37) & 1.012 (4/152/40) & 1.007 (1/70/9) & 1.013 (14/228/33) & 1.011 (30/378/105) \\
recompilation & \ref{sec:postroute} & 1.017 (0/118/92) & 1.013 (0/151/45) & 1.002 (0/74/6) & 1.030 (0/201/74) & 1.022 (0/286/227) \\
beam candidate & \ref{sec:beam} & 1.001 (4/197/9) & 1.004 (1/181/14) & 1.003 (0/74/6) & 0.998 (2/270/3) & 1.001 (6/488/19) \\
\midrule
all final-round components & & 1.229 / 1.146 & 1.080 / 1.006 & 1.049 / 1.035 & 1.269 / 1.133 & 1.207 / 1.115 \\
\quad CNOT--phase resynthesis (5) & & 1.178 / 1.153 & 1.035 / 1.023 & 1.037 / 0.999 & 1.135 / 1.112 & 1.123 / 1.120 \\
\quad instantiation (2) & & 1.013 / 1.003 & 1.016 / 0.999 & -- & 1.078 / 0.999 & 1.023 / 1.002 \\
\quad plain-SABRE and beam (2) & & 1.012 / 1.008 & 1.018 / 1.007 & 1.011 / 1.033 & 1.011 / 1.004 & 1.013 / 1.002 \\
\midrule
candidate search (level 3 $\to$ 2) & \ref{sec:candidates} & 1.098 (1.081) & 1.085 (1.074) & -- & 1.051 (1.025) & -- \\
routing group (7) & \ref{sec:routing} & 1.074 (1.138) & 1.061 (1.087) & -- & 1.039 (1.049) & 1.057 \\
\quad commutation-aware family & \ref{sec:routing} & 1.033 (1.044) & 1.020 (1.028) & -- & 1.028 (1.012) & -- \\
\quad window LNS & \ref{sec:lns} & 1.009 (1.014) & 1.000 (1.004) & -- & 1.009 (1.007) & -- \\
synthesis group (4) & \ref{sec:synthesis} & 1.078 (1.079) & 1.007 (1.006) & -- & 1.002 (1.043) & 1.040 \\
\quad three-qubit QSD & \ref{sec:synthesis} & 1.074 (1.070) & 1.000 (1.000) & -- & 0.997 (1.027) & -- \\
\quad phase folding & \ref{sec:synthesis} & 1.002 (1.007) & 1.001 (1.001) & -- & 1.005 (1.010) & -- \\
scheduling & \ref{sec:sched} & 1.002 (1.000) & 1.001 (1.000) & 1.000 (1.000) & 1.002 (1.000) & 1.005 \\
\bottomrule
\end{tabular}
\end{table*}

Removing all final-round components at once increases the two-qubit count by 22.9\% on Standard-42, 8.0\% on the synthetic suite, 4.9\% on Large-20, 26.9\% on the arXiv set and 20.7\% on \texttt{guoq\_dev}, and the depth by 0.6--14.6\%. Individually, the largest contributions come from the pre-routing second start (6.2\% on Standard-42, mostly from all-to-all entangling layers: without it, \texttt{real\_amplitudes\_full\_12} on restricted topologies needs 11 times as many two-qubit gates; 1.9\% on arXiv), from numerical instantiation (3.4\% each for whole three-qubit circuits and for routed blocks on arXiv, 7.8\% together, because the arXiv set contains many three- and four-qubit circuits) and from recompilation (0.2--3.0\%, and no case is better without it). The small and large CNOT--phase regions contribute 0.2--1.9\% and 0.8--1.3\%, and the plain-SABRE family 0.7--1.3\%. Two-level scoring and the beam candidate are close to zero on average. The final-round components substitute for each other even more than the routing components of the previous version: the ten single effects add up to about 14.5\% on Standard-42 and 15\% on arXiv, against 23\% and 27\% for the group, and the five CNOT--phase switches add up to 10.4\% on Standard-42 against 17.8\% for their group. Region resynthesis, instantiation and recompilation partly substitute for each other. The candidate-search ablation (Table~\ref{tab:ablation-final}, bottom) reduces candidate generation while retaining level-3 synthesis. These grouped interventions characterize the integrated pipeline; the marginal effect of the ranking objective on an identical candidate pool remains to be measured.

\begin{figure*}[t]
\centering
\includegraphics[width=\textwidth]{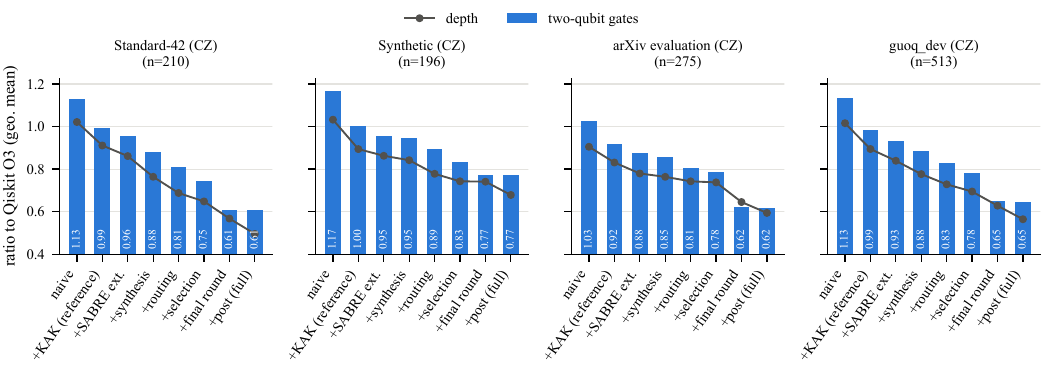}
\caption{Contribution ladder of the final compiler, as in Fig.~\ref{fig:ladder} with one more step (``+final round'': all components of Sec.~\ref{sec:postroute}, the plain-SABRE family, recompilation and the beam candidate) placed after candidate selection; seed~0, \texttt{guoq\_dev} restricted to its 513 cases with at least six qubits. Large-20 was not run for the ladder.}
\label{fig:ladder-final}
\end{figure*}

Figure~\ref{fig:ladder-final} extends the contribution ladder by a final-round step. The reference pipeline has two-qubit ratios to Qiskit O3 of 0.992, 1.001, 0.917 and 0.985 on Standard-42, the synthetic suite, arXiv and \texttt{guoq\_dev}, close to the values for the previous version. Full O3 is now at 0.607, 0.770, 0.618 and 0.647 (seed~0). Thus, on the development suites, nearly the whole lead comes from the components added beyond the reference pipeline, while on arXiv the reference already accounts for 18\% of the log-gap from Qiskit O3 to full level~3. Added after candidate selection, the final-round step multiplies the two-qubit count by 0.815, 0.928, 0.790 and 0.833 on these suites, corresponding to 42\%, 28\%, 60\% and 43\% of the log-gap between the reference pipeline and full O3. Placed directly after +SABRE ext., the step multiplies the two-qubit count by 0.801, 0.927, 0.767 and 0.814 (45\%, 29\%, 67\% and 49\%). The step thus has similar shares in these two insertion orders; other orders were not run. For the other groups, which substitute for one another, a ladder share is an attribution in one order, not an independent contribution. On Large-20, the order matters (0.954 after selection, 0.983 directly after +SABRE ext.); a possible reason, which we did not test, is that on large circuits the final-round components act mainly by changing which routing candidate wins.

The new components also change the contributions of the earlier ones (Table~\ref{tab:ablation-final}, bottom; seed~0 compared with the median over three seeds for the previous version). Candidate search matters more (5.1--9.8\% instead of 2.5--8.1\%), because the plain-SABRE family, the beam candidate and two-level scoring run only inside it. The routing group matters less (3.9--7.4\% instead of 4.9--13.8\%): the plain-SABRE family and recompilation supply some of the routings that the specialized families found before, although the commutation-aware family gained on arXiv. On arXiv the earlier synthesis components are almost replaced: the synthesis group gives 0.2\% instead of 4.3\%, and removing three-qubit QSD now gives slightly fewer gates, because numerical instantiation and post-routing resynthesis do the same work better on three-qubit blocks. On Standard-42, whose QPE and multi-controlled circuits have larger blocks, QSD still saves 7.4\%. The other components (low-dilation layout, mirror gates, LNS, approximation, Kendall-$\tau$ and rollout routing, VF2) contribute about as much as in the previous version. A single scheduling pass still leaves gate counts unchanged, but removing scheduling now changes the final two-qubit count by up to $\pm0.5\%$ (on Standard-42, 30 cases are better and 62 worse without it), because recompilation starts from the scheduled circuit. Data: \qs{} \texttt{be4ecda} with switches \texttt{c82472d} (tag \texttt{paper-r21-ablation}); benchmark commit \texttt{76a5256}.

During development, the stacked steps of the final round reduced the two-qubit ratio on \texttt{guoq\_dev} from 0.783 to 0.721, 0.666 and 0.621 (Sec.~\ref{sec:devprocess}); these are increments on a moving baseline, not independent attributions.

\section{Stage-level gaps and routing-cost diagnostic}
\label{sec:optimality}

The ablation shows how much each component contributes; it does not show how much room for improvement remains. For each stage for which an exact solution or a valid lower bound can be computed, we measured the gap of the previous version of the compiler (Table~\ref{tab:optimality}). Unless stated otherwise, the numbers in this section come from experiments on that version, quspect \texttt{8ec73d8} (tag \texttt{paper-r20}; benchmark commit \texttt{7355da5}); the compiler was not modified, and the LNS was switched off at run time. The final round did not change the algorithms of the stages measured here, but the final compiler selects a different routing candidate and resynthesizes the routed circuit afterwards, so the routing gaps refer to the candidate selected by the previous version. The routing gap of the final compiler is re-measured at the end of the routing paragraph.

\begin{table*}[t]
\caption{Gaps of individual stages of quspect O3 in the previous version, \texttt{8ec73d8} (CZ basis unless noted). ``Same-layout A*'' is an exhaustive A* over the routing of the selected candidate from its initial layout. Data: \texttt{paper/optimality/}, benchmark commit \texttt{7355da5}.}
\label{tab:optimality}
\footnotesize
\setlength{\tabcolsep}{4pt}
\begin{tabular}{@{}p{30mm}p{27mm}p{24mm}p{46mm}p{38mm}@{}}
\toprule
Stage & Instances & Reference & \qs{} & Comparison \\
\midrule
Routing, merge-aware cost & 158 routed, $\le 8$ qubits & same-layout A* & $+90$ CZ in total (1.2\%); 136 at 0, 156 within 6 CZ & LNS off: $+229$ CZ (3.0\%) \\
Routing, SWAP count & 150, same set & same-layout A* & $1.134\times$ (geo.\ mean), $+17.9\%$ in total & LNS off: $1.151\times$ \\
Exact layout & QUEKO, 248 & 0 SWAP & 0 SWAP in 241 & Qiskit O3 247, tket O2 161 \\
Routing, SWAP count & QUBIKOS, 60 & known optimum & $2.51\times$ & Qiskit O3 $4.10\times$, tket O2 $9.83\times$ \\
Layered routing (Kendall-$\tau$) & QV on lines, $n\le10$, 15 & aligned DP optimum & optimal in 14, $+1$ SWAP in 1 & per-layer exact: $+29$--$37\%$ \\
CX orientation (IBM basis) & 611 & ILP optimum & energy $1.049\times$; 1q gates $+3.3\%$ & greedy flips $1.046\times$; local only $+5.3\%$ \\
Depth scheduling & 350 & commutation-DAG bound & $1.003\times$ (2q depth $1.013\times$) & default order $1.115\times$ \\
Depth scheduling & 147, $\le 400$ gates & time-indexed MILP & $1.002\times$, optimal in 93\% & bound $=$ optimum in 98\% \\
\bottomrule
\end{tabular}
\end{table*}

\paragraph*{Routing.} We used 60 circuits with 6--8 qubits (the synthetic families, the small Standard-42 circuits and 4--8-qubit arXiv circuits) on a line, a ring, a grid and heavy-hex (240 instances), 173 of which need SWAPs; the exhaustive A* finished on 158 of these under the merge-aware cost and on 150 under the SWAP count. For the routing that level~3 selects, an exhaustive A* from the same initial layout gives the minimum SWAP count and a merge-aware routing in which each SWAP costs $\min(c_{\text{before}},c_{\text{after}},3)$ CZ (Sec.~\ref{sec:lns}). The level-3 routings use 18\% more SWAPs than the minimum, yet the SWAP-minimal routings are more expensive after merging in 94 of 150 cases, by 811 CZ in total. Under the merge-aware cost, the gap of the \qs{} compiler is 90 CZ over 158 instances, 1.2\% of their two-qubit gates: 136 instances have no gap and 156 have a gap of at most 6 CZ. The two exceptions are a 6-qubit Givens-transform circuit on a line (77 against 43 CZ, where a block-family candidate was selected) and a 7-qubit reciprocal circuit on heavy-hex (37 against 32). Switching the window LNS off raises the total gap to 229 CZ (3.0\%) and the final two-qubit count by 1.7\% in the geometric mean, with 39 cases worse and none better. Layout contributes little: on 124 instances where both are available, the minimum SWAP count from the compiler's layout is 912 and the best known from a free layout (OLSQ2~\cite{lin2023olsq2} with a 600~s limit, or the same-layout A*) is 889.

\emph{Routing-cost diagnostic for the final compiler.} We repeated the same-layout comparison on the same 240 instances with the final compiler (\texttt{be4ecda}), with recompilation, the pre-routing second start and whole-circuit instantiation switched off so that the selected routing candidate is well defined (instantiation of routed blocks stays on). The selected routings use 13.6\% more SWAPs than the same-layout optimum in total (1446 against 1273; geometric mean $1.117\times$ over 152 instances, with equal counts in 85), against 17.9\% for the previous version. Under the merge-aware cost, the gap is 556 CZ over 159 instances, against 90 CZ over 158 for the previous version: 94 instances have no gap and 137 are within 6 CZ. Of the 556 CZ, 445 come from the 45 instances in which a plain gate-level SABRE routing (Sec.~\ref{sec:postroute}) was selected directly and 28 from the 25 in which an LNS refinement of such a routing was selected; the other 89 instances, with block-node, commutation-aware, layered or beam routings and their LNS refinements, are together 83 CZ above the A* optimum, similar to the previous version. The final compiler ranks candidates by the two-qubit count after post-routing CNOT--phase resynthesis and instantiation, and plain routings that cost more before resynthesis often yield fewer gates after it: in this configuration the final two-qubit count on these instances is 8.0\% lower than that of the previous version (26{,}143 against 28{,}431, over the 173 instances that need SWAPs). This comparison demonstrates disagreement between the routing-cost proxy and final circuit quality. Compiler components also differ in this comparison, so it does not isolate the selection rule. Switching the LNS off increases the final two-qubit count by 1.9\% in the geometric mean (36 cases worse, 7 better), against 1.7\% (39 worse, none better) for the previous version. We did not repeat the free-layout comparison with OLSQ2, because the routed DAGs changed. Data: \url{paper/optimality/routing/results/small_exact_r21.jsonl}, benchmark commit \texttt{76a5256}.

On QUEKO~\cite{tan2021queko}, whose circuits admit a SWAP-free layout by construction, the compiler finds it in 241 of 248 circuits, against 247 for Qiskit O3 and 161 for tket O2. The seven misses (1--8 SWAPs each, on Aspen-4, Rochester and Sycamore) are caused by the deterministic budget of $10^5$ states for the VF2 search (Sec.~\ref{sec:layout}). On QUBIKOS~\cite{ping2025qubikos}, whose optimal SWAP count is known, the compiler inserts 2.51 times the optimum (Qiskit O3 4.10, tket O2 9.83). On the 16--19-qubit devices, Qiskit's SABRE often recovers the planted optimum when that optimum is ten or more SWAPs, and the compiler inserts 10--21\% more SWAPs than Qiskit there; on 54-qubit Sycamore it inserts 43\% of Qiskit's SWAPs but still 5.3 times the optimum. Because the compiler selects routings by the gate count after merging and resynthesis, not by SWAP count, it still uses 0.712 times as many two-qubit gates as Qiskit O3 on QUBIKOS (fewer in 59 of 60 cases). This is in line with the argument of NASSC~\cite{liu2022nassc} that a router should be judged by its post-merge cost.

\paragraph*{Layered routing.} On quantum-volume circuits on a line we compared the Kendall-$\tau$ router with an exact dynamic program over all $n!$ arrangements of the aligned model ($n\le10$). The layered candidate reaches the aligned optimum in 14 of 15 instances and is one SWAP above it in the last. A per-layer exact router in the style of Zulehner and Wille~\cite{zulehner2019su4} needs 29--37\% more paid SWAPs, so the gain comes from optimizing the whole chain of layers. On larger instances, where no optimum is available, Qiskit O3 uses 1.09 ($n=6$) to 1.31 ($n=50$) times as many two-qubit gates as the \qs{} compiler, and a fixed odd--even swap network and a per-layer application of 2QAN~\cite{lao20222qan} are further behind. All 35 six-qubit instances that we solved asynchronously, whose layers are perfect matchings, have the same optimum as the aligned model; we have no proof that this holds in general. When consecutive blocks on one pair merge, the layers have slack, and in five of ten 8--10-qubit instances the final output of the compiler (after the window LNS) is one SWAP \emph{below} the ASAP-aligned optimum.

\paragraph*{CX orientation.} We recorded the 645 orientation problems that the level-3 pipeline solves on Standard-42 and the synthetic suite under the IBM basis (2--3811 CX gates) and solved 611 of them exactly as integer programs over the per-wire chain graphs. Measured by the objective of Sec.~\ref{sec:gauge}, the dual decomposition is 4.9\% above the optimum in the geometric mean and reaches it in 216 instances; greedy single-flip hill climbing is 4.6\% above it on the 562 instances where it ran, and keeping the input orientation yields a value 7.3\% above it. The two heuristics are complementary (157 wins, 154 ties and 251 losses for the dual method against greedy). On QAOA-SK-10 the dual method finds the optimum (214 single-qubit gates), where greedy flipping is stuck at 379. When we count actual single-qubit gates after the full single-qubit minimization, the compiler is 3.3\% above the optimal orientation, against 5.3\% for local minimization alone. The method fails on Clifford-like circuits (QEC, hidden linear function, graph states), where about half of the CX gates should be reversed and the dual method reverses none, leaving 28--58\% more single-qubit gates. With the beam in the chain DP, the dual value is not a lower bound: it exceeds the true optimum in 237 of 611 instances. We therefore regard the orientation stage as a heuristic that is about as good as hill climbing, not as an optimality result.

\paragraph*{Scheduling.} As a lower bound, we use the critical path of the commutation DAG (using the same strict-commutation definition as the compiler) with each qubit relaxed to a single machine with release times; on 400 random small circuits it never exceeded the exhaustive optimum. On 350 cases (Standard-42 and the synthetic families with 6--20 qubits on five topologies), the final depth of the compiler's output is within 0.3\% of this bound in the geometric mean (1.3\% for two-qubit depth) and at or below it in 87\% of the cases; on the 280 cases that are not all-to-all, the geometric means are 0.998 (depth) and 1.000 (two-qubit depth). For the 147 cases small enough for an exact time-indexed MILP, the compiler is optimal in 93\% and 0.2\% above the optimum on average, and the bound equals the optimum in 98\%. The same gates in their order before scheduling have a depth 11.5\% above the bound. The remaining gaps are on all-to-all targets with long commuting chains, with a worst case of 22\% (QFT-10: depth 89 against an optimum of 73). Block-level commutation takes 19 cases below the gate-level bound.

\paragraph*{Summary.} Measured against a computed optimum or bound for the same stage, routing under the merge-aware cost (in the previous version), layered routing on paths (in the aligned model) and scheduling are within a few percent of it on the instances where this can be checked. In the final compiler, which ranks candidates after post-routing resynthesis, the selected routings are further from the merge-aware optimum while the final circuits have fewer gates. The SWAP counts are well above the minimum (13.6--17.9\% on the small instances, 2.51 times on QUBIKOS); the selection objective is the gate count after merging and resynthesis. The orientation stage is a heuristic. These results indicate the gains that better algorithms for these stages could still achieve on similar instances; they say nothing about stages whose optimum we cannot compute, such as multi-qubit resynthesis, where BQSKit's numerical synthesis still finds savings (Sec.~\ref{sec:failure}).

\section{Global beam-search routing}
\label{sec:beam}

This section describes the global beam-search router at level~3. It was added to the previous version (commit \texttt{a7ba770}) before the final round, so it is part of the evaluated compiler and of all its results in Sec.~\ref{sec:eval}; its code is unchanged between \texttt{a7ba770} and the evaluated commit \texttt{be4ecda}. The routing-only experiments below measure the router alone and do not depend on the rest of the pipeline. The end-to-end measurement and the ablation were performed using the pipeline of the previous version, before the final round. The caveats are listed in Sec.~\ref{sec:beam-caveats}.

\subsection{Algorithm}
\label{sec:beam-alg}

The router takes the two-qubit routing nodes of a circuit (blocks, Sec.~\ref{sec:routing}) and a start layout, inserts only SWAPs, and minimizes the number of SWAPs. It keeps a single beam of up to $B$ partial routings from the start of the circuit to its end. Unlike rolling-horizon schemes, it never commits a SWAP before the search has reached the end of the circuit.

\paragraph*{State.} A state is the layout $\pi$ together with one pointer per logical qubit into the ordered list of blocks on that qubit. A block is in the front layer exactly when it is the next block on both its qubits. The state therefore has $O(|V_P|)$ words and is cheap to copy. The executed SWAPs are stored once, in a shared tree of parent pointers, and the SWAP sequence is recovered by backtracking from the first state that completes the circuit.

\paragraph*{Expansion.} At every step, each state in the beam is expanded using every coupling edge incident to a physical qubit that holds a front-layer qubit, except the edge of the SWAP that produced the state (no immediate undo). After the SWAP, all front-layer blocks that have become adjacent are executed greedily, together with any blocks that their execution unblocks. Every state in the beam has therefore inserted exactly the same number of SWAPs, and the first state that executes all blocks has the fewest SWAPs among the routings the beam explored.

\paragraph*{Score.} Children are ranked by
\begin{equation}
  \label{eq:beamcost}
  c(\sigma)=\tfrac12\,F(\sigma)+\tfrac{w}{2}\,X(\sigma)-\alpha\,\mathrm{exec}(\sigma)+P(\sigma),
\end{equation}
where $F$ is the sum of $d-1$ over front-layer blocks, $X=\sum_q\sum_{k=1}^{K}\gamma^{k}(d_{q,k}-1)$ sums the distances of the next $K$ blocks on every qubit $q$ with geometric decay, $\mathrm{exec}$ is the number of blocks executed so far, and $P$ is a decay penalty. Both $F$ and $X$ are counted from both endpoints, hence the factors $\tfrac12$. The structure of Eq.~\eqref{eq:beamcost}, a reward for executed gates plus a decayed multi-gate lookahead window, follows Qiu, Zhu and Wei~\cite{qiu2025beam}. The decay penalty is the SABRE decay~\cite{li2019sabre,zou2024lightsabre} transplanted into the beam: since the last executed block, each further SWAP on a physical qubit costs one more unit, and the counters reset when a block executes. Without it, the beam drifts between SWAPs of equal score on crowded devices. On a 50-qubit random circuit on a line, an early prototype without decay inserted about 128{,}000 SWAPs, against 9{,}304 for SABRE and about 12{,}000 with decay at $B=16$. Deterministic noise of order $10^{-6}$, drawn from a seeded generator, breaks ties. The parameters are $B=128$, $K=8$, $\gamma=0.7$, $w=0.5$ and $\alpha=3$, the same for all devices.

\paragraph*{Deduplication.} We use a 64-bit hash to ensure that a state (layout and pointers) that has already entered the beam is never admitted again. A state reached later has used more SWAPs, so dropping it loses nothing that the search could not reach more cheaply. This rule also prevents the beam from cycling through equivalent states.

\paragraph*{Release valve.} If the largest number of executed blocks in the beam has not increased for $4|V_P|$ steps, or if every child has been seen before, the best state routes its nearest front-layer block along a shortest path and executes it, and the beam restarts from that single state. This is the release valve of LightSABRE~\cite{zou2024lightsabre} applied to the beam. It guarantees termination.

\paragraph*{Layout.} The router is started from up to four distinct layouts: the best-ranked candidates of the level-3 SABRE search, which include VF2, low-dilation and randomized SABRE layouts (Sec.~\ref{sec:candidates}). From each start, it performs two forward--backward iterations in which every pass, forward and backward, is itself a beam search. The final layout of a backward pass becomes the next start. The forward routing with the fewest SWAPs over all starts and iterations is returned.

\paragraph*{Integration into level~3 and budget.} The beam router optimizes the SWAP count, not the merge-aware cost of Sec.~\ref{sec:lns}. Its routing therefore enters level~3 only as one more candidate. It is scored by the same proxy as the other candidates and, if it ranks among the best, fully compiled. Level~3 then keeps whichever candidate has the fewest two-qubit gates after synthesis, so the beam candidate can only help, up to the non-monotonicity discussed in Sec.~\ref{sec:candidates}. The router is skipped for DAGs with three-qubit nodes. A routing pass is abandoned once it exceeds $1.5\times$ the fewest SWAPs among the SABRE candidates, because it could not win. The cost of one pass is bounded by (steps)$\times B\times$(children per state)$\times$(cost of a child). Under a deterministic work budget of $6\times10^9$ elementary operations per start, the largest $B\le128$ that fits is used. If even $B=16$ does not fit, the layout iterations are dropped; if the router still exceeds the budget, it is skipped. The budget scales with the user's work limit, and the router is not run under a wall-clock time limit, which keeps the output deterministic.

\subsection{Relation to prior work}
\label{sec:beam-related}

Beam and tree search for qubit routing are established. Nishio et al.~\cite{nishio2020extracting} ran a global beam search that expands gate by gate and ranks partial mappings by estimated success probability. Qiskit's LookaheadSwap~\cite{qiskitlookahead} is a bounded beam lookahead (width and depth 4 by default) that commits one SWAP per round. Monte Carlo tree search~\cite{zhou2022mcts,sinha2022qroute}, filtered depth-limited search~\cite{li2021fidls} and dynamic lookahead~\cite{zhu2020dynamic} follow the same pattern of looking ahead and then committing a prefix, and A*-based mappers~\cite{zulehner2019astar} search over SWAP layers. Closest to our router is the beam-search framework of Qiu, Zhu and Wei~\cite{qiu2025beam}. In that framework, a state consists of the mapping, the physical circuit so far and the remaining logical circuit. Each expansion inserts one SWAP that touches the front layer and then removes the executable gates. The score adds a reward for executed two-qubit gates to the value of a decayed lookahead window (window 7, decay 0.7). The beam (width 8) is grown to depth 15, after which only the first SWAP of the best leaf is committed. Their main results use the identity initial mapping and IBM Q Tokyo (20 qubits). \emph{Our scoring function is taken from that work}, and we do not claim beam-search routing or this score as new.

The differences are as follows. (a) A single beam runs to the end of the circuit, so no SWAP is committed on partial information. (b) Compact states with deduplication make a wide global beam affordable on 105- and 156-qubit devices and on circuits with thousands of blocks; the release valve and the decay penalty guarantee termination and guard against drift. (c) The beam itself performs forward--backward layout iteration from several SABRE, VF2 and low-dilation starts. (d) Inside a full compiler, the router is one candidate selected by post-synthesis gate count. Table~\ref{tab:beam-ablation} measures each of (a)--(c).

\subsection{Routing-only comparison}
\label{sec:beam-routing}

\paragraph*{Protocol.} We follow the routing benchmark of SimpleTES~\cite{ye2026simpletes}, which evolves a Rust routing policy (an initial layout and a SWAP scoring function) with a large language model. Only layout and SWAP insertion are compared. The output consists of the original gates, unchanged, plus the inserted SWAPs, and the metric is the number of inserted SWAPs. There is no final-SWAP elimination, gate merging or resynthesis. The SimpleTES benchmark has 24 SABRE benchmark circuits~\cite{li2019sabre} on three devices, IBM Q20 (20 qubits), Google Willow (105 qubits) and IBM Heron (Fez, 156 qubits), for 72 instances. We add a second set of 117 instances: the 39 circuits of Standard-42 (Sec.~\ref{sec:suites}) that have at most 20 qubits and at least 20 two-qubit gates, on the same three devices. The SimpleTES numbers are for its best published policy (commit \texttt{47d3413}) run by its own evaluator with a 14{,}400~s budget. The LightSABRE baseline is Qiskit~2.6.1 with 4 layout iterations, 20 layout trials and 20 swap trials. The beam numbers are from the native implementation in \qs{} (commit \texttt{a7ba770}) with the parameters of Sec.~\ref{sec:beam-alg} on all devices. Every beam output passed an independent replay check (every two-qubit gate on a coupling edge, the gate sequence on every logical qubit unchanged) and the SimpleTES equivalence checker, with no failures.

\begin{table*}[t]
\caption{Routing only: inserted SWAPs (lower is better). $\Delta$ is the change for the beam router relative to SimpleTES (S) and LightSABRE (L). W/T/L counts instances on which the beam's SWAP count is lower than, equal to or higher than that of SimpleTES.}
\label{tab:beam-routing}
\footnotesize
\begin{tabular}{lrrrrrc}
\toprule
Device & LightSABRE & SimpleTES & beam & $\Delta_S$ & $\Delta_L$ & W/T/L vs S \\
\midrule
\multicolumn{7}{l}{\emph{SimpleTES set, 72 instances}} \\
Q20    & 20{,}063  & 13{,}980 & 5{,}019  & $-64.1\%$ & $-75.0\%$ & 17/7/0 \\
Willow & 36{,}802  & 32{,}050 & 26{,}864 & $-16.2\%$ & $-27.0\%$ & 18/6/0 \\
Heron  & 45{,}827  & 42{,}032 & 39{,}986 & $-4.9\%$  & $-12.7\%$ & 17/6/1 \\
Total  & 102{,}692 & 88{,}062 & 71{,}869 & $-18.4\%$ & $-30.0\%$ & 52/19/1 \\
\midrule
\multicolumn{7}{l}{\emph{Compiler-suite set, 117 instances}} \\
Q20    & 1{,}685 & 1{,}383 & 991     & $-28.3\%$ & $-41.2\%$ & 27/11/1 \\
Willow & 3{,}137 & 3{,}123 & 2{,}439 & $-21.9\%$ & $-22.3\%$ & 33/5/1 \\
Heron  & 4{,}780 & 5{,}000 & 4{,}317 & $-13.7\%$ & $-9.7\%$  & 33/2/4 \\
Total  & 9{,}602 & 9{,}506 & 7{,}747 & $-18.5\%$ & $-19.3\%$ & 93/18/6 \\
\bottomrule
\end{tabular}
\end{table*}

\paragraph*{Results.} Table~\ref{tab:beam-routing} gives the totals. The beam router inserts 18.4\% fewer SWAPs than SimpleTES on the 72 instances and 18.5\% fewer on the 117 instances. Against LightSABRE, it inserts 30.0\% and 19.3\% fewer SWAPs, respectively. It never inserts more on the 72 instances (56/16/0) and inserts more on 10 of the 117 (85/22/10). The gain is largest on the small Q20 device and smallest on Heron.

\paragraph*{SimpleTES and its own benchmark.} SimpleTES evolved its policy on the 72 instances, on which it inserts 14.2\% fewer SWAPs than LightSABRE. On the 117 instances, it inserts only 1.0\% fewer. Per device, it is 17.9\% better on Q20, equal on Willow ($-0.4\%$) and 4.6\% worse on Heron, and it loses to LightSABRE on 65 of the 117 instances (30/22/65). We read this as overfitting of an evolved heuristic to the instances it was evolved on. The beam router's parameters were not tuned on the 117 instances, and its advantage over LightSABRE shrinks from 30.0\% to 19.3\% but does not disappear. The 117-instance set was not used to tune the router, but it is development data for the \qs{} compiler as a whole (Sec.~\ref{sec:beam-caveats}).

\subsection{Ablation}
\label{sec:beam-ablation}

To assess the effects of the design choices (a)--(c) of Sec.~\ref{sec:beam-related}, we added environment switches to the router that disable deduplication, disable the release valve, or replace the global beam with a rolling scheme that commits one SWAP after a lookahead of a given depth. The switches exist only in a separate patch to commit \texttt{a7ba770}; with all of them off, the router is unchanged. Every variant starts from the same four SABRE layouts as the default, and only the listed setting differs. Table~\ref{tab:beam-ablation} reports results for the 72 and 117 instances of Sec.~\ref{sec:beam-routing} and the same 39 circuits on a 20-qubit line. No variant failed on any instance.

\begin{table*}[t]
\caption{Ablation of the beam router: inserted SWAPs (sums over instances; lower is better). $\Delta$ is relative to the default on the same instances; W/T/L counts instances on which the variant's SWAP count is lower than, equal to or higher than the default's. ``Line-20'' is the 39 circuits of the 117-instance set on a 20-qubit line. ``No release valve'' runs with a SWAP cap of $3\times$ the best SABRE count and would report a failure if the beam ever stalled. ``Rolling commit'' looks 15 steps ahead and commits one SWAP per round, keeping all other default settings. ``Qiu-style'' is described in Sec.~\ref{sec:beam-qiu}. n/r: not run.}
\label{tab:beam-ablation}
\footnotesize
\begin{tabular}{lrrcrrcr}
\toprule
 & \multicolumn{3}{c}{72 instances} & \multicolumn{3}{c}{117 instances} & Line-20 \\
Variant & SWAPs & $\Delta$ & W/T/L & SWAPs & $\Delta$ & W/T/L & $\Delta$ \\
\midrule
Default ($B=128$, dedup, valve, decay, 2 layout iterations) & 71{,}869 & -- & -- & 7{,}747 & -- & -- & -- \\
\midrule
Beam width $B=1$  & 107{,}798 & $+50.0\%$ & 0/21/51 & 10{,}203 & $+31.7\%$ & 1/26/90 & $+19.9\%$ \\
Beam width $B=8$  & 83{,}101  & $+15.6\%$ & 2/23/47 & 8{,}403  & $+8.5\%$  & 1/47/69 & $+4.2\%$ \\
Beam width $B=32$ & 74{,}780  & $+4.1\%$  & 1/26/45 & 7{,}945  & $+2.6\%$  & 11/53/53 & n/r \\
No deduplication  & 99{,}571  & $+38.5\%$ & 1/23/48 & 8{,}996  & $+16.1\%$ & 4/48/65 & $+13.5\%$ \\
No layout iteration & 72{,}518 & $+0.9\%$ & 0/31/41 & 7{,}997  & $+3.2\%$  & 0/43/74 & $+1.2\%$ \\
Identity start, 2 iterations & 74{,}229 & $+3.3\%$ & 5/21/46 & 8{,}739 & $+12.8\%$ & 2/21/94 & $+1.2\%$ \\
Identity start, no iteration & 77{,}814 & $+8.3\%$ & 0/3/69 & 10{,}493 & $+35.4\%$ & 0/2/115 & n/r \\
No lookahead ($w=0$) & 72{,}722 & $+1.2\%$ & 8/25/39 & 7{,}781 & $+0.4\%$ & 27/61/29 & $+0.8\%$ \\
No decay penalty & 71{,}634 & $-0.3\%$ & 26/29/17 & 7{,}736 & $-0.1\%$ & 27/71/19 & $-0.3\%$ \\
No release valve & 71{,}869 & $0.0\%$ & 0/72/0 & 7{,}747 & $0.0\%$ & 0/117/0 & $0.0\%$ \\
No valve, no decay & 71{,}634 & $-0.3\%$ & 26/29/17 & 7{,}736 & $-0.1\%$ & 27/71/19 & $-0.3\%$ \\
No dedup, no decay & 101{,}579 & $+41.3\%$ & 0/26/46 & 8{,}884 & $+14.7\%$ & 5/50/62 & $+12.7\%$ \\
\midrule
Rolling commit, $B=8$ & 79{,}120 & $+10.1\%$ & 1/22/49 & 8{,}297 & $+7.1\%$ & 3/51/63 & $+3.4\%$ \\
Qiu-style & 91{,}468 & $+27.3\%$ & 0/3/69 & 11{,}863 & $+53.1\%$ & 0/2/115 & $+16.0\%$ \\
\bottomrule
\end{tabular}
\end{table*}

\paragraph*{What matters.} Beam width and deduplication account for most of the gain. Going from $B=128$ to a greedy single path ($B=1$) costs 50.0\% more SWAPs on the 72 instances and 31.7\% on the 117 instances. $B=32$ already recovers most of the gain ($+4.1\%$, $+2.6\%$). Without deduplication, the beam fills with permutations of the same states, and the SWAP count rises by 38.5\% and 16.1\%. Layout is next: removing the forward--backward iterations costs 0.9\% and 3.2\%, starting from the identity layout with iterations costs 3.3\% and 12.8\%, and the identity layout without iterations costs 8.3\% and 35.4\%. The lookahead term $X$ in Eq.~\eqref{eq:beamcost}, the part of the score taken from Qiu et al., contributes only 0.4--1.2\% on these sets, where a wide deduplicated beam already captures most of the information the window would provide; on long, crowded chains it matters more (Table~\ref{tab:beam-stress}).

\paragraph*{Global beam versus rolling commit.} At equal width, committing one SWAP after a 15-step lookahead is better than a global beam: with $B=8$ it inserts 79{,}120 SWAPs against 83{,}101 on the 72 instances and 8{,}297 against 8{,}403 on the 117 instances. It also needs about ten times the work of the global beam at $B=8$, because it reruns the lookahead for every committed SWAP. The global beam at $B=128$ inserts 9.2\% and 6.6\% fewer SWAPs than the rolling scheme at $B=8$. We therefore do not claim that a global beam is better than rolling commitment as such. The global beam allows the width to be raised to 128 at a cost that fits the level-3 budget.

\paragraph*{Decay penalty and release valve.} On the 72, 117 and Line-20 sets, neither changes the result much. Removing the decay penalty changes the total by $-0.3\%$, $-0.1\%$ and $-0.3\%$, with mixed wins and losses. With the release valve disabled, the beam never stalled, and the output was identical to the default; even with both disabled, every instance was routed.

\paragraph*{Stress set.} The decay penalty and the release valve were added for long, crowded chains. We therefore also ran eight synthetic circuits (random, QFT, QAOA and Heisenberg, 32 and 50 qubits, from the synthetic suite of Sec.~\ref{sec:suites}) on a line of the same width, with a cap of $5\times$ the best SABRE count (``no release valve'' keeps its $3\times$ cap). Table~\ref{tab:beam-stress} lists all 14 variants.

\begin{table}[t]
\caption{Stress set: eight synthetic circuits with 32 and 50 qubits on a line of the same width, inserted SWAPs. Sums are over the instances a variant routed; $\Delta$ is relative to the default on those instances. A failure is a run that exceeded the SWAP cap or stalled. The two Heisenberg circuits need no SWAPs. $^{a}$Fewest SWAPs among the SABRE candidates of the \qs{} compiler, which serve as starting points for the beam.}
\label{tab:beam-stress}
\footnotesize
\setlength{\tabcolsep}{2.4pt}
\begin{tabular}{lrrrrc}
\toprule
Variant & Line-32 & Line-50 & total & $\Delta$ & failed \\
\midrule
SABRE$^{a}$ & 2{,}983 & 11{,}073 & 14{,}056 & $-18.1\%$ & -- \\
Default & 3{,}385 & 13{,}780 & 17{,}165 & -- & 0 \\
$B=1$ & 3{,}755 & 16{,}418 & 20{,}173 & $+17.5\%$ & 0 \\
$B=8$ & 3{,}488 & 15{,}020 & 18{,}508 & $+7.8\%$ & 0 \\
$B=32$ & 3{,}487 & 14{,}190 & 17{,}677 & $+3.0\%$ & 0 \\
No deduplication & 4{,}052 & 16{,}397 & 20{,}449 & $+19.1\%$ & 0 \\
No decay penalty & 8{,}279 & 4{,}248 & 12{,}527 & $+107.5\%$ & 1 \\
No release valve & 3{,}385 & 13{,}780 & 17{,}165 & $0.0\%$ & 0 \\
No valve, no decay & 1{,}187 & 2{,}587 & 3{,}774 & $+23.3\%$ & 3 \\
No dedup., no decay & 1{,}952 & 4{,}023 & 5{,}975 & $+95.3\%$ & 3 \\
No layout iteration & 3{,}399 & 13{,}979 & 17{,}378 & $+1.2\%$ & 0 \\
Identity start & 3{,}431 & 13{,}851 & 17{,}282 & $+0.7\%$ & 0 \\
No lookahead ($w=0$) & 4{,}271 & 17{,}809 & 22{,}080 & $+28.6\%$ & 0 \\
Rolling commit, $B=8$ & 3{,}551 & 15{,}073 & 18{,}624 & $+8.5\%$ & 0 \\
Qiu-style & 1{,}556 & 0 & 1{,}556 & $+106.4\%$ & 5 \\
\bottomrule
\end{tabular}
\end{table}

Width and deduplication behave as on the other sets. Without the decay penalty, the beam drifts: \texttt{random\_32} needs 7{,}136 SWAPs instead of 2{,}334, \texttt{qaoa\_50} 1{,}267 instead of 644, and \texttt{random\_50} exceeds the cap of $5\times$ the SABRE count, so it fails; over the seven instances the beam routes, it inserts 107.5\% more SWAPs than the default. With both the decay penalty and the release valve removed, three of the eight instances fail. Unlike on the device sets, the lookahead term matters here: without it, the beam inserts 28.6\% more SWAPs, whereas the starting layout matters little (identity start $+0.7\%$, no layout iteration $+1.2\%$). The rolling-commit variant at width 8 is slightly worse than the global beam at the same width ($+8.5\%$ against $+7.8\%$), and the Qiu-style scheme, which has no decay penalty, fails on five of the eight instances and needs 106\% more SWAPs on the three it routes. With the decay penalty on, the valve still never fires. This set also shows a limit of the router: on long lines, it inserts 22\% more SWAPs than the best SABRE candidate of the \qs{} compiler (17{,}165 against 14{,}056), which is one reason it enters level~3 only as a candidate under a SWAP cap. We therefore describe the decay penalty and the release valve as mechanisms that prevent drift and guarantee termination, not as sources of quality on typical instances.

\subsection{Comparison with the scheme of Qiu et al.}
\label{sec:beam-qiu}

The ``Qiu-style'' row of Table~\ref{tab:beam-ablation} reimplements the search scheme of Qiu, Zhu and Wei~\cite{qiu2025beam} in our code: identity initial layout, width 8, a 15-step lookahead after which one SWAP is committed, lookahead window 7 with $\gamma=0.7$, and no decay penalty, deduplication or layout iteration. The score is given by Eq.~\eqref{eq:beamcost}, and the release valve is kept so that every run terminates. This is our reading of their method, not their code. On the 24 SABRE circuits on Q20, it inserts 10{,}223 SWAPs, 49\% fewer than LightSABRE (20{,}063), in line with the 44\% reduction against SABRE that they report on IBM Q Tokyo. On the larger devices, the advantage shrinks or reverses: 17.4\% fewer than LightSABRE on Willow and 11.0\% more on Heron (50{,}855 against 45{,}827). On the 117 instances, it inserts 11{,}863 SWAPs, 23.5\% more than LightSABRE (9{,}602), and it is worse than our default on 115 of the 117 instances and never better. The difference between the two routers is the combination measured above (width, deduplication, layout), not the score.

\subsection{End-to-end effect in the \qs{} compiler}
\label{sec:beam-e2e}

In the compiler, the beam is only one more candidate at level~3 (Sec.~\ref{sec:beam-alg}). Before the final round, we compiled the three development suites twice with quspect O3 under the CZ gate set, using the pipeline of the previous version with the beam candidate enabled and disabled and everything else held equal. We compared per-case medians over seeds (Table~\ref{tab:beam-e2e}). In the final compiler, whose other candidates and post-routing resynthesis are stronger, the effect is about zero (Table~\ref{tab:ablation-final}, seed~0): removing the beam candidate changes the geometric-mean two-qubit count by $+0.1\%$ on Standard-42, $+0.4\%$ on the synthetic suite, $+0.3\%$ on Large-20, $-0.2\%$ on the arXiv set and $+0.1\%$ on the 513 \texttt{guoq\_dev} cases. The corresponding counts of cases with fewer / equal / more two-qubit gates without the beam candidate are 4/197/9, 1/181/14, 0/74/6, 2/270/3 and 6/488/19. The depth changes by $-0.5\%$ to $+0.4\%$. The beam therefore stays in the compiler as a candidate that occasionally wins rather than as a source of the average lead.

\begin{table}[t]
\caption{End-to-end effect of the beam candidate in quspect O3 on the pipeline of the previous version (CZ gate set): geometric mean over cases of the ratio with/without beam, per-case two-qubit win/tie/loss, and total two-qubit count without and with beam.}
\label{tab:beam-e2e}
\footnotesize
\setlength{\tabcolsep}{2.4pt}
\begin{tabular}{lrrcrr}
\toprule
 & \multicolumn{2}{c}{ratio on/off} & & \multicolumn{2}{c}{two-qubit total} \\
Suite (cases) & two-qubit & depth & W/T/L & off & on \\
\midrule
Standard-42 (210) & 0.9953 & 0.9964 & 24/186/0 & 76{,}268 & 75{,}582 \\
Synthetic (196)   & 0.9921 & 1.0074 & 23/173/0 & 98{,}095 & 96{,}940 \\
Large-20 (80)     & 0.9958 & 0.9967 & 5/75/0   & 878{,}504 & 877{,}746 \\
\bottomrule
\end{tabular}
\end{table}

The geometric-mean two-qubit count falls by 0.4--0.8\%, and no case gets worse. The largest single gains are 12--13\%: \texttt{qft\_32} on heavy-hex (1{,}612 $\to$ 1{,}401 two-qubit gates) and \texttt{poly\_pauli\_rot\_9} on heavy-hex (2{,}026 $\to$ 1{,}783). On the synthetic suite, the depth ratio is slightly above one (1.0074), because level~3 selects by two-qubit count first. No run hit a time limit. The end-to-end gain is much smaller than the routing-only gain for two reasons. The SABRE, LNS and mirror-gate pipeline of the \qs{} compiler is already much stronger than LightSABRE, and the final gate count depends on gate merging and synthesis, which are not reflected in the SWAP count. These runs were performed on a working tree whose router is behaviorally identical to \texttt{a7ba770}.

\subsection{Caveats}
\label{sec:beam-caveats}

\paragraph*{Exposure to arXiv results.} The beam router was designed, and the final round was carried out, after the results of the previous version on the arXiv set had been computed and inspected. The final commit, which includes the beam router, was evaluated on the set once; later runs on it (ablations and bit-identity checks) reproduced that evaluation and did not feed back into development. Knowledge of the earlier results, for example which opponents won where, informed the final development direction, so the arXiv numbers of the final compiler include development-time exposure to aggregate arXiv results (Sec.~\ref{sec:evalprotocol}).

\paragraph*{The 117 instances are development data for the \qs{} compiler.} The 117-instance set was not used to tune the router, but its 39 circuits are from Standard-42, a development suite of the compiler. It is new data for the router only, and the end-to-end comparison of Sec.~\ref{sec:beam-e2e} runs entirely on development suites.

\paragraph*{SWAP count is not gate count.} Sections~\ref{sec:beam-routing}--\ref{sec:beam-qiu} count inserted SWAPs with the gates unchanged. A router that wins on this metric can lose after gate merging, mirror gates and resynthesis, and the 18--30\% routing-only gains correspond to 0.4--0.8\% fewer two-qubit gates end to end. The routing-only comparison with SimpleTES and LightSABRE should not be read as a comparison of compilers.

\section{Hardware experiments}
\label{sec:hw}

\paragraph*{Device and access.}
We used a 17-qubit superconducting processor, designated QZ01 by its cloud provider, which exposes two coupling subgraphs of the same chip. The \emph{chain} has 16 couplers forming a Hamiltonian path, and the \emph{grid} has 24 couplers in a surface-code layout with degree at most 4. The chain's edges are a subset of the grid's. The average CZ fidelity was about 0.994, with $T_1\approx37~\mu$s, $T_2\approx9~\mu$s and $T_2^*\approx3~\mu$s. The native gates are CZ, arbitrary single-qubit SU(2) pulses and virtual $R_z$. Circuits were submitted with the provider's compiler disabled. Before submission we decoded the serialized payload, simulated it, and found it identical to the logical circuit (gate counts equal; total variation distance below $2\times10^{-15}$ for the v0.6.0 main run and at most $2.3\times10^{-13}$ for the runs with the final circuits). Readout correction and dynamical decoupling were off, and we report raw counts.

\paragraph*{Compilers.} The main run reported below uses the final compiler (commit \texttt{be4ecda}). The experiments were first run with the earlier \qs{} release v0.6.0 (commit \texttt{3b849e1}), which on the six main-run scenarios produces the same CZ gates, initial layout and final mapping as the previous version \texttt{8ec73d8} (we did not check other circuits). When recompiled with the final compiler, five of these six circuits changed, so the main run was repeated with the final circuits, while the Qiskit and tket circuits were kept byte-identical. The extended study (Sec.~\ref{sec:hw-ext}) was run with v0.6.0; its 16 scenarios for which the final compiler changes the seed-averaged CZ count by at least 14\% were re-run with the final circuits. In the other 20, the change is at most 11\% (never an increase). quspect O3, Qiskit O3 (2.6.1) and tket O2 (pytket 2.18.5) each compiled for the full coupling graph without calibration data and chose their own physical qubits. Adjacent single-qubit gates were fused into one $u_3$ (one SU(2) pulse or a virtual $R_z$) with the same rule for all compilers, and every compiled circuit passed the equivalence check.

\paragraph*{Statistics.} The versions of a scenario were submitted back to back, the submission order was rotated between rounds, the two coupling graphs were interleaved, and every job had 4000 shots. The success probability $p$ is the frequency of the ideal string, with a 95\% Wilson interval~\cite{wilson1927}, which reflects shot noise only and ignores drift between rounds. To assess significance, we use a hierarchical bootstrap~\cite{efron1979bootstrap} with 20{,}000 resamples (4000 for QAOA), which first resamples rounds and then uses binomial (for QAOA multinomial) resampling of shots, so drift between rounds contributes to the error. A difference is significant if its two-sided 95\% interval excludes zero. For multiple comparisons, we apply the Benjamini--Hochberg procedure~\cite{benjamini1995fdr} at $q=0.05$ to the two-sided bootstrap tail probabilities, $2\min(p,1-p)$ for the one-sided bootstrap probability $p$. We treat all seed-averaged comparisons of a study as one family (12 in the main run, 72 in the extended study, 32 in its re-run). The dates of hardware runs are the UTC dates in the provider's job records. In the main run we also report readout-corrected values and the estimated success probability (ESP), the product of the CZ fidelities of the couplers used, the single-qubit XEB fidelity raised to the number of SU(2) pulses, and the readout fidelities.

\subsection{Main run}
We scanned 18 candidate circuits on both graphs and kept three circuits whose interaction graphs require routing, whose compiled CZ counts are between about ten and eighty, and whose ideal output is a single bit string. These are BV-10 (all-ones secret, a 9-to-1 fan-in; ideal output \texttt{1111111111}), IQFT-6 (a Fourier state of 45 followed by an inverse QFT with SWAP stages; \texttt{101101}), and an 8-qubit Cuccaro adder with Toffolis (\texttt{10000010}). A chain GHZ circuit was excluded because all three compilers produced identical circuits. All compilers used seed 0, fixed in advance. We ran three rounds on each coupling graph. The run with the final circuits took place on 2026-11-08; all 54 jobs ran within two minutes under one calibration.

\begin{table*}[t]
\caption{Main hardware run on QZ01 (2026-11-08; seed 0, 3 rounds $\times$ 4000 shots). \qs{}: final version (commit \texttt{be4ecda}); the Qiskit and tket circuits are identical to those of the earlier runs. $p$: success probability with 95\% Wilson interval; $p_{\mathrm{ro}}$: readout-corrected (reference only); ESP: calibration-based estimate. Bold: highest $p$ in the scenario. Uniform guessing gives $2^{-10}$, $2^{-6}$ and $2^{-8}$.}
\label{tab:hw}
\scriptsize
\setlength{\tabcolsep}{2.8pt}
\begin{tabular}{lllrrrrlccc}
\toprule
Chip & Circuit & Compiler & CZ & CZ depth & SU(2) & depth & $p$ [95\% Wilson] & per-round $p$ & $p_{\mathrm{ro}}$ & ESP \\
\midrule
chain & BV-10 & quspect O3 & 16 & 11 & 21 & 22 & \textbf{0.579} [0.571, 0.588] & 0.584/0.571/0.583 & 0.735 & 0.703 \\
chain & BV-10 & Qiskit O3 & 30 & 30 & 61 & 60 & 0.039 [0.036, 0.043] & 0.053/0.033/0.032 & 0.048 & 0.626 \\
chain & BV-10 & tket O2 & 25 & 24 & 38 & 48 & 0.266 [0.258, 0.274] & 0.280/0.259/0.259 & 0.337 & 0.660 \\
\addlinespace[1pt]
chain & IQFT-6 & quspect O3 & 39 & 24 & 72 & 48 & \textbf{0.451} [0.443, 0.460] & 0.458/0.441/0.455 & 0.488 & 0.686 \\
chain & IQFT-6 & Qiskit O3 & 57 & 39 & 90 & 69 & 0.192 [0.185, 0.199] & 0.217/0.176/0.183 & 0.207 & 0.603 \\
chain & IQFT-6 & tket O2 & 45 & 28 & 80 & 53 & 0.378 [0.369, 0.386] & 0.401/0.388/0.344 & 0.413 & 0.640 \\
\addlinespace[1pt]
chain & Adder-8 & quspect O3 & 45 & 37 & 81 & 72 & \textbf{0.247} [0.240, 0.255] & 0.255/0.250/0.238 & 0.263 & 0.639 \\
chain & Adder-8 & Qiskit O3 & 82 & 75 & 123 & 139 & 0.090 [0.085, 0.096] & 0.092/0.094/0.085 & 0.095 & 0.475 \\
chain & Adder-8 & tket O2 & 71 & 63 & 120 & 125 & 0.047 [0.043, 0.051] & 0.059/0.051/0.030 & 0.049 & 0.528 \\
\addlinespace[1pt]
grid & BV-10 & quspect O3 & 11 & 11 & 23 & 19 & \textbf{0.438} [0.429, 0.447] & 0.439/0.470/0.404 & 0.530 & 0.756 \\
grid & BV-10 & Qiskit O3 & 18 & 18 & 37 & 30 & 0.304 [0.296, 0.313] & 0.359/0.240/0.313 & 0.382 & 0.703 \\
grid & BV-10 & tket O2 & 24 & 18 & 40 & 31 & 0.316 [0.307, 0.324] & 0.321/0.323/0.303 & 0.397 & 0.680 \\
\addlinespace[1pt]
grid & IQFT-6 & quspect O3 & 32 & 28 & 51 & 52 & \textbf{0.382} [0.373, 0.391] & 0.401/0.381/0.363 & 0.412 & 0.742 \\
grid & IQFT-6 & Qiskit O3 & 45 & 40 & 63 & 70 & 0.265 [0.257, 0.273] & 0.295/0.237/0.263 & 0.301 & 0.671 \\
grid & IQFT-6 & tket O2 & 51 & 47 & 72 & 88 & 0.157 [0.150, 0.163] & 0.154/0.157/0.159 & 0.169 & 0.653 \\
\addlinespace[1pt]
grid & Adder-8 & quspect O3 & 43 & 39 & 85 & 79 & 0.089 [0.084, 0.094] & 0.102/0.094/0.069 & 0.092 & 0.683 \\
grid & Adder-8 & Qiskit O3 & 73 & 66 & 102 & 121 & \textbf{0.166} [0.159, 0.172] & 0.162/0.167/0.168 & 0.174 & 0.575 \\
grid & Adder-8 & tket O2 & 72 & 59 & 113 & 114 & 0.045 [0.041, 0.049] & 0.046/0.047/0.042 & 0.047 & 0.562 \\
\bottomrule
\end{tabular}

\end{table*}

\begin{figure*}[t]
\centering
\includegraphics[width=0.86\textwidth]{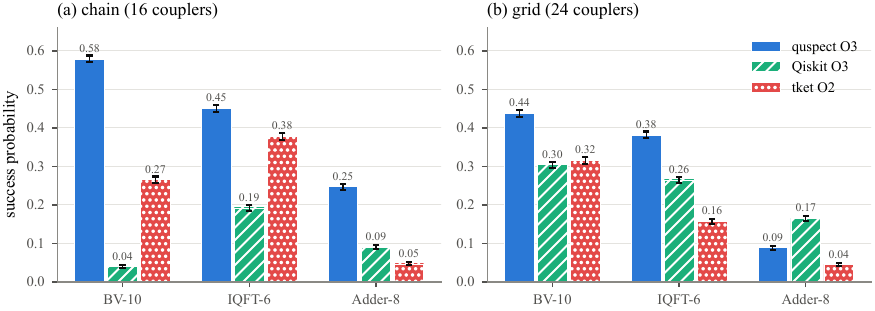}
\caption{Main run (final circuits, 2026-11-08): success probability on the (a) chain and (b) grid coupling graphs of QZ01. Error bars are 95\% Wilson intervals over 12{,}000 shots.}
\label{fig:hw}
\end{figure*}

In these six scenarios, the \qs{} compiler produced the fewest CZ gates (Table~\ref{tab:hw}): 29--47\% fewer than Qiskit O3 (mean 39\%) and 13--54\% fewer than tket O2 (mean 36\%). It also had the lowest CZ depth, SU(2) count and total depth in each of them. Within this session, its success probability was significantly higher in 11 of 12 pairwise comparisons and significantly lower in one, under the hierarchical bootstrap and also after the two-sided Benjamini--Hochberg procedure over the 12 comparisons (Table~\ref{tab:hwpairs}). With three rounds of about two minutes, the bootstrap has only three clusters at its top level, and $p_{\mathrm{boot}}<5\times10^{-5}$ is its resolution limit rather than an estimate. The largest gap is for BV-10 on the chain: the \qs{} compiler needed 16 CZ with CZ depth 11, whereas Qiskit O3's 30 CZ were fully serial (CZ depth 30). The success probabilities were 0.579 for \qs{}, 0.039 for Qiskit O3 and 0.266 for tket O2. On the chain, IQFT-6 reached 0.451 against 0.192 and 0.378, and Adder-8, now compiled with 45 instead of 65 CZ gates, reached 0.247 against 0.090 and 0.047. The exception is Adder-8 on the grid, for which the success probability was significantly lower than that of Qiskit O3 (0.089 against 0.166, $\Delta=-0.077$, CI $[-0.098,-0.058]$) despite 30 fewer CZ gates and a higher ESP. Submitting the v0.6.0 and the final \qs{} circuits back to back under the same calibration (55 jobs) showed that the v0.6.0 circuit loses to Qiskit O3 in this scenario as well ($-0.052$), while the final circuit is slightly better than the v0.6.0 one ($+0.015$). The loss is therefore not caused by the final-round changes, and we did not determine whether qubit selection or the state of the chip is responsible. In the same control, the final circuit was significantly better than the v0.6.0 circuit in three scenarios (up to $+0.091$ on chain Adder-8), worse in one (grid IQFT-6, $-0.060$, with three fewer CZ gates but three more layers) and not significantly different in two (in one of them, chain BV-10, the circuit is the same). One of the 11 wins did not replicate in the control: on the grid, IQFT-6 was no longer significantly better than Qiskit O3 ($+0.007$, CI $[-0.039,+0.046]$). The success probability of the same circuit can change by about 0.1 within half an hour: grid IQFT-6 reached 0.382 in the main run and 0.265 in the control 25 minutes later.

\paragraph*{Earlier runs.} With the v0.6.0 circuits on 2026-11-02, the \qs{} compiler used 21--47\% fewer CZ gates than Qiskit O3 (mean 28\%) and 8--46\% fewer than tket O2 (mean 25\%), and its success probability was significantly higher in 11 of 12 comparisons; the exception was the same scenario, grid Adder-8 against Qiskit O3, which was then a tie (0.119 against 0.131, CI $[-0.029,+0.007]$). Chain BV-10, whose circuit has not changed, reached 0.543 against 0.061 and 0.249. A run on 2026-10-30 with the previous version \texttt{8ec73d8} and an older calibration gave the same qualitative picture (11 of 12 higher, 10 significant). The Qiskit and tket circuits were identical in all three runs, and their success probabilities moved by at most $\pm0.064$ between 2026-11-02 and 2026-11-08, so the chip was in a similar state on both days.

\begin{table*}[t]
\caption{Main run: pairwise comparisons, \qs{} minus baseline. CI: two-sided 95\% hierarchical-bootstrap interval; $p_{\mathrm{boot}}$: one-sided bootstrap probability $P(\Delta\le0)$. The two-sided tail probability $2\min(p_{\mathrm{boot}},1-p_{\mathrm{boot}})$ is below the bootstrap resolution of $10^{-4}$ in all 12 rows, so all 12 differences remain significant under the two-sided Benjamini--Hochberg procedure at $q=0.05$.}
\label{tab:hwpairs}
\footnotesize
\begin{tabular}{lllrrcrc}
\toprule
Chip & Circuit & vs. & $\Delta p$ & ratio & bootstrap 95\% CI & $p_{\mathrm{boot}}$ & rounds won \\
\midrule
chain & BV-10 & Qiskit O3 & $+0.540$ & 14.70 & $[+0.527, +0.553]$ & $<5\times10^{-5}$ & 3/3 \\
chain & BV-10 & tket O2 & $+0.314$ & 2.18 & $[+0.299, +0.328]$ & $<5\times10^{-5}$ & 3/3 \\
chain & IQFT-6 & Qiskit O3 & $+0.259$ & 2.35 & $[+0.240, +0.277]$ & $<5\times10^{-5}$ & 3/3 \\
chain & IQFT-6 & tket O2 & $+0.074$ & 1.20 & $[+0.046, +0.110]$ & $<5\times10^{-5}$ & 3/3 \\
chain & Adder-8 & Qiskit O3 & $+0.157$ & 2.74 & $[+0.147, +0.167]$ & $<5\times10^{-5}$ & 3/3 \\
chain & Adder-8 & tket O2 & $+0.201$ & 5.28 & $[+0.190, +0.211]$ & $<5\times10^{-5}$ & 3/3 \\
grid & BV-10 & Qiskit O3 & $+0.133$ & 1.44 & $[+0.075, +0.226]$ & $<5\times10^{-5}$ & 3/3 \\
grid & BV-10 & tket O2 & $+0.122$ & 1.39 & $[+0.098, +0.148]$ & $<5\times10^{-5}$ & 3/3 \\
grid & IQFT-6 & Qiskit O3 & $+0.117$ & 1.44 & $[+0.095, +0.143]$ & $<5\times10^{-5}$ & 3/3 \\
grid & IQFT-6 & tket O2 & $+0.225$ & 2.44 & $[+0.203, +0.247]$ & $<5\times10^{-5}$ & 3/3 \\
grid & Adder-8 & Qiskit O3 & $-0.077$ & 0.54 & $[-0.098, -0.058]$ & 1 & 0/3 \\
grid & Adder-8 & tket O2 & $+0.044$ & 1.97 & $[+0.029, +0.057]$ & $<5\times10^{-5}$ & 3/3 \\
\bottomrule
\end{tabular}

\end{table*}

\subsection{Extended study}
\label{sec:hw-ext}
The main run covers three circuits with one input each and seed~0 only. To broaden its scope, we repeated the experiment on the same chip on 2026-11-07 with the v0.6.0 compiler and the same design, but with nine circuit families: BV-10, IQFT-6 and Adder-8 as before, QAOA-8 ($p=1$, scored by the Hellinger fidelity to the ideal distribution), GHZ-10 prepared on a random tree, Grover-3, a Draper QFT adder (3+3 bits), W states prepared on a binary tree ($n=8$ and 7) and a 6-qubit hidden-shift circuit. Each family has two inputs tested on both coupling graphs, giving 36 scenarios; each compiler used seeds 0--2 (202 distinct circuits after removing duplicates). The inputs a and b are the secret strings 111111111 and 110101111 (BV-10), $x=101101$ and $110010$ (IQFT-6), adder inputs from seeds 1 and 3 (Adder-8), random 3-regular graphs from seeds 0 and 1 (QAOA-8), random trees from seeds 3 and 11 (GHZ-10), marked strings 101 and 110 (Grover-3), $(a,b)=(3,6)$ and $(5,7)$ (Draper adder), $n=8$ and $7$ (W state) and shifts 101101 and 011010 (hidden shift). A mirrored random Clifford circuit was excluded because tket reduced it to zero CZ gates. Every circuit was run in six rounds of 4000 shots: three planned rounds (batch A) and three replication rounds (batch B), added after batch A had been analyzed because the round-to-round variation for a single circuit was large. The provider keeps only the most recent jobs, about 200 per account, so the results of the first two planned rounds (405 executed jobs) were lost; these rounds were re-run with the same design, and the 8 lost jobs from the third planned round were re-run together with all versions of their scenario. The 1225 analyzed jobs completed within one hour under unchanged calibration data. Data: \texttt{hardware/results/20261107\_p8-qz01-expanded/} (benchmark commit \texttt{93f86f7}).

\begin{table}[t]
\caption{Extended study with the v0.6.0 circuits (2026-11-07): 72 pairwise comparisons (36 scenarios $\times$ 2 baselines) of the seed-averaged success metric, hierarchical bootstrap; sig.\,$\pm$: the two-sided 95\% interval excludes zero. BH: improvements / regressions that remain after the Benjamini--Hochberg procedure at $q=0.05$, applied to the two-sided bootstrap tail probabilities of the 72 comparisons in the row as one family. Per-scenario results in Table~\ref{tab:hwext}.}
\label{tab:hwsum}
\footnotesize
\setlength{\tabcolsep}{3pt}
\begin{tabular}{lccccc}
\toprule
Rounds & higher & sig.\,$+$ & sig.\,$-$ & n.s. & BH $+$/$-$ \\
\midrule
A: 1--3 (planned) & 59 & 51 & 7 & 14 & 49/7 \\
B: 4--6 (replication) & 60 & 43 & 2 & 27 & 43/2 \\
all six & 60 & 49 & 6 & 17 & 48/6 \\
all six, seed 0 & 58 & 51 & 7 & 14 & 47/7 \\
\bottomrule
\end{tabular}
\end{table}

\paragraph*{Compilation.} Averaged over seeds, the \qs{} compiler never used more CZ gates than either baseline: it used fewer than Qiskit O3 in 32 of 36 scenarios and the same number in 4, fewer than tket O2 in 31 and the same number in 5, on average 22\% fewer. Per seed, it used fewer CZ gates in 93 and the same number in 15 of 108 comparisons with each baseline. In four scenarios (W-7 on the chain, W-8, W-7 and GHZ-10 a on the grid), all three compilers use the same seed-averaged CZ count. The total depth of the \qs{} circuits exceeded that of a baseline in one of 216 per-seed comparisons, by one layer. Recompiled with the final compiler, 73 of the 108 \qs{} circuits use fewer CZ gates than the v0.6.0 ones and none uses more (26 of the 36 scenarios change on average), so the final circuits never use more CZ gates than the baselines either.

\paragraph*{Success probability.} Averaged over seeds, the success metric of the \qs{} circuits was higher in 60 of 72 pairwise comparisons. Two-sided 95\% hierarchical-bootstrap intervals excluded zero in 49 improvements and six regressions; applying Benjamini--Hochberg at $q=0.05$ to all 72 two-sided bootstrap tail probabilities leaves 48 improvements and six regressions (Table~\ref{tab:hwsum}; the improvement that does not survive is QAOA-8 b on the chain against Qiskit O3, $+0.055$ with interval $[+0.001,+0.095]$). Against Qiskit O3, the record is 25 higher, 2 lower and 9 not significant; against tket O2 it is 24, 4 and 8. Within single rounds, the \qs{} circuits scored higher in 182 of 216 back-to-back pairs against Qiskit O3 and 166 of 216 against tket O2. The planned and the replication rounds agree (Table~\ref{tab:hwsum}), and the conclusion does not depend on the seed: for seeds 0, 1 and 2 separately, the \qs{} circuits have the higher point estimate in 30, 30 and 28 of 36 scenarios against Qiskit O3 and in 28, 27 and 29 against tket O2. In the 7 W-state comparisons with equal seed-averaged CZ counts, where the circuits differ only in layout, single-qubit gates and scheduling, the \qs{} circuits were significantly better in 6.

\paragraph*{Counterexamples.} With the v0.6.0 circuits, the six significant losses were concentrated on two families; only grid Grover-3 b against tket was significant in both batches, whereas the other five were significant in one batch only. On the chain, all compilers were close to random guessing on the hidden-shift circuits ($p\approx0.02$--$0.06$ against $1/64$), and the significant differences there were of order 0.01--0.04. On the grid, tket reached 0.17--0.19 on hidden shift with 27 CZ, against 0.06 for \qs{} with 22 CZ, and on Grover-3 with input b, tket reached 0.416 against 0.280 for \qs{} ($\Delta=-0.136$, CI $[-0.209,-0.067]$) despite using six more CZ gates and having a lower ESP. Neither difference was explained by CZ count, depth or ESP. Qubit selection by compilers that ignore noise is a possible cause, but the calibration data do not show it: on grid Grover-3 b the qubits used by the \qs{} circuits had the higher ESP (0.80 against 0.73 for tket). Adder-8 against Qiskit O3 was a tie on both graphs; its success probability (0.02--0.05) was close to the noise floor. How these cases changed with the final circuits is described below.

\paragraph*{Drift.} The chip was in a worse and less stable state than in the main run: the standard deviation for the same circuit between adjacent rounds was 0.125 on average. For example, seed-0 BV-10 on the chain reached 0.292 for \qs{} and 0.000 for Qiskit O3, against 0.543 and 0.061 in the v0.6.0 main run five days earlier. Absolute success probabilities from different days should therefore not be compared, which is why each comparison in this study uses back-to-back submissions within a round.

\paragraph*{Re-run with the final circuits.} For the 16 scenarios in which the final compiler changed the seed-averaged CZ count by at least 14\% (Grover-3 and Adder-8 on both graphs, QAOA-8 except input b on the chain, the Draper adder on the grid, and hidden shift except input a on the grid; Table~\ref{tab:hwextfinal}), we re-ran the study on 2026-11-08 with the final \qs{} circuits and the unchanged baseline circuits, using the same design (seeds 0--2, back-to-back, six rounds, 535 jobs within about 20 minutes under one calibration, the same as for the main run). Of the 32 comparisons, the \qs{} circuits scored significantly higher in 27 and significantly lower in 3 (the same counts after the two-sided Benjamini--Hochberg procedure over the 32 comparisons), compared with 18 and 5 for the v0.6.0 circuits on the same scenarios (Table~\ref{tab:hwsumfinal}); against tket O2, they scored higher in all 16 and significantly higher in 15 (with seed~0 only, 2 of these 16 comparisons, both QAOA-8 b on the grid, were significantly lower). Grover-3 now uses 9--10 CZ gates instead of 29, and the counterexample of Grover-3 on the grid against tket is reversed (0.764 against 0.488, $\Delta=+0.276$, CI $[+0.252,+0.302]$); all four Grover-3 scenarios show leads of 0.17--0.48 over both baselines. On the chain, the four hidden-shift comparisons changed from significantly lower to significantly higher (0.34--0.47 against 0.06--0.18), but the chip was in a much better state than on 2026-11-07: the unchanged baseline circuits gained 0.10 on average, and the standard deviation between adjacent rounds fell from 0.125 to 0.040. On 2026-11-07 all three compilers were close to random guessing there, so this reversal cannot be credited to the circuits alone; comparisons across sessions are confounded by changes in device state. On the grid, hidden shift against tket remained a tie (0.383 against 0.367). The three significant losses are all small and occur against Qiskit O3 on the grid: Adder-8 with both inputs ($-0.028$ and $-0.014$, as in the main run) and QAOA-8 with input a ($-0.019$ in Hellinger fidelity). The 20 scenarios that were not re-run retain their v0.6.0 results in Table~\ref{tab:hwsum} and Appendix~\ref{app:hwext}.

\begin{table}[t]
\caption{Re-run of the extended study with the final circuits on 2026-11-08: 16 scenarios $\times$ 2 baselines, seed-averaged success metric, hierarchical bootstrap (sig.\,$+$ / sig.\,$-$ / n.s.). The first row gives the v0.6.0 results of 2026-11-07 on the same 16 scenarios; the two rows compare results from different days, on which the unchanged baseline circuits differed by $+0.102$ on average, so comparisons between the first row and the others are confounded by the device state. BH as in Table~\ref{tab:hwsum}, with the 32 comparisons of a row as one family. Per-scenario results in Table~\ref{tab:hwextfinal}. Data: benchmark commits \texttt{5cecf2c} and \texttt{8a181b7}.}
\label{tab:hwsumfinal}
\footnotesize
\setlength{\tabcolsep}{2.6pt}
\begin{tabular}{lcccc}
\toprule
Circuits, rounds & vs Qiskit O3 & vs tket O2 & higher & BH $+$/$-$ \\
\midrule
v0.6.0, all six & 9/2/5 & 9/3/4 & 21 & 18/5 \\
final, all six & 12/3/1 & 15/0/1 & 29 & 27/3 \\
final, seed 0 & 13/3/0 & 14/2/0 & 27 & 27/5 \\
\bottomrule
\end{tabular}
\end{table}

\begin{figure}[t]
\centering
\includegraphics[width=\linewidth]{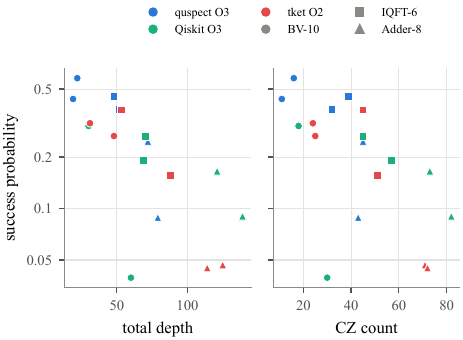}
\caption{Main run (final circuits, 2026-11-08): success probability against total depth (left) and CZ count (right) for the 18 data points.}
\label{fig:hwdepth}
\end{figure}

\paragraph*{Depth or gate count?}
In both studies $\log p$ correlates slightly more strongly with total depth than with CZ count, but the strength depends on the day and the circuit mix, so we only claim the direction. In the main run (18 points) the Spearman coefficients are $-0.80$ and $-0.64$ (Fig.~\ref{fig:hwdepth}; $-0.81$ and $-0.72$ with the v0.6.0 circuits). In the extended study (175 points: the circuits scored by success probability, without QAOA and without points with $p=0$) they are $-0.42$ and $-0.40$, a difference that is not significant, and $-0.61$ and $-0.53$ after centering within each scenario. Depth and CZ count are strongly correlated across these circuits. The same 18 configurations measured in the extended study show almost no correlation, because grid BV-10 fell to about 0.01 for all three compilers. The measured $p$ in the main run was only 0.06--0.82 of the ESP, and the shortfall grew with the length and seriality of the CZ chains. Together with the W-state results, this is compatible with idle decoherence ($T_2^*\approx3~\mu$s) as an error source beyond the CZ gates, but our data cannot separate it from gate errors. The counterexamples, in particular grid Adder-8 against Qiskit O3, show that qubit quality matters as well.

\section{Discussion and limitations}
\label{sec:limitations}

\paragraph*{What the results say about the design.} Four observations from the ablations and the gap measurements shape how we read the results; the ablations assess contributions within the integrated pipeline and do not isolate the selection rule on a fixed candidate pool. First, in the previous version the lead over Qiskit O3 was broad rather than concentrated: no component contributed more than about 8\% on its own, and the largest share came from searching over many routings and judging each by the gate count after synthesis. Second, the stages for which we could measure the gap were close to the computed optimum or bound on small instances in the previous version, so further gains on similar circuits had to come from changing the objective of a stage (for example, routing for merged cost rather than SWAP count) or from stages whose optimum we had not addressed. Third, the final round added resynthesis across blocks after routing, where GUOQ and BQSKit had found their savings. Its components are strongly redundant: together they are worth 23--27\% on Standard-42 and the arXiv set, but the largest single contributions come from the pre-routing second start, numerical instantiation and recompilation (up to 6.2\%, 7.8\% and 3.0\%), whereas the A* search on small CNOT--phase regions contributes only 0.2--1.9\% on its own. With selection after this stage, the selected routings are further from the merge-aware optimum while the final gate count is lower; the comparison changes other compiler components as well. The remaining losses to BQSKit are on small circuits where its $\epsilon$-resynthesis of three-qubit blocks still finds savings of a few gates. Fourth, the more specialized techniques (low-dilation layout, Kendall-$\tau$ and rollout routing, QSD, phase folding, approximation) contribute little on average but are decisive for the circuit families they target, which is what a general compiler needs from them.

\paragraph*{Limitations.}
(i) The compiler is noise-unaware; layout ignores calibration data, unlike noise-adaptive mappers~\cite{murali2019noise}. The hardware counterexamples (grid Adder-8 against Qiskit O3 with 30 fewer CZ gates, and, with the v0.6.0 circuits, hidden shift and Grover-3 against tket on the grid) show that factors other than the CZ count, possibly qubit selection, can outweigh a difference of several CZ gates.
(ii) Formal verification does not scale to the whole large suite. The benchmark's state-vector check stops at 20 physical qubits, so we re-checked all 220 \qs{} O3 outputs of the previous version with more than 20 physical qubits (Large-20, the wide synthetic cases and the larger Standard-42 cases in all three gate sets), which were re-compiled from commit \texttt{8ec73d8} and were identical in all reported metrics. With QCEC (decision-diagram and ZX checkers, trace threshold $10^{-13}$, relaxed to $10^{-10}$ for four outputs whose residual was at the floating-point level) and exact stabilizer simulation for Clifford circuits, 102 outputs were proven equivalent up to global phase, 25 more passed dense state-vector tests with random product inputs ($1-F\le10^{-12}$), and none was found non-equivalent; fault-injection tests confirmed that the checks detect single-gate deletions and $10^{-5}$ rotation errors. The other 93 checks, including 66 of the 80 Large-20 outputs (quantum-volume, arithmetic, QPE and Trotter circuits with 40--100 qubits), did not finish within 600--1800\,s or ran out of memory, so most of the large suite remains formally unverified. We repeated the check for the 220 outputs of the final version with more than 20 physical qubits, re-compiled from \texttt{be4ecda} and identical to the reported results in all metrics (only 17 are byte-identical to their counterparts from the previous version): 92 were proven equivalent (6 of them at the relaxed threshold $10^{-10}$), 28 passed dense state-vector tests, none was found non-equivalent, and 100 were inconclusive, including 65 of the 80 Large-20 outputs (15 proven). For one output, \texttt{qram\_20} on heavy-hex under CZ, QCEC's simulation checker reported non-equivalence at both thresholds while its other checkers were inconclusive. Dense state vectors with random product and random entangled inputs gave $1-F\le3.9\times10^{-13}$ (a single $10^{-5}$ rotation error gave $2.5\times10^{-11}$ in the fault-injection tests), and the CX and IBM outputs and the outputs for the other seeds of the same circuit were proven equivalent. We count it among the 28 that passed the state-vector tests and attribute the report to numerical error in the decision-diagram simulation on the generic angles produced by instantiation. Clifford outputs on the 127- and 156-qubit devices were verified exactly in both rounds (Sec.~\ref{sec:scale}).
(iii) The hardware study covers one 17-qubit chip and one calibration window per run. The main run took three rounds within about two minutes, so its bootstrap has only three top-level clusters, and one of its 11 wins did not replicate in a control 25 minutes later. The extended study (nine circuit families, two inputs, seeds 0--2, two coupling graphs) addresses the narrow scope of the main run but not the dependence on the device. Only 16 of its 36 scenarios were re-run with the final circuits, on a day when the chip was in a much better state, so comparisons between the two days mix circuit and chip changes. Dynamical decoupling was off, and the provider's pulse scheduling is opaque.
(iv) BQSKit was tested at level~2, without its PAS/PAM workflow. The gate-level superoptimizers GUOQ, QUESO and Quartz were evaluated only on the arXiv set, with seed~0 and a 300~s (Quartz: 120~s) limit per case, on top of Qiskit O3's routing. GUOQ and QUESO were restricted to the qubit pairs of the matched pattern or input block, a stricter constraint than the device couplings. We did not evaluate applying these superoptimizers to the outputs of the \qs{} compiler itself or before routing, and GUOQ did not finish Standard-42 in the time available (it was run on the 37 Standard-42 circuits with at most 16 qubits as part of \texttt{guoq\_dev}, which is development data for the final round). The AI transpiler passes were not run under the CX basis, and their routing is not seeded. QMAP's exact mode ran only on the arXiv set.
(v) Level~3 depends on a numerical optimizer, whose results are deterministic only because its budget counts optimizer iterations rather than wall-clock time.
(vi) Cross-platform reproducibility is a design goal but has been only partly tested. The layout kernel uses integer arithmetic only, and the routing kernels reproduce their Python references bit for bit. In the evaluated version, the numerical parts of KAK, single-qubit and QSD synthesis, CNOT--phase resynthesis and three-qubit instantiation run on a deterministic numerical layer without BLAS, LAPACK, the platform math library or fused multiply--add, and are bit-identical across thread counts. Using a separate build on the same architecture, we reproduced the quspect O3 results of the final evaluation bit for bit (Standard-42 (CZ), the synthetic suite and Large-20); we did not test other architectures. The Python fallbacks are numerically equivalent to the native kernels but not bit-identical, so an installation without the native module can differ in individual cases.
(vii) Known weak spots: the VF2 state budget misses 7 of 248 SWAP-free QUEKO layouts; on 16--19-qubit QUBIKOS devices, the previous version inserts 10--21\% more SWAPs than Qiskit; the orientation stage fails on Clifford-like circuits; on large devices, the \qs{} compiler loses on unstructured random Clifford circuits and, under the IBM basis, on some VQE and QAOA circuits; the candidate with the fewest gates can be deeper than Qiskit's (up to 1.55 times on QFT-50 on the all-to-all target); numerical instantiation covers only whole three-qubit circuits and selected three-qubit blocks after routing, so BQSKit's $\epsilon$-resynthesis still wins some small cases; and circuit-size caps restrict the contribution of region resynthesis on large outputs.
(viii) The main ablation and gap measurements in Secs.~\ref{sec:ablation} and~\ref{sec:optimality} refer to the pipeline of the previous version (\texttt{8ec73d8}). The ablation of the final compiler used one seed, only key configurations on Large-20 and 513 of the 785 development cases (none with fewer than six qubits), and it could not remove stabilizer selection; the routing gap of the final compiler was measured only on the small instances of the earlier study. Leave-one-out ablation understates the contributions of components that substitute for each other; the grouped ablation of Sec.~\ref{sec:ablation} measures the routing components jointly, but it does not separate every pairwise interaction, and its reference pipeline uses the compiler's own LightSABRE kernel rather than Qiskit's.
(ix) The evaluation includes prior exposure to aggregate arXiv results (Sec.~\ref{sec:evalprotocol}), and BQSKit and GUOQ were the development targets of the final round. The development set for that round overlaps the other development suites (37 circuits of Standard-42 and 21 synthetic instances), so the final-round gains on Standard-42 and the synthetic suite are partly in-sample. The numerical instantiation is checked per block (each replaced block is within $10^{-9}$ per matrix element), not per circuit; the largest whole-circuit deviation we measured was $1.02\times10^{-9}$ (Sec.~\ref{sec:correctness}).
(x) Direct comparisons with HOPPS, PhasePoly, MIRAGE and PAS/PAM, equal-budget evaluation, and a fixed-candidate-pool study of the ranking objective remain open.

\section{Conclusion}
The \qs{} compiler produces circuits with 23--41\% fewer two-qubit gates and 32--50\% lower depth than Qiskit O3 on the development suites, and with 38\% fewer two-qubit gates and 41\% lower depth on the arXiv evaluation set. Among the tested configurations, which include BQSKit at level~2 and superoptimizers applied after Qiskit's routing, the compiler has the lowest geometric-mean two-qubit ratio; per case, the best tested opponent configuration is better on 11 of the 275 arXiv cases, by a median of one gate, and GUOQ is better on 2. The lead persists, though smaller, on 127- and 156-qubit devices, except on random Clifford and Clifford+T circuits and some circuits under the IBM basis. Within single sessions on a superconducting chip, it mostly carried over to higher success probabilities; there, the compiler never used more CZ gates than the baselines but lost some comparisons for reasons that CZ count and ESP do not explain. The integrated pipeline evaluates screened routing candidates by their cost after coupling-constrained resynthesis. The routing-cost diagnostics illustrate how resynthesis can alter which candidate is preferred; the ablations quantify contributions within the complete pipeline. Open items are direct comparisons with the closest methods, equal-budget evaluation, a fixed-candidate-pool study of the ranking objective, the small cases where BQSKit's $\epsilon$-resynthesis still saves gates, and noise-aware qubit selection. The evaluated version was completed on 8 November 2026 and released in \qs{} v0.6.3~\cite{quspect2026}; the code is available at \url{https://github.com/QuSpect/quspect}.

\begin{acknowledgments}
We thank Quanfeng Lu for checking the conclusions of this paper. GPT-6 Astra and Claude Opus 5.5 assisted with language revisions to the manuscript and development of the engineering code. AI tools also assisted with the review of the arXiv circuits.
\end{acknowledgments}

\section*{Reproducibility statement}
The numbers reported in the evaluation (Sec.~\ref{sec:eval} except Secs.~\ref{sec:competitors} and~\ref{sec:scale}, and the main hardware run) are computed from a shared data package that is regenerated from raw benchmark results by a single script (\texttt{final/make\_all.py -{}-version r21}, which writes the package to \texttt{final/data\_r21/}; without the option, it regenerates the package for the previous version, \texttt{final/data/}). The tables and figures in these sections are regenerated from that package by \texttt{publish/paper/tables/make\_tables.py -{}-version r21} and \texttt{publish/paper/figures/make\_figures.py -{}-version r21} (into \texttt{tables/r21/} and \texttt{figures/r21/}), and derived quantities are written to \texttt{figures/r21/derived.json}. Each result directory records the benchmark and compiler commits, the command line, the machine and the time stamp. The arXiv evaluation set is public with its manifest, review log and extraction tools at \url{https://github.com/QuSpect/arxiv-circuits} (CC BY 4.0; no archived release yet). Hardware job identifiers, raw counts and payload checks are archived with the results. The benchmark data (raw results, scripts and hardware records; ``benchmark'' commits below) are at \url{https://github.com/QuSpect/quspect-compiler-benchmark}.

\emph{Code.} The compiler is the \texttt{quspect.transpiler} package of the \qs{} SDK, so the commits cited by their hashes below are \qs{} commits. The SDK is public at \url{https://github.com/QuSpect/quspect} under the Apache-2.0 license and archived at Zenodo~\cite{quspect2026}: v0.6.3 (released 2026-11-12) has the version DOI \texttt{10.5281/zenodo.24157320}, v0.6.4 (2026-11-13) has \texttt{10.5281/zenodo.24184333}, and the concept DOI of all versions is \texttt{10.5281/zenodo.24157319}. The numbers in this paper were produced by commit \texttt{be4ecda} (2026-11-08), which is in the history of the main branch; v0.6.3 is the first release that contains it. To reproduce the numbers exactly, check out \texttt{be4ecda}: the releases contain later changes to the compiler, and we did not re-run the benchmark with them. Of the changes to the compiler merged after \texttt{be4ecda} and before v0.6.3, one alters outputs on circuits such as ours: since pull request \#49, levels~2 and~3 also compile at level~1 and return that result when it has strictly fewer two-qubit gates, which changes the output of level~2 (``quspect O1'' in our tables) on some circuits (18 of the 360 records of the compiler's golden test, all adders at level~2, according to the commit message). The other changes fix inputs that our benchmark does not contain or are reported by their commit messages to leave the golden outputs unchanged (barriers and noise channels kept through routing; requested layouts with idle top qubits; single-qubit synthesis in u3-only, $\{r_x,r_y\}$ and u2 bases; truncated replacements in basis-aware single-qubit fusion; interpreter-independent summation in three-qubit synthesis; native \texttt{rxx}, \texttt{ryy} and \texttt{rzz} gates); v0.6.4 changes no compiler source relative to v0.6.3. The commits of the final round (pull requests \#14--\#16: \texttt{7b3224c}, \texttt{5dde602}, \texttt{be4ecda}) are in the main history. The ablation switches for the final compiler (\texttt{c82472d}, \texttt{f2c5dca}) are preserved by the tag \texttt{paper-r21-ablation} (\texttt{f2c5dca}). The previous version \texttt{8ec73d8} is preserved by the tag \texttt{paper-r20} (\texttt{69a6991}), whose \texttt{quspect/}, \texttt{cpp/} and \texttt{tests/} directories are identical to those of \texttt{8ec73d8}; its ablation switches (\texttt{46b640a}, \texttt{16fbb94}) are preserved by the tag \texttt{paper-r20-ablation} (\texttt{6935e64}), again with identical sources under other hashes. The original commits \texttt{8ec73d8}, \texttt{46b640a} and \texttt{16fbb94} are not in the public repository, and \texttt{8ec73d8} shares no common ancestor with v0.6.0 or \texttt{be4ecda}: the final round was developed on top of the v0.6.2 release. The beam-router commit \texttt{a7ba770} is contained in the release v0.6.2, and \texttt{3b849e1} is the release v0.6.0. The later experiments have the following sources (\qs{} commit; benchmark commit and data):
\begin{itemize}
\item Final evaluation (2026-11-08): \qs{} \texttt{be4ecda}; benchmark \texttt{d350f87}, \texttt{final/data\_r21/}, \url{compilation/results/20261108-01*_r21-final-*}, summary in \texttt{paper/r21\_final/}. Only the \qs{} compiler was run; the other compilers' results come from the evaluation of the previous version and the later experiments below.
\item Final-round development: development set \texttt{compilation/qasm\_guoqdev/} (benchmark \texttt{6b2eedd}, with manifest and the overlap check against the arXiv set), baselines, diagnosis and per-step evaluations in benchmark \texttt{33fec84}, \texttt{21f9c6d} and \texttt{e7a1150} (\texttt{paper/guoq\_evolve/}).
\item Evaluation of the previous version: \qs{} \texttt{8ec73d8} (tag \texttt{paper-r20}); \texttt{final/data/}.
\item Ablation (Sec.~\ref{sec:ablation}): \qs{} \texttt{8ec73d8} with ablation switches (\texttt{46b640a}); benchmark \texttt{f64c9ae}, \texttt{compilation/results/20261106-*\_ablation-*}, \texttt{20261107-001*\_ablation-*}; bit-identity evidence in \texttt{20261106-212259\_ablation-bitident-m3u}. Grouped ablation: switches \texttt{16fbb94}; benchmark \texttt{f3bbd3b}, \url{compilation/results/20261107-0*_ablation-grp-*}, bit-identity evidence in \url{20261107-024000_ablation-grp-bitident-m3u}.
\item Optimality (Sec.~\ref{sec:optimality}): \qs{} \texttt{8ec73d8}; benchmark \texttt{7355da5}, \texttt{paper/optimality/}.
\item Scaling (Sec.~\ref{sec:scale}): \qs{} \texttt{be4ecda} (previous version: \texttt{8ec73d8}); benchmark \texttt{0620898} and \texttt{d350f87}, \texttt{compilation/qasm\_xl/}, \texttt{compilation/results/20261107-*\_p7-xl-*} and \url{20261108-014*_r21-final-xl-*}, \texttt{paper/scale/}.
\item Beam router (Sec.~\ref{sec:beam}): \qs{} \texttt{a7ba770} (router code identical in \texttt{be4ecda}); benchmark \texttt{612adbc}, \texttt{17d9615}, \texttt{27aca4a}, \texttt{routing/results/}.
\item Hardware: main run with the final circuits (\qs{} \texttt{be4ecda}), the re-run of the extended study with the final circuits and the control run (all 2026-11-08), benchmark \texttt{5cecf2c} and \texttt{8a181b7}, \url{hardware/results/20261108_r21-main-rerun/}, \url{20261108_r21-p8-subset/}, \url{20261108_r21-main-control/}; earlier main run (2026-11-02) \qs{} v0.6.0 (\texttt{3b849e1}; its equality with the previous version \texttt{8ec73d8} in CZ gates, initial layout and final mapping was checked only on the six main-run scenarios), benchmark \texttt{9bfc849}, \url{hardware/results/v0.6.0/}; the run of 2026-10-30 with \texttt{8ec73d8}, benchmark \texttt{1df9c0f}, \texttt{hardware/results/} (top level) and \texttt{hardware/report.md}; extended study (2026-11-07) benchmark \texttt{93f86f7}, \texttt{hardware/results/}\allowbreak\texttt{20261107\_p8-qz01-expanded/}.
\item Verification of outputs with more than 20 qubits (limitation (ii)): previous version, benchmark \texttt{3b3beea}, \url{paper/verification/} (re-compiled outputs, QCEC and stabilizer results, fault-injection tests); final version, benchmark \texttt{76a5256}, \texttt{paper/verification/r21/} (re-compiled outputs, QCEC, state-vector and full-unitary results, including \texttt{numeric/} for the 420 outputs with numerical instantiation).
\item Ablation of the final-round components and routing gap of the final compiler: \qs{} \texttt{be4ecda} with ablation switches (branch \texttt{paper/ablation-r21}, pull request \#17, \texttt{c82472d} and \texttt{f2c5dca}); benchmark \texttt{76a5256}, \url{compilation/results/20261108-0*_r21-ablation-*}, bit-identity evidence in \url{20261108-010401_r21-ablation-bitident}, summary \texttt{paper/ablation/r21\_summary.\{md,json\}}; routing gap \url{paper/optimality/routing/results/small_exact_r21.jsonl}.
\item Further opponents (Sec.~\ref{sec:competitors}): \qs{} \texttt{be4ecda} (previous version: \texttt{8ec73d8}), summary \texttt{paper/competitors/summary\_r21.md} (benchmark \texttt{d350f87}) and \url{compilation/results/20261108-014209_r21-final-benchpress-cz}; opponents' runs in benchmark \texttt{82f5caa}, \texttt{b7fae92}, \texttt{cae668c}; adapters in \texttt{compilation/compilers/} (\texttt{superopt\_compiler.py}, \texttt{aitp\_compiler.py}), tool builds and patches in \texttt{paper/competitors/build/}, summary and convergence traces in \texttt{paper/competitors/}, Benchpress circuits in \texttt{compilation/qasm\_benchpress/}, results in \url{compilation/results/20261106-235117_p4-bqskit-qasm-cx}, \url{20261107-*_p4-bqskit-qasm-ibm*}, \url{20261107-*_p4-{guoq,queso,quartz,aitp}-*} and \url{20261107-165740_p4-benchpress-cz}.
\end{itemize}

\appendix

\clearpage
\onecolumngrid
\section{arXiv evaluation circuits}
\label{app:arxiv}
\begin{table}[h!]
\caption{The 55 circuits of the arXiv evaluation set: arXiv identifier, circuit name (as in the data set), topic, qubits $n$, and CZ count after unoptimized translation to the CZ basis.}
\label{tab:arxivlist}
\scriptsize
\setlength{\tabcolsep}{3pt}
\begin{tabular}[t]{llcrr}
\toprule
arXiv & circuit & topic & $n$ & CZ \\
\midrule
2607.01180 & \texttt{cz02\_routed\_transpiled} & decomp. & 3 & 7 \\
2607.06230 & \texttt{full\_entangler\_pqc\_layer} & ansatz & 4 & 6 \\
2607.09976 & \texttt{vqe\_style\_diagnostic\_kernel} & ansatz & 4 & 26 \\
2607.11665 & \texttt{index\_transposition\_v3} & other & 4 & 42 \\
2607.16800 & \texttt{hea\_ryrzrx\_cnot\_ring\_n4\_d2} & ansatz & 4 & 8 \\
2607.20943 & \texttt{vqc\_ry\_rz\_cz\_full} & ansatz & 4 & 6 \\
2607.27247 & \texttt{harmonic\_frame\_povm\_naimark} & alg. & 3 & 10 \\
2607.27915 & \texttt{mvc\_constrained\_walk\_4vertex} & alg. & 4 & 56 \\
2607.28499 & \texttt{wave2d\_bright\_flux\_rotation\_Rx} & sim. & 5 & 40 \\
2607.28499 & \texttt{wave2d\_givens\_transform\_G} & sim. & 6 & 96 \\
2607.28499 & \texttt{wave2d\_phase\_transform\_P} & sim. & 8 & 104 \\
2608.03962 & \texttt{poor\_mans\_cat\_parity\_sampler} & prep. & 9 & 8 \\
2608.06570 & \texttt{qlb\_streaming\_increment\_3bit} & arith. & 3 & 7 \\
2608.09509 & \texttt{boson\_sampling\_4mode\_2photon} & sim. & 5 & 81 \\
2608.09509 & \texttt{boson\_sampling\_optimized\_1010} & sim. & 4 & 10 \\
2608.09509 & \texttt{three\_photon\_bs\_unit\_gray} & sim. & 4 & 120 \\
2608.09578 & \texttt{highest\_bit\_const\_adder\_dirty} & arith. & 7 & 28 \\
2608.09578 & \texttt{highest\_bit\_const\_adder\_uncompute} & arith. & 6 & 41 \\
2608.09578 & \texttt{mch\_dirty\_ancilla} & decomp. & 9 & 36 \\
2608.11733 & \texttt{double\_excitation\_12cx} & decomp. & 4 & 12 \\
2608.11733 & \texttt{double\_excitation\_12cx\_depth7\_relabel} & decomp. & 4 & 12 \\
2608.11733 & \texttt{double\_excitation\_c3ry} & decomp. & 4 & 26 \\
2608.17136 & \texttt{coin\_controlled\_increment\_3q} & arith. & 4 & 21 \\
2608.20443 & \texttt{quantum\_switch\_fredkin\_wiring} & alg. & 6 & 32 \\
2609.01904 & \texttt{incrementer\_shift\_3q} & arith. & 3 & 7 \\
2609.03583 & \texttt{qft5\_parity\_twine\_linear\_dcnot} & alg. & 5 & 24 \\
2609.06179 & \texttt{d2q4\_lgca\_collision} & sim. & 5 & 36 \\
2609.08233 & \texttt{global\_cz\_16} & other & 16 & 120 \\
\bottomrule
\end{tabular}
\hfill
\begin{tabular}[t]{llcrr}
\toprule
arXiv & circuit & topic & $n$ & CZ \\
\midrule
2609.16721 & \texttt{d2q5\_sbm\_amplitude\_encoding} & prep. & 3 & 20 \\
2609.21153 & \texttt{periodic\_shift\_increment\_4q} & sim. & 4 & 21 \\
2610.00547 & \texttt{faq\_running\_example\_c3} & other & 3 & 6 \\
2610.00579 & \texttt{shift\_basis\_unitary} & other & 3 & 18 \\
2610.01446 & \texttt{matchgate\_ansatz\_layer\_n4} & ansatz & 4 & 6 \\
2610.01446 & \texttt{off\_diagonal\_ansatz\_layer\_n4} & ansatz & 4 & 12 \\
2610.03686 & \texttt{aim\_onebody\_target\_action} & sim. & 6 & 6 \\
2610.03686 & \texttt{complementary\_pair\_cswap\_network\_n5} & alg. & 10 & 40 \\
2610.03686 & \texttt{cswap\_swap\_tree\_10leaf} & alg. & 14 & 72 \\
2610.04731 & \texttt{free\_abs\_value\_q3} & arith. & 7 & 6 \\
2610.04731 & \texttt{phase\_gradient\_injection\_l3} & arith. & 7 & 48 \\
2610.05086 & \texttt{jaqsi\_pulse\_benchmark\_crx\_ring\_n3} & other & 3 & 6 \\
2610.05222 & \texttt{bcfe\_bell\_part\_3q} & alg. & 6 & 7 \\
2610.06448 & \texttt{hubbard\_pair\_spin\_init} & prep. & 12 & 9 \\
2610.09141 & \texttt{cognitive\_encoder\_readout\_cry\_star} & ansatz & 6 & 10 \\
2610.12148 & \texttt{qnn\_angle\_enc\_allpairs\_crx} & ansatz & 4 & 24 \\
2610.17470 & \texttt{su3\_glueball\_interaction\_givens} & sim. & 3 & 10 \\
2610.24426 & \texttt{iqp\_embedding\_layer\_4q} & other & 4 & 6 \\
2610.25350 & \texttt{su3\_singlet\_prep} & prep. & 6 & 11 \\
2610.28349 & \texttt{beam\_splitter\_v\_even\_block} & sim. & 4 & 10 \\
2610.28349 & \texttt{beam\_splitter\_v\_odd\_block} & sim. & 4 & 8 \\
2610.28349 & \texttt{truncated\_beam\_splitter\_u1\_block} & sim. & 4 & 22 \\
2610.28993 & \texttt{all\_to\_all\_rzz\_phi2\_5q} & sim. & 5 & 20 \\
2610.30315 & \texttt{deutsch\_jozsa\_n4\_clifford\_t} & alg. & 4 & 7 \\
2610.31330 & \texttt{qsvt\_sign\_phase\_gadget} & alg. & 6 & 42 \\
2610.33188 & \texttt{hamming\_weight\_one\_hot\_counter} & arith. & 7 & 48 \\
2610.37247 & \texttt{elastic\_wave\_directional\_block\_3d} & sim. & 4 & 152 \\
\bottomrule
\end{tabular}

\end{table}

\clearpage
\section{Extended hardware study: per-scenario results}
\label{app:hwext}
\begin{table}[h!]
\caption{Extended hardware study on QZ01 (2026-11-07, \qs{} v0.6.0): CZ count and total depth averaged over seeds 0--2 (rounded), seed-averaged success metric $p$ (Hellinger fidelity for QAOA-8), and the difference of \qs{} relative to each baseline with its 95\% hierarchical-bootstrap interval over six rounds. $+$/$-$: significantly higher/lower; 0: not significant. q, Q, t: quspect O3, Qiskit O3, tket O2. Inputs a and b are as defined in Sec.~\ref{sec:hw-ext}; ``QFT adder'' is the Draper adder. Data: benchmark commit \texttt{93f86f7}.}
\label{tab:hwext}
\scriptsize
\setlength{\tabcolsep}{2.4pt}
\begin{tabular}{llccclclc}
\toprule
Graph & Scenario & CZ q/Q/t & Depth q/Q/t & $p$ q / Q / t & $\Delta$ vs Qiskit O3 [95\% CI] & & $\Delta$ vs tket O2 [95\% CI] & \\
\midrule
chain & BV-10 a & 16/30/25 & 22/60/48 & 0.186/0.000/0.040 & $+$0.185 [$+$0.155, $+$0.217] & $+$ & $+$0.146 [$+$0.126, $+$0.167] & $+$ \\
chain & BV-10 b & 12/24/14 & 18/40/28 & 0.312/0.023/0.134 & $+$0.290 [$+$0.251, $+$0.329] & $+$ & $+$0.178 [$+$0.073, $+$0.269] & $+$ \\
chain & IQFT-6 a & 40/61/45 & 48/77/53 & 0.243/0.078/0.109 & $+$0.165 [$+$0.128, $+$0.215] & $+$ & $+$0.134 [$+$0.110, $+$0.160] & $+$ \\
chain & IQFT-6 b & 40/61/45 & 48/77/53 & 0.196/0.110/0.115 & $+$0.086 [$+$0.047, $+$0.131] & $+$ & $+$0.081 [$+$0.051, $+$0.100] & $+$ \\
chain & Adder-8 a & 65/82/71 & 115/139/125 & 0.042/0.037/0.021 & $+$0.005 [$-$0.008, $+$0.017] & 0 & $+$0.021 [$+$0.014, $+$0.030] & $+$ \\
chain & Adder-8 b & 65/82/71 & 115/139/125 & 0.033/0.039/0.013 & $-$0.006 [$-$0.025, $+$0.012] & 0 & $+$0.020 [$+$0.015, $+$0.026] & $+$ \\
chain & QAOA-8 a & 41/51/59 & 31/43/67 & 0.746/0.650/0.660 & $+$0.095 [$+$0.019, $+$0.188] & $+$ & $+$0.086 [$+$0.047, $+$0.122] & $+$ \\
chain & QAOA-8 b & 41/57/63 & 42/61/82 & 0.687/0.632/0.641 & $+$0.055 [$+$0.001, $+$0.095] & $+$ & $+$0.046 [$-$0.015, $+$0.102] & 0 \\
chain & GHZ-10 a & 12/20/16 & 13/21/21 & 0.248/0.247/0.222 & $+$0.001 [$-$0.030, $+$0.024] & 0 & $+$0.026 [$-$0.027, $+$0.078] & 0 \\
chain & GHZ-10 b & 18/22/42 & 29/31/51 & 0.184/0.141/0.047 & $+$0.043 [$+$0.025, $+$0.065] & $+$ & $+$0.137 [$+$0.044, $+$0.234] & $+$ \\
chain & Grover-3 a & 29/45/31 & 58/84/62 & 0.424/0.236/0.440 & $+$0.188 [$+$0.168, $+$0.209] & $+$ & $-$0.016 [$-$0.124, $+$0.092] & 0 \\
chain & Grover-3 b & 29/45/31 & 59/84/62 & 0.326/0.236/0.314 & $+$0.090 [$+$0.049, $+$0.128] & $+$ & $+$0.012 [$-$0.093, $+$0.115] & 0 \\
chain & QFT adder a & 31/45/48 & 40/63/73 & 0.425/0.371/0.266 & $+$0.054 [$-$0.064, $+$0.167] & 0 & $+$0.159 [$+$0.093, $+$0.219] & $+$ \\
chain & QFT adder b & 31/45/48 & 40/63/73 & 0.384/0.305/0.224 & $+$0.079 [$+$0.043, $+$0.112] & $+$ & $+$0.160 [$+$0.104, $+$0.226] & $+$ \\
chain & W-8 & 20/21/20 & 19/21/19 & 0.540/0.371/0.476 & $+$0.168 [$+$0.128, $+$0.208] & $+$ & $+$0.064 [$+$0.030, $+$0.095] & $+$ \\
chain & W-7 & 15/15/15 & 19/19/19 & 0.650/0.573/0.639 & $+$0.077 [$+$0.021, $+$0.137] & $+$ & $+$0.011 [$-$0.052, $+$0.107] & 0 \\
chain & Hidden shift a & 33/42/59 & 48/52/83 & 0.024/0.057/0.033 & $-$0.033 [$-$0.062, $-$0.009] & $-$ & $-$0.009 [$-$0.015, $-$0.001] & $-$ \\
chain & Hidden shift b & 33/42/59 & 49/53/83 & 0.023/0.063/0.034 & $-$0.040 [$-$0.084, $-$0.002] & $-$ & $-$0.011 [$-$0.017, $-$0.005] & $-$ \\
\midrule
grid & BV-10 a & 12/16/24 & 20/26/31 & 0.032/0.013/0.004 & $+$0.019 [$+$0.009, $+$0.030] & $+$ & $+$0.028 [$+$0.016, $+$0.038] & $+$ \\
grid & BV-10 b & 9/10/16 & 15/16/27 & 0.224/0.175/0.072 & $+$0.049 [$+$0.012, $+$0.087] & $+$ & $+$0.153 [$+$0.125, $+$0.183] & $+$ \\
grid & IQFT-6 a & 35/45/51 & 49/70/88 & 0.239/0.096/0.071 & $+$0.143 [$+$0.110, $+$0.173] & $+$ & $+$0.168 [$+$0.155, $+$0.184] & $+$ \\
grid & IQFT-6 b & 35/45/51 & 49/70/88 & 0.240/0.131/0.089 & $+$0.109 [$+$0.063, $+$0.150] & $+$ & $+$0.151 [$+$0.085, $+$0.208] & $+$ \\
grid & Adder-8 a & 58/74/72 & 106/118/114 & 0.028/0.029/0.015 & 0.000 [$-$0.022, $+$0.013] & 0 & $+$0.013 [$+$0.006, $+$0.022] & $+$ \\
grid & Adder-8 b & 58/74/72 & 106/118/114 & 0.046/0.050/0.022 & $-$0.004 [$-$0.029, $+$0.018] & 0 & $+$0.023 [$+$0.014, $+$0.034] & $+$ \\
grid & QAOA-8 a & 29/33/37 & 24/38/41 & 0.821/0.712/0.682 & $+$0.110 [$+$0.059, $+$0.181] & $+$ & $+$0.139 [$+$0.071, $+$0.225] & $+$ \\
grid & QAOA-8 b & 28/33/37 & 26/40/36 & 0.832/0.683/0.760 & $+$0.150 [$+$0.073, $+$0.236] & $+$ & $+$0.072 [$+$0.047, $+$0.119] & $+$ \\
grid & GHZ-10 a & 9/9/9 & 11/12/12 & 0.294/0.286/0.268 & $+$0.008 [$-$0.001, $+$0.018] & 0 & $+$0.026 [$-$0.040, $+$0.097] & 0 \\
grid & GHZ-10 b & 10/12/15 & 13/16/19 & 0.293/0.190/0.244 & $+$0.103 [$+$0.060, $+$0.150] & $+$ & $+$0.049 [$+$0.022, $+$0.083] & $+$ \\
grid & Grover-3 a & 29/45/35 & 58/82/65 & 0.394/0.244/0.422 & $+$0.150 [$+$0.105, $+$0.208] & $+$ & $-$0.028 [$-$0.165, $+$0.126] & 0 \\
grid & Grover-3 b & 29/45/35 & 59/82/65 & 0.280/0.225/0.416 & $+$0.055 [$-$0.003, $+$0.108] & 0 & $-$0.136 [$-$0.209, $-$0.067] & $-$ \\
grid & QFT adder a & 26/33/36 & 37/59/55 & 0.461/0.354/0.203 & $+$0.107 [$+$0.062, $+$0.163] & $+$ & $+$0.258 [$+$0.207, $+$0.295] & $+$ \\
grid & QFT adder b & 26/33/36 & 37/59/55 & 0.389/0.301/0.230 & $+$0.088 [$+$0.038, $+$0.137] & $+$ & $+$0.159 [$+$0.086, $+$0.237] & $+$ \\
grid & W-8 & 14/14/14 & 13/13/13 & 0.529/0.387/0.411 & $+$0.142 [$+$0.079, $+$0.214] & $+$ & $+$0.119 [$+$0.073, $+$0.156] & $+$ \\
grid & W-7 & 12/12/12 & 12/13/13 & 0.610/0.381/0.461 & $+$0.230 [$+$0.171, $+$0.283] & $+$ & $+$0.149 [$+$0.095, $+$0.205] & $+$ \\
grid & Hidden shift a & 22/28/27 & 37/43/41 & 0.060/0.044/0.189 & $+$0.016 [$-$0.016, $+$0.052] & 0 & $-$0.129 [$-$0.235, $-$0.023] & $-$ \\
grid & Hidden shift b & 22/28/27 & 37/43/41 & 0.066/0.033/0.172 & $+$0.034 [$+$0.009, $+$0.062] & $+$ & $-$0.105 [$-$0.222, $+$0.007] & 0 \\
\bottomrule
\end{tabular}

\end{table}

\begin{table}[h!]
\caption{Re-run of the extended study on QZ01 with the final circuits (2026-11-08, \qs{} \texttt{be4ecda}, unchanged baseline circuits) for the 16 scenarios of Sec.~\ref{sec:hw-ext}; these rows supersede the corresponding rows of Table~\ref{tab:hwext} for the final compiler. CZ count for \qs{} (averaged over seeds 0--2 for the v0.6.0 and the final circuits) and for the baselines; seed-averaged success metric $p$ for this run (Hellinger fidelity for QAOA-8); difference of \qs{} relative to each baseline with its 95\% hierarchical-bootstrap interval over six rounds. Notation as in Table~\ref{tab:hwext}. Data: benchmark commit \texttt{8a181b7}, \texttt{hardware/results/20261108\_r21-p8-subset/}.}
\label{tab:hwextfinal}
\scriptsize
\setlength{\tabcolsep}{2.4pt}
\begin{tabular}{llcclclc}
\toprule
Graph & Scenario & CZ q (v0.6.0$\to$final)/Q/t & $p$ q / Q / t & $\Delta$ vs Qiskit O3 [95\% CI] & & $\Delta$ vs tket O2 [95\% CI] & \\
\midrule
chain & Grover-3 a & 29$\to$9/45/31 & 0.709/0.376/0.539 & $+$0.333 [$+$0.303, $+$0.368] & $+$ & $+$0.170 [$+$0.123, $+$0.230] & $+$ \\
chain & Grover-3 b & 29$\to$9/45/31 & 0.760/0.375/0.513 & $+$0.385 [$+$0.354, $+$0.416] & $+$ & $+$0.247 [$+$0.192, $+$0.306] & $+$ \\
chain & Adder-8 a & 65$\to$45/82/71 & 0.153/0.051/0.020 & $+$0.102 [$+$0.072, $+$0.136] & $+$ & $+$0.133 [$+$0.103, $+$0.171] & $+$ \\
chain & Adder-8 b & 65$\to$45/82/71 & 0.120/0.061/0.015 & $+$0.059 [$+$0.029, $+$0.088] & $+$ & $+$0.105 [$+$0.075, $+$0.140] & $+$ \\
chain & QAOA-8 a & 41$\to$33/51/59 & 0.764/0.755/0.628 & $+$0.009 [$-$0.008, $+$0.027] & 0 & $+$0.136 [$+$0.121, $+$0.152] & $+$ \\
chain & Hidden shift a & 33$\to$28/42/59 & 0.335/0.164/0.063 & $+$0.171 [$+$0.109, $+$0.223] & $+$ & $+$0.272 [$+$0.199, $+$0.318] & $+$ \\
chain & Hidden shift b & 33$\to$26/42/59 & 0.471/0.180/0.102 & $+$0.291 [$+$0.228, $+$0.350] & $+$ & $+$0.369 [$+$0.342, $+$0.398] & $+$ \\
grid & Grover-3 a & 29$\to$10/45/35 & 0.774/0.340/0.536 & $+$0.433 [$+$0.420, $+$0.448] & $+$ & $+$0.237 [$+$0.208, $+$0.265] & $+$ \\
grid & Grover-3 b & 29$\to$9/45/35 & 0.764/0.284/0.488 & $+$0.480 [$+$0.453, $+$0.504] & $+$ & $+$0.276 [$+$0.252, $+$0.302] & $+$ \\
grid & Adder-8 a & 58$\to$40/74/72 & 0.064/0.092/0.020 & $-$0.028 [$-$0.039, $-$0.017] & $-$ & $+$0.044 [$+$0.033, $+$0.058] & $+$ \\
grid & Adder-8 b & 58$\to$40/74/72 & 0.071/0.085/0.014 & $-$0.014 [$-$0.021, $-$0.007] & $-$ & $+$0.057 [$+$0.049, $+$0.065] & $+$ \\
grid & QAOA-8 a & 29$\to$25/33/37 & 0.831/0.849/0.793 & $-$0.019 [$-$0.034, $-$0.008] & $-$ & $+$0.038 [$+$0.021, $+$0.057] & $+$ \\
grid & QAOA-8 b & 28$\to$24/33/37 & 0.873/0.853/0.842 & $+$0.020 [$+$0.007, $+$0.038] & $+$ & $+$0.031 [$+$0.014, $+$0.050] & $+$ \\
grid & QFT adder a & 26$\to$22/33/36 & 0.628/0.514/0.302 & $+$0.114 [$+$0.095, $+$0.129] & $+$ & $+$0.326 [$+$0.305, $+$0.353] & $+$ \\
grid & QFT adder b & 26$\to$21/33/36 & 0.514/0.482/0.344 & $+$0.032 [$+$0.007, $+$0.059] & $+$ & $+$0.171 [$+$0.119, $+$0.242] & $+$ \\
grid & Hidden shift b & 22$\to$19/28/27 & 0.383/0.216/0.367 & $+$0.167 [$+$0.122, $+$0.206] & $+$ & $+$0.016 [$-$0.042, $+$0.071] & 0 \\
\bottomrule
\end{tabular}

\end{table}

\end{document}